\documentclass[sigconf]{acmart}
\usepackage[english]{babel}

\renewcommand\footnotetextcopyrightpermission[1]{} % removes footnote with conference info
\setcopyright{none}
\acmDOI{}

\acmISBN{}

\acmPrice{}

\usepackage[linesnumbered,ruled,vlined]{algorithm2e}
\usepackage[noend]{algpseudocode}
\usepackage{array}
\newcolumntype{x}[1]{>{\centering\arraybackslash\hspace{0pt}}p{#1}}
\usepackage{amsmath}
\usepackage{amssymb}
\usepackage[inline]{enumitem}
\usepackage{epsfig}
\usepackage{fancyvrb}
\usepackage{graphicx}
\usepackage{booktabs, tabularx}
\usepackage[flushleft]{threeparttable}
\usepackage{latexsym, color}
\usepackage{listings}
\usepackage{tikz}
\usepackage{url}
\usepackage[font=normalsize,labelfont=bf]{caption}
\usepackage{hyperref}
\usepackage{cleveref}
\usepackage{tabularx}
\usepackage{subcaption}
\usepackage{xcolor}
\usepackage{multirow}
\usepackage{multicol}
\usepackage{float}
\newcommand{\var}[1]{\text{\textit{#1}}}

\crefformat{figure}{Figure~#2#1#3}
\SetKwProg{Fn}{Function}{}{end}
\SetFuncSty{textsc}

\definecolor{commentgreen}{RGB}{2,112,10}
\definecolor{eminence}{RGB}{108,48,130}
\definecolor{weborange}{RGB}{255,165,0}
\definecolor{frenchplum}{RGB}{129,20,83}

\long\def\comment#1{}
\newtheorem{theorem}{\textbf{Theorem}}
\newtheorem{definition}{\textbf{Definition}}
\newtheorem{proposition}{\textbf{Proposition}}

\newtheorem{property}{\textbf{Property}}
\usepackage{cuted}
\usepackage{float}
\usepackage{xspace}
\newcommand{\ie}{{\em i.e.}}
\newcommand{\eg}{{\em e.g.}}

\newcommand{\system}{Coral}
\newcommand{\dvnet}{{\em DVNet}}
\newcommand{\lnet}{NGClos}

\newcommand{\para}[1]{\noindent {\bf #1}}%remove \smallskip

\newcommand{\qiao}[1]{\textcolor{red}{[qiao: #1]}}

\newcommand{\dduan}[1]{\textcolor{orange}{[dduan: #1]}}

\newcommand{\squishlist}{
   \begin{list}{$\bullet$}
    { \setlength{\itemsep}{0pt}      \setlength{\parsep}{3pt}
      \setlength{\topsep}{3pt}       \setlength{\partopsep}{0pt}
      \setlength{\leftmargin}{3.5mm} \setlength{\labelwidth}{1em}
      \setlength{\labelsep}{0.5em} } }

\newcommand{\squishlisttwo}{
   \begin{list}{$\bullet$}
    { \setlength{\itemsep}{0pt}    \setlength{\parsep}{0pt}
      \setlength{\topsep}{0pt}     \setlength{\partopsep}{0pt}
      \setlength{\leftmargin}{2em} \setlength{\labelwidth}{1.5em}
      \setlength{\labelsep}{0.5em} } }

\newcommand{\squishend}{
    \end{list}  }

\usepackage{array}
\usepackage{setspace}

\begin{document}

\title[Distributed, On-Device Data Plane Verification]{\huge
	%Toward Optimal End-to-End Interdomain Route Control as a Service
	%Toward An Elastic Framework for Data Plane Verification
	Switch as a Verifier: Toward Scalable Data Plane Checking via Distributed, On-Device Verification
}

%\numberofauthors{1}
\author{Qiao Xiang}
\affiliation{
	\institution{Xiamen University}
}
\author{Ridi Wen}
\affiliation{
	\institution{Xiamen University}
}
\author{Chenyang Huang}
\affiliation{
	\institution{Xiamen University}
}
\author{Yuxin Wang}
\affiliation{
	\institution{Xiamen University}
}
% \author{Jiwu Shu}
% \affiliation{
% 	\institution{Xiamen University}
% }
\author{Franck Le}
\affiliation{
	\institution{IBM}
}

%\author{
  %  To appear in INFOCOM 2020\\
%\large	Paper ID: \#310,
%	Number of Pages: 12 pages + References (3 pages) + Appendices (5 pages)
%}%%%end of a

%\renewcommand{\shortauthors}{Anonymous authors}
\renewcommand{\shortauthors}{Xiang et al.}

%\begin{spacing}{1}
\begin{abstract}

	Data plane verification (DPV) is important for finding network
	errors. Current DPV
	tools
employ a centralized architecture,
	where a server collects the data planes of all devices and verifies them.
	Despite substantial efforts on accelerating DPV, this centralized
	architecture is inherently unscalable. 
	In this paper, to tackle the scalability challenge of DPV, we circumvent
	the scalability bottleneck of centralized design and
	design \system{}, a distributed, on-device DPV framework.  
	The key
	insight of \system{} is that DPV 
	can be transformed into a counting problem on a directed acyclic graph,
	which can be naturally decomposed into lightweight tasks executed
	at network devices, enabling scalability.  
	\system{} consists of (1) a declarative requirement specification language,  
	(2) a planner that employs a novel data structure \dvnet{} to
	systematically decompose global verification into on-device counting tasks,
	%constructs a novel data structure
	%\dvnet{} to compactly represent all valid paths in the network, and uses
	%it to systematically decompose a global counting into smaller
	%counting tasks executed at devices, 
	and (3) a distributed verification
	(DV) protocol that specifies how on-device verifiers communicate task
	results efficiently to collaboratively verify the requirements. 
	We implement a
	prototype of \system{}. Extensive experiments with real-world datasets
	(WAN/LAN/DC)
	show that \system{}
consistently achieves scalable DPV under various networks and DPV scenarios,
	\ie, up to $1250\times$ speed up in the scenario of burst update, and
	up to 202$\times$ speed up on 80\% quantile of incremental
	verification,  
\iffalse{
	and conduct extensive experiments with real
	world datasets. Results show that for a real, large operational network,
	\system{} achieves at least
	$1.25$K$\times$ speed up, in the scenario of burst update, and 
at least $16.42\times$ speed up on 80\%
	quantile of
	incremental verification, 
}\fi
	than state-of-the-art DPV tools, 
	with little overhead on commodity network devices.\footnote{If you are interested in learning more about \system{}, please contact Qiao Xiang (xiangq27@gmail.com).} 
\end{abstract}

%\iffalse{
% One exception is RCDC, in which each switch performs local verification, but it
% can only verify a very specific requirement.}
%\fi

\maketitle

\vspace{-1em}
\section{Introduction}\label{sec:intro}
Network errors such as forwarding loops, undesired blackholes and waypoint violations are
the outcome of various issues (\eg, software bugs, faulty hardware, protocol
misconfigurations, and oversight negligence). They can happen in all types of
networks (\eg, enterprise networks, wide area networks and data center
networks), and have both disastrous financial and social
consequences~\cite{amazon-aug-2019, amazon-nov-2020, google-dec-2020,
google-aug-2021, facebook-2021}. How to detect and prevent such errors efficiently is a fundamental challenge
for the network community.
A major advance for this problem has been network verification, which
analyzes the control plane~\cite{bagpipe, batfish, minesweeper,
ali-sigcomm20, era-osdi16, bonsai, shapeshifter, gember2016fast, arc-sosp17,
tiramisu, plankton-nsdi20, netdice-sigcomm20, nv-pldi20, spider, groot,
rcc-nsdi05, fsr, ball2014vericon} and data plane~\cite{xie2005static, flowchecker,
anteater, hsa, libra, symmetry, ap-icnp13, ap-ton16, practical-ap-conext15,
practical-ap-ton17, scalable-ap-ton17, netkat-popl14, veriflow, deltanet-nsdi17,
apkeep-nsdi20, azure} of network devices to identify errors.

%This paper focuses on data plane verification (DPV), because it can find a wider
%range of network errors by checking the actual data plane at network devices
%(\eg, switch OS bugs). There has been a long line of research on DPV. Earlier tools focus on verifying
%a snapshot of the complete data plane of the network~\cite{xie2005static,
%flowchecker, anteater, hsa, libra, symmetry, nod, ap-icnp13, ap-ton16,
%practical-ap-conext15, practical-ap-ton17, scalable-ap-ton17, netkat-popl14}.
%Recent studies focus on incremental verification (\ie, verifying forwarding rule
%updates)~\cite{netplumber, veriflow, deltanet-nsdi17, apkeep-nsdi20, azure}.  A
%state-of-the-art DPV tool~\cite{apkeep-nsdi20} has reported to achieve an
%incremental verification time of tens of microseconds per rule update. 

There has been a long line of research on data plane verification (DPV), because this approach can detect a wider range of network errors by checking the actual data plane at the network devices (\eg, switch OS bugs). More specifically, earlier tools analyzed
a snapshot of the complete data plane of the network~\cite{xie2005static,
flowchecker, anteater, hsa, libra, symmetry, nod, ap-icnp13, ap-ton16,
practical-ap-conext15, practical-ap-ton17, scalable-ap-ton17, netkat-popl14}; and more
recent solutions focus on incremental verification (\ie, verifying forwarding rule
updates)~\cite{netplumber, veriflow, deltanet-nsdi17, apkeep-nsdi20, azure}. State-of-the-art DPV tools (e.g., ~\cite{apkeep-nsdi20}) can achieve
incremental verification times of tens of microseconds per rule update. 

Despite the substantial progress in accelerating DPV, existing tools employ
a centralized architecture, which lacks the scalability needed for deployment in
large networks. Specifically, they use a centralized server to collect the data
plane from each network device and verify the requirement. Such a design
requires a management network to provide reliable connections between the server
and network devices, which is hard to build itself. Moreover, the server becomes
the performance bottleneck and the single point of failure of DPV tools. To
scale up DPV, Libra~\cite{libra} partitions the data plane into disjoint packet
spaces and uses MapReduce to achieve parallel verification in a cluster; Azure
RCDC~\cite{azure} partitions the data plane by device to verify the availability
of all shortest paths with a higher level of parallelization in a cluster.
However, both are still centralized designs with the limitations above, and RCDC
can only verify that particular requirement.

\iffalse{
Azure~\cite{azure} deploys a local verification system for devices to verify its
own data plane.  However, the former is still a centralized design with the
limitations above, and the latter can only verify the particular requirement of
the availability and fault tolerance of shortest paths.
}\fi

In this paper, we systematically tackle the important problem of how to scale
the data plane verification to be applicable in real, large networks. Not only
can a scalable DPV tool quickly find network errors in large networks, it can
also support novel services such as fast rollback and switching among multiple
data planes~\cite{vissicchio2015co, fboss, le2010theory}, and data plane
verification across administrative domains~\cite{cove, dhamdhere2018inferring}.

\iffalse{
To this end, instead of continuing squeezing the space of performance improvement of
centralized DPV, we embrace a distributed design to circumvent the inherent scalability
bottleneck of centralized designs, and develop a generic, efficient distributed data plane
verification (\system{}) framework, to achieve scalable data plane checkups on a wide
range of requirements (\eg, reachability, waypoint, multicast and anycast). 
}\fi

To this end, instead of continuing to squeeze incremental performance
improvements out of centralized DPV, we embrace a distributed design to
circumvent the inherent scalability bottleneck of centralized design.
Azure~\cite{azure} takes a first step along this direction by partitioning
verification into local contracts of devices.  It gives an interesting analogy
between such local contracts and program verification using annotation with
inductive loop invariants, but stops at designing communication-free local
contracts for the particular all-shortest-path availability requirement and
validating them in parallel on a centralized cluster.  In contrast, we go beyond
this point and show that for a wide range of requirements (\eg, reachability,
waypoint, multicast and anycast), with lightweight tasks running on commodity
network devices and limited communication among them, we can verify these
requirements in a compositional way, achieving scalable data plane checkups in
generic settings. 

\begin{figure}[t]
\setlength{\abovecaptionskip}{0cm}
\setlength{\belowcaptionskip}{-0.cm}
\centering
\includegraphics[width=0.9\columnwidth]{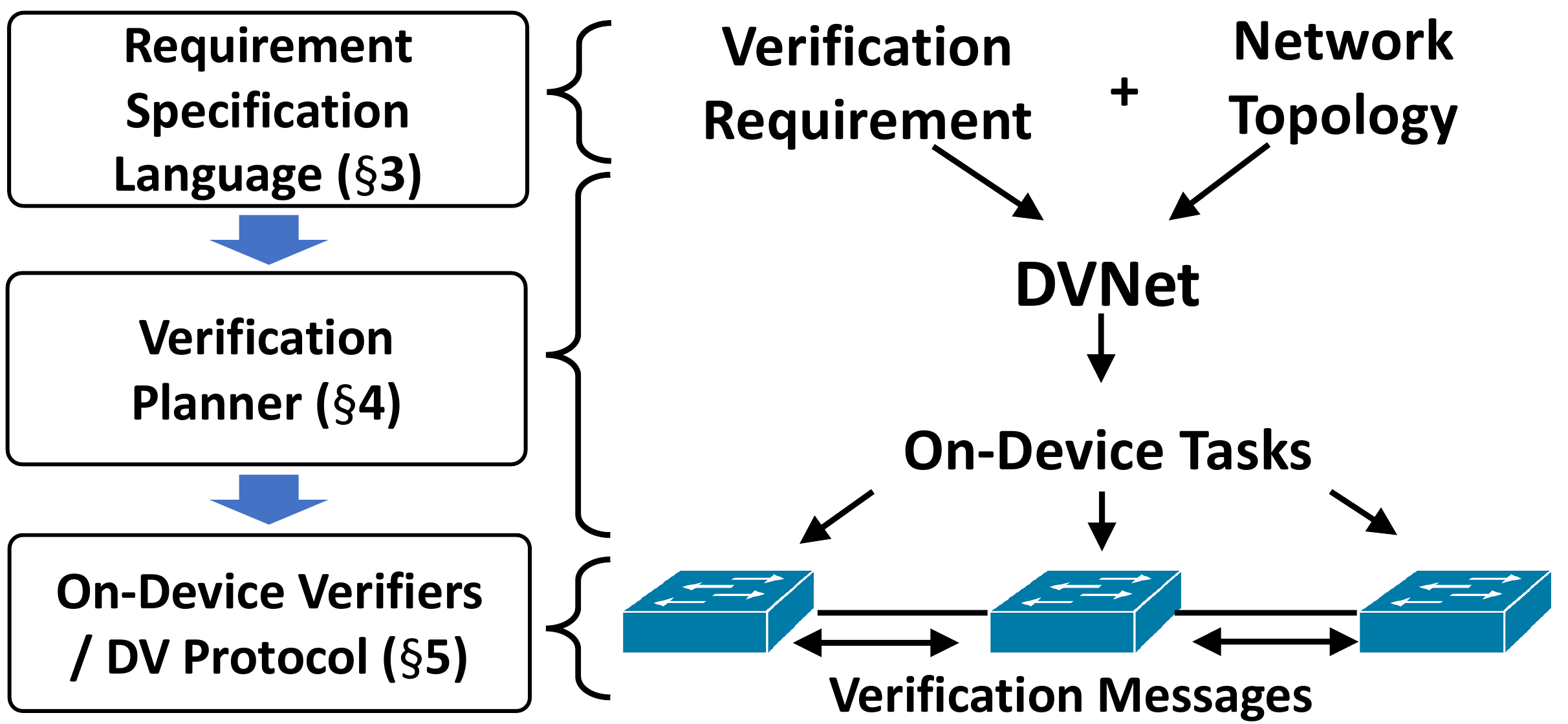}
\caption{The architecture and workflow of \system{}.}
\label{fig:arch}
%\reducespace
%\vspace{-1em}
\end{figure}

To be concrete, we design \system{}, a generic, distributed, on-device DPV
framework with a key insight: the problem of DPV can be transformed
into a counting problem in a directed acyclic graph (DAG)
representing all valid paths in the network;
the latter can then be decomposed into small tasks at nodes on the DAG, which
can be distributively executed at corresponding network devices, enabling scalability.
Figure~\ref{fig:arch} gives the architecture and basic workflow of \system{},
which consists of three novel components:

%\para{A declarative language for specifying flexible verification requirements.}
\para{A declarative requirement specification language (\S\ref{sec:lang}).}  
This language abstracts a requirement as a tuple of packet space, ingress
devices and behavior, where a behavior is a predicate on whether the paths of
packets match a pattern specified in a regular expression.  This design allows
operators to flexibly specify common requirements studied by existing DPV tools
(\eg, reachability, blackhole free, waypoint and isolation), and more advanced,
yet understudied requirements (\eg, multicast, anycast, no-redundant-delivery
and all-shortest-path availability).

\iffalse{
This language abstracts a requirement as a tuple of packet space, ingress
devices, regular expression based path pattern and requirement context.  It
allows operators to flexibly specify two classes of network requirements:
existence requirements that check whether packets are delivered to intended
destinations along a certain number of paths of certain patterns (\eg,
reachability, waypoint, multicast, anycast and 1+1 routing), and equivalence
requirements that check all paths of certain patterns in the network should be
available (\eg, shortest path reachability and fault tolerance~\cite{azure}). 
}\fi

%\para{A verification planner for deciding on-device verification tasks.} 
\para{A verification planner (\S\ref{sec:dvnet}).} 
Given a requirement, the planner decides the tasks to be executed on devices to
verify it. The core challenge is how to make these tasks lightweight,
because commodity network devices have little computation power to spare.
To this end, the planner first uses the requirement and the network topology
to compute a novel data structure called \dvnet{}, a DAG compactly representing
all paths in the network that satisfies the path patterns in the requirement. It
then transforms the DPV problem into a counting problem on \dvnet{}.
The latter can be solved by a reverse topological traversal along \dvnet{}.
In its turn, each node in \dvnet{} takes as input the data plane of its
corresponding device and the counting results of its downstream nodes to compute
for different packets, how many copies of them can be delivered to the intended
destinations along downstream paths in \dvnet{}. This traversal can be naturally
decomposed to on-device counting tasks, one for each node in \dvnet{}, and
distributed to the corresponding network devices by the planner.  
\iffalse{
and decomposes the latter into small on-device counting
tasks. A task at a node in \dvnet{} takes in the data plane of its residing device and the counting
results of tasks on its downstream nodes in \dvnet{} to compute, for different
packets, how many copies of them can be delivered to the intended destinations
along valid paths in the network. 
}\fi
We also design optimizations to compute the minimal counting information of each
node in \dvnet{} to send to its upstream neighbors, and prove that for
requirements such as all-shortest-path availability, their minimal counting
information is an empty set, making
the local contracts in RCDC~\cite{azure} a special case of \system{}. 

\para{On-device verifiers equipped with a DV protocol 
(\S\ref{sec:dvp}).}
On-device verifiers execute the on-device counting tasks specified by the
planner, and share their results with neighbor devices to collaboratively verify the
requirements. In particular, we are inspired by vector-based routing protocols
~\cite{RIP, BGP} to design a DV protocol that specifies how neighboring
on-device verifiers communicate counting results in an efficient, correct way.

\para{Experiment results (\S\ref{sec:eval}).} 
We implement a prototype of \system{} and will open-source it upon the
publication of this paper. We evaluate it extensively using real-world
datasets, in both testbed and simulations.  \system{} consistently outperforms
state-of-the-art DPV tools under various networks (WAN/LAN/DC) and DPV scenarios, \ie, up to
$1250\times$ speed up in the scenario of burst update, and up to
$202\times$ speed up on 80\% quantile of incremental verification, with
little overhead on commodity network devices.
\iffalse{
For a large operational network, \system{} achieves at least $1.25$K$\times$
than state-of-the-art DPV tools when a large number of rule updates are
installed all at once, and at least $16.24\times$ speed up on 80\% quantile of
incremental verification, while incurring small overhead on commodity network
devices. 
speed up on verification time 
than
APKeep, the state-of-the-art centralized DPV tool, when a large number of data plane updates
arrive in a short time, and a 90\% quantile $13\times$ speed up on incremental
verification, while incurring small overhead on devices.
}\fi
%addition, the incurred computation and communication overhead at devices are reasonably small.
\system{} achieves scalable DPV for two reasons.  First, by decomposing
verification into lightweight tasks executed on devices, the
performance of \system{} achieves a scalability approximately linear to the network diameter. 
Second, when a data plane update happens, only devices whose
counting results may change need to incrementally update their results, and
send them to needed neighbors incrementally. As such, its
verification time can be substantially shortened in large networks.

\para{Updates from v2.}
Compared with the previous version, we have redone the incremental update experiment on the datacenter topologies of the two data plane verification tools, APKeep and AP, and updated the experimental results here. The reasons are as follows. For APKeep, in our previous incremental update experiment on datacenter topologies, we mistakenly used only 1000 rule updates, but 10000 for other tools. We correct this mistake to make the results fair and consistent, by using 10,000 rule updates in the experiments of APKeep. For the same experiments on the datacenter topologies of AP, we accidently misconfigured the rules in one switch, making the datasets inconsistent with other tools.
%\newpage

\section{Overview}\label{sec:overview}
This section introduces some key concepts in \system{}, and illustrates its basic
workflow using an example.

\subsection{Basic Concepts}\label{sec:basic-concepts}
\para{Data plane model.} 
For ease of exposition, given a network device, we model its data plane as a
match-action table.
\iffalse{
For ease of exposition, we explain the design of \system{} assuming the data
plane of each network device is a match-action table. We discuss how \system{}
works with devices with a pipeline of match-action tables
~\cite{kohler2000click, bosshart2014p4, apkeep-nsdi20} in Section~\ref{sec:ext}.
}\fi
Entries in the table are ordered in descending
priority. Each entry has a match field to match packets on packet headers (\eg,
TCP/IP 5-tuple) and an action field to process packets. Possible actions include
modifying the headers of the packet and forwarding the packet to a \emph{group}
of next-hops~\cite{azure, openflow1-5}. An empty group means the action is to
drop the packet. If an action forwards the packet to all next-hops in a
non-empty group, we call it an $ALL$-type action. If it forwards the packet to
one of the next-hops in a non-empty group, we call it an $ANY$-type action.
Given an $ANY$-type action, we do not assume any knowledge on how the device
selects one next-hop from the group. This is because this selection algorithm is
vendor-specific, and sometimes a blackbox~\cite{openflow1-5}.

\para{Packet traces and universes.} 
Inspired by NetKAT~\cite{netkat-popl14}, we introduce the concept of packet
trace to record the state of a packet as it travels from device to device, and
use it to define the network behavior of forwarding packets. When $p$ enters a
network from an ingress device $S$, a \textit{packet trace} of $p$ is defined 
as a non-empty \textit{sequence} of devices visited by $p$ until it is delivered
to the destination device or dropped. 

\begin{figure}[t]
\setlength{\abovecaptionskip}{0.cm}
\setlength{\belowcaptionskip}{-0.cm}
\centering
\includegraphics[width=0.9\columnwidth]{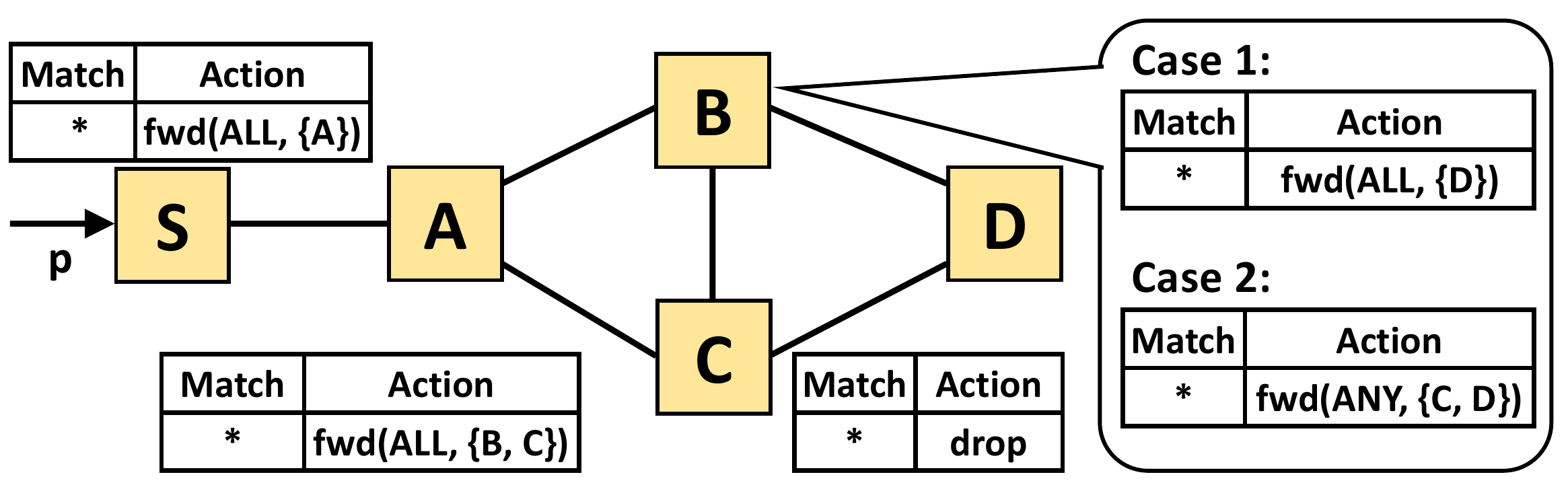}
    \caption{An example to demonstrate the data plane model, packet traces and universes. 
	Case 1: $p$ has 1 universe of 2 traces,
			$[S, A, B, D]$ and $[S, A, C]$. Case 2: $p$ has 2
			universes of 2 traces: $\{[S, A, B, C], [S, A, C]\}$ and
			$\{[S, A, B, D], [S, A, C]\}$.}
\label{fig:model-example}
%\reducespace
%\vspace{-1em}
\end{figure}

However, due to ALL-type actions, a packet may not be limited to one packet
trace each time it enters a network. For example, in
Figure~\ref{fig:model-example}, Case 1,  
%suppose the action of $B$ is $fwd(ALL, \{D\})$,
the network forwards $p$ along a set of two traces $\{[S, A, B, D], [S, A,
C]\}$ because $A$ forwards $p$ to both $B$ and $C$. We denote this set to be a
\textit{universe} of packet $p$ from ingress $S$. In addition, with the
existence of ANY-type actions, a packet may traverse one of a number of
different \textit{sets} of packet traces (universes) each time it enters a
network. For example, in Figure~\ref{fig:model-example}, if the action of $B$ is
changed  to $fwd(ANY, \{C, D\})$, when $p$ enters the network in different
instances, the network may forward $p$ according to the universe $\{[S, A, B,
C], [S, A, C]\}$ or the universe $\{[S, A, B, D], [S, A, C]\}$, because $B$
forwards $p$ to either $C$ or $D$. These universes (each being a set of traces)
can be thought of as a "multiverse" - should the packet enter the network
multiple times, it may experience different fates each time.

\iffalse{
However, due to ALL-type actions, a packet may not be limited to one packet
trace each time it enters a network. For example, in Figure~\ref{fig:fib-all},
the network forwards $p$ along a set of two traces $\{[S, A, B, D], [S, A, C]\}$
because $A$ forwards $p$ to both $B$ and $C$. We denote this set to be a
\textit{universe} of packet $p$ from ingress $S$. In addition, with the
existence of ANY-type actions, a packet may traverse one of a number of
different \textit{sets} of packet traces (universes) each time it enters a
network. For example, in Figure~\ref{fig:fib-any}, when $p$ enters the network
in different instances, the network may forward $p$ according to the universe
$\{[S, A, B, D]\}$ or the universe $\{[S, A, C]\}$, because $A$ forwards $p$ to
either $B$ or $C$. These universes (each being a set of traces) can be thought
of as a "multiverse" - should the packet enter the network multiple times, it
may face experience different fates each time. To continue our example, in
Figure~\ref{fig:fib-universe}, $p$ has 2 universes: $\{[S, A, B, C], [S, A,
C]\}$ and $\{[S, A, B, D], [S, A, C]\}$
}\fi

The notion of universes is a foundation of \system{}. We are inspired by
multipath consistency~\cite{batfish}, a property where a packet is either
accepted on all paths or dropped on all paths, but go beyond. In particular, for
each requirement, we verify whether it is satisfied across all universes.

\begin{figure*}[h]
\setlength{\abovecaptionskip}{0.1cm}
\setlength{\belowcaptionskip}{-0.cm}
    \centering
	\begin{subfigure}[t]{0.36\linewidth}
\setlength{\abovecaptionskip}{0.1cm}
\setlength{\belowcaptionskip}{-0.cm}
	    \centering\includegraphics[width=0.95\linewidth]{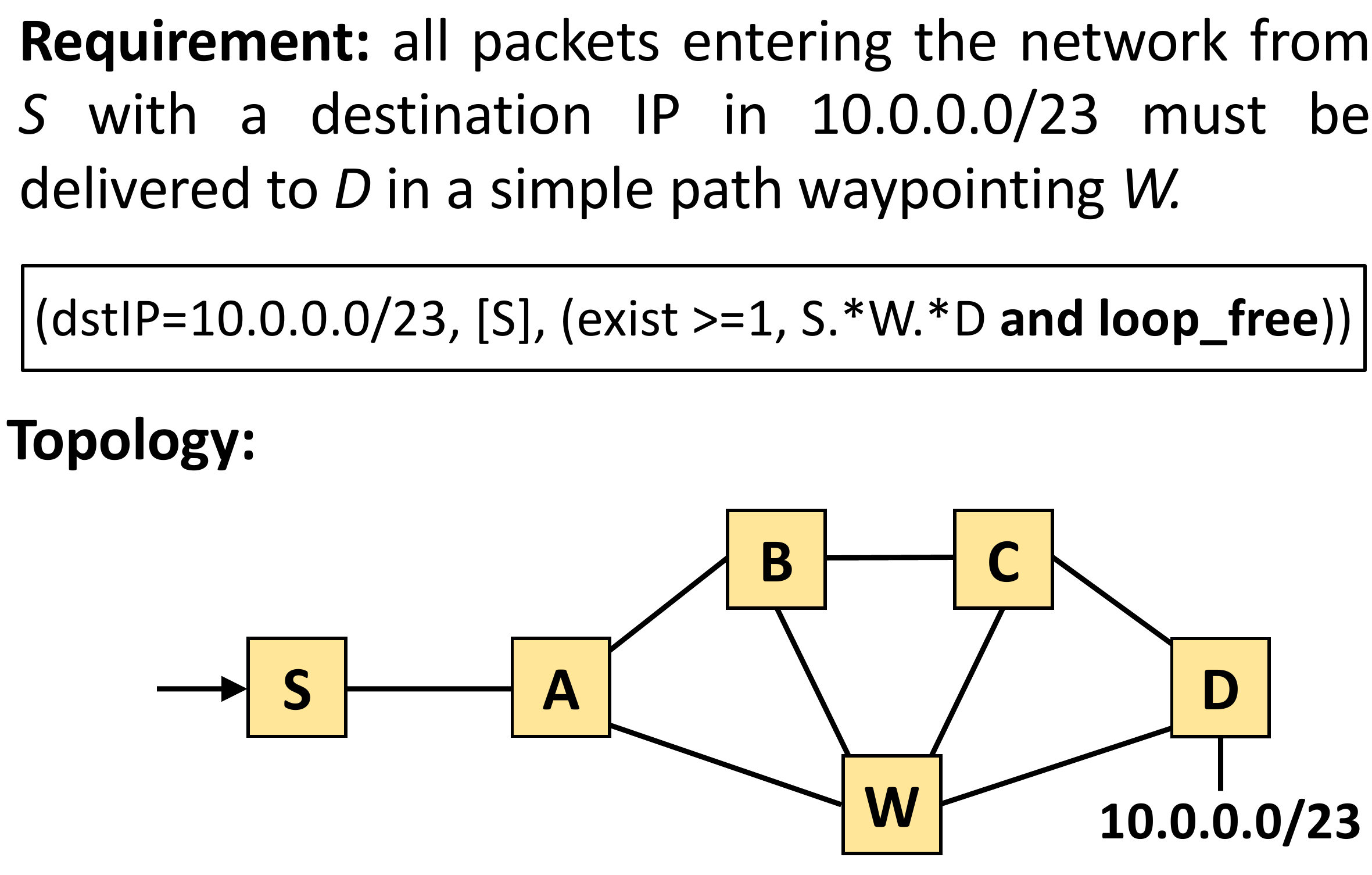}
		\caption{\label{fig:workflow-topo} \footnotesize An example topology
		and requirement.}
%\vspace{-0.5em}
	\end{subfigure}
\hfill
	\begin{subfigure}[t]{0.23\linewidth}
\setlength{\abovecaptionskip}{0.1cm}
\setlength{\belowcaptionskip}{-0.cm}
	    \centering\includegraphics[width=0.95\linewidth]{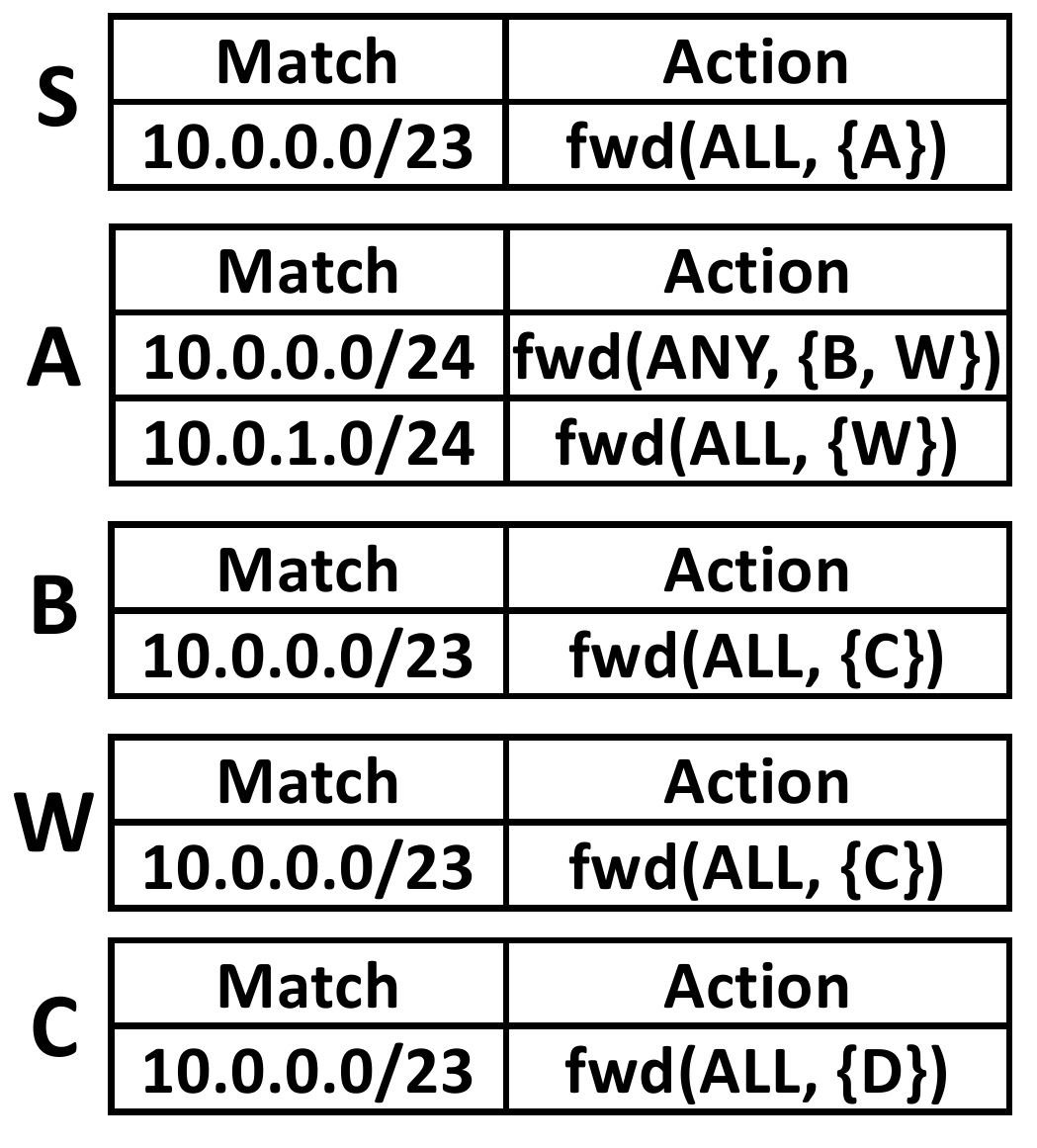}
		\caption{\label{fig:workflow-fib} \footnotesize The network data plane.}
%\vspace{-0.5em}
	\end{subfigure}
\hfill
	\begin{subfigure}[t]{0.38\linewidth}
\setlength{\abovecaptionskip}{0.1cm}
\setlength{\belowcaptionskip}{-0.cm}
	    \centering\includegraphics[width=0.95\linewidth]{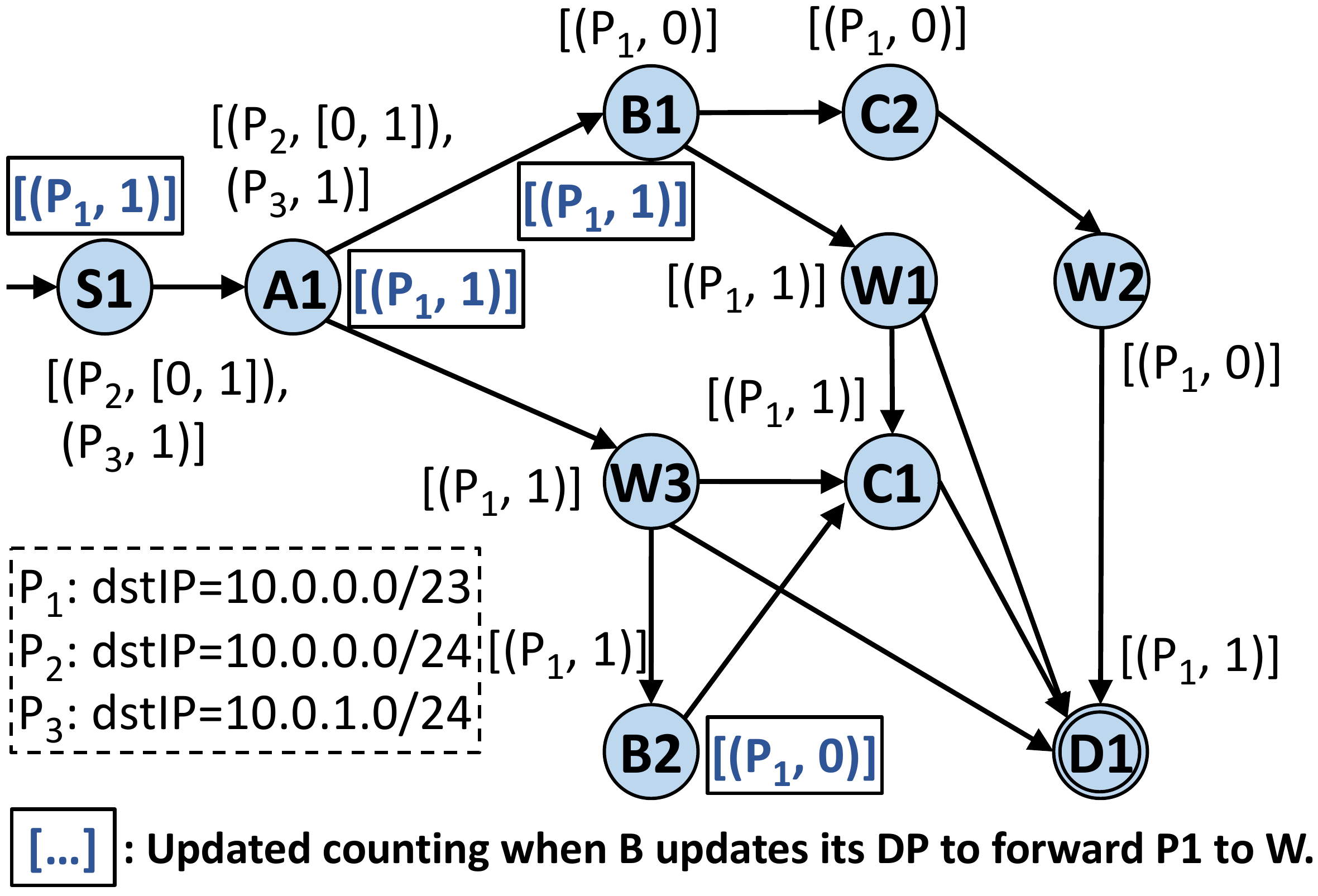}
		\caption{\label{fig:workflow-dvnet} \footnotesize The \dvnet{} and the counting process.}
%\vspace{-0.5em}
	\end{subfigure}
    \caption{An illustration example to demonstrate the workflow of \system{}.}
    \label{fig:workflow-example}
\vspace{-1.5em}
\end{figure*}

\subsection{Workflow}\label{sec:workflow}
\iffalse{
Given a requirement using the \system{} specification language,
the verification planner first compiles the requirement into \dvnet{}, and
transform the verification problem to a counting problem in \dvnet{}. It then
decomposes the counting problem into smaller on-device counting tasks, one for
each node in \dvnet{}, and sends these tasks to the corresponding devices in the
network. 
}\fi

We demonstrate the basic workflow of \system{} using an example in
Figure~\ref{fig:workflow-example}. We consider the network in
Figure~\ref{fig:workflow-topo} and the following requirement: for all packets
destined to 10.0.0.0/23, when they enter the network from $S$, they must be able
to reach $D$, the device with an external port reachable to 10.0.0.0/23, via a
simple path passing $W$. \system{} verifies this requirement in three phases.

\subsubsection{Requirement Specification}\label{sec:overview-spec}
In \system{}, operators specify verification requirements using a declarative
language. A requirement is specified as a $(packet\_space,$ $ingress\_set,$
$behavior)$ tuple. The \textbf{semantic} means: for each packet $p$ in $packet\_space$
entering the network from any device in $ingress\_set$, the traces of $p$ in all
its universes must satisfy the constraint specified in $behavior$, which is
specified as a tuple of a regular expression of valid paths $path\_exp$ and a
match operator.  Figure~\ref{fig:workflow-topo} gives the program of the example
requirement, where $\mathbf{loop\_free}$ is a shortcut in the language for a
regular expression that accepts no path with a loop.  It specifies that when any
$p$ destined to 10.0.0.0/23 enters from $S$, at least 1 copy of it will be
delivered to $D$ along a simple path waypointing $W$.

%\qiao{Remove this sentence because
%it is repetitive? 
%This program specifies that: given any packet $p$ destined to 10.0.0.0/23, when
%it enters from $S$, in each universe of $p$, at least 1 copy of $p$ will be
%delivered to $D$ along a simple path waypointing $W$}.
\vspace{-0.5em}
\subsubsection{Verification Decomposition and Distribution}
Given a requirement, \system{} uses a planner to decide the tasks
to be executed distributively on devices to verify it.  The core challenge is
how to make these on-device tasks lightweight, because a network device
typically runs multiple protocols (\eg, SNMP, OSPF and BGP) on a low-end CPU,
with little computation power to spare. To this end, the \system{} planner
employs a data structure called \dvnet{} to decompose the DPV problem into small
on-device verification tasks, and distribute them to on-device verifiers for
distributed execution.

\iffalse{
takes as input user-specified verification
requirements and the topology information, decomposes the DPV problem into many
small on-device verification tasks for devices to execute
distributively, and distribute such tasks to the on-device verifiers.
}\fi
%and device-IP prefix mapping information of the topology, 

\para{From requirement and topology to \dvnet{}.}
The planner first leverages the automata theory~\cite{lewis1998elements} to take
the product of the regular expression $path\_exp$ in the requirement and the
topology, and get a DAG called \dvnet{}. Similar to the logical topology in
Merlin~\cite{merlin} and the product graph in Propane~\cite{propane} and
Contra~\cite{contra}, a \dvnet{} compactly represents all paths in the topology
that match the pattern $path\_exp$. It is decided only by $path\_exp$ 
and the topology, and is independent of the actual data plane of the
network. 
%multiplies $S.^*W.^*D$ $\mathbf{and}$
% $\mathbf{loop\_free}$ and the topology and generates the \dvnet{} 

Figure~\ref{fig:workflow-dvnet} gives the computed \dvnet{} in our example.  
Note the devices in the network and the nodes
in \dvnet{} have a 1-to-many mapping. For each node $u$ in \dvnet{}, we assign a
unique identifier, which is a concatenation of $u.dev$ and an integer.  For
example, device $C$ in the network is mapped to two nodes $C1$ and $C2$ in
\dvnet{}, because the regular expression allows packets to reach $D$ via $[C, W,
D]$ or $[W, C, D]$.
 
%\dduan{In general, will this always be many-1, or might it sometimes be many-many?}

\para{Backward counting along \dvnet{}.} 
With \dvnet{}, a DPV problem is transformed into a counting problem on \dvnet{}:
\textit{given a packet $p$, can the network deliver a satisfactory number of
copies of $p$ to the destination node along paths in the DVNet in each
universe?} In our example, the problem of verifying whether the data plane of
the network (Figure~\ref{fig:workflow-fib}) satisfies the requirement is
transformed to the problem of counting whether at least 1 copy of each $p$
destined to 10.0.0.0/23 is delivered to $D1$ in Figure~\ref{fig:workflow-dvnet}
in all of $p$'s universes. 

This counting problem can be solved by a centralized algorithm that traverses
the nodes in \dvnet{} in reverse topological order. At its turn, each node $u$
takes as input (1) the data plane of $u.dev$ and (2) for different $p$ in
$packet\_space$, the number of copies that can be delivered from each of $u$'s
downstream neighbors to the destination, along \dvnet{}, by the network data
plane, to compute the number of copies that can be delivered from $u$ to the
destination along \dvnet{} by the network data plane. In the end, the source
node of \dvnet{} computes the final result of the counting problem.

\iffalse{
and the number of actual paths from each of
$u$'s downstream neighbors to the destination that are provided by the network
data plane, to compute the number of actual paths from $u$ to the destination
that are provided by the network data plane. In the end, the number of actual
paths from the source node to the destination node is the number of copies of
$p$ delivered to the destination along paths in \dvnet{}.
}\fi

\iffalse{
sort all the nodes in \dvnet{} in a topological order, and then scan the nodes
from the destination backwards to the source, 
}\fi

%The planner
%then decomposes this job into a series of counting tasks, which are
%executed on corresponding devices. 
%\qiao{whether or not emphasize bakcwards?}

%\para{Backward, distributed counting along \dvnet{}.} 
Figure~\ref{fig:workflow-dvnet} illustrates this
algorithm. For simplicity, we use $P_1, P_2, P_3$ to represent
the packet spaces with destination IP prefixes of $10.0.0.0/23$, $10.0.0.0/24$,
and $10.0.1.0/24$, respectively. Note that $P_2 \cap P_3 = \emptyset$  and
$P_1 = P_2 \cup P_3$. Each $u$ in \dvnet{} initializes a $packet~space \mapsto
count$ mapping, $(P_1, 0)$, except for $D1$ that initializes the mapping as
$(P_1, 1)$ (\ie, one copy of any packet in $P_1$ will be sent to the correct
external ports). Afterwards, we traverse all the nodes in
\dvnet{} in reverse topological order to update their mappings. 
Each node $u$ checks the data plane of $u.dev$ to find the set of next-hop
devices $u.dev$ will forward $P_1$ to. If the action of forwarding to this
next-hop set is of $ALL$-type, the mapping at $u$ can be updated by
adding up the count of all downstream neighbors of $u$ whose corresponding device
belongs to the set of next-hops of $u.dev$ for forwarding $P_1$. For example,
node $C1$ updates its mapping to $(P_1, 1)$ because device $C$ forwards to $D$,
but node $W2$'s mapping is still $(P_1, 0)$ because $W$ does not forward $P_1$
to $D$. Similarly, although $W1$ has two downstream neighbors
$C1$ an $D1$, each with an updated mapping $(P_1, 1)$. At its turn, we update
its mapping to $(P_1, 1)$ instead of $(P_1, 2)$, because device $W$ only
forwards $P_1$ to $C$, not $D$.

Given a node $u$ in \dvnet{}, if the action of forwarding is of $ANY$-type, the
count may vary at different universes. As such, we update the mapping at $u$ to
record distinct counts at different universes. Consider the mapping update at
$A1$. $A$ would forward $P_2$ to either $B$ or $W$. As such, in one universe
where $A$ forwards $P_2$ to $B$, the mapping at $A1$ is $(P_2, 0)$, because
$B1$'s updated mapping is $(P_1, 0)$ and $P_2 \subset P_1$. In the other
universe where $A$ forwards $P_2$ to $W$, the mapping at $A1$ is $(P_2, 1)$
because $W3$'s updated mapping is $(P_1, 1)$. Therefore, the updated mapping for
$P_2$ at $A1$ is $(P_2, [0, 1])$, indicating the different counts at different
universes. In the end, the updated mapping of $S1$
$[(P_2, [0, 1]), (P_3, 1)]$ reflects the final counting results, indicating that
the data plane in Figure~\ref{fig:workflow-fib} does not satisfy the
requirements in Figure~\ref{fig:workflow-fib} in all universes. In other words,
the network data plane is erroneous.

%Given a node in \dvnet{}, the algorithm uses the data plane of its corresponding device in
%the network and the results of the counting tasks from all its downstream
%neighbor nodes as input, and 

\iffalse{
For a given packet space, this
mapping records the number of packet copies in all
universes that can reach from the current node to the destination node in
\dvnet{}.

Each node stores its computed  mapping in the CIB of its corresponding device,
and sends it to the devices of upstream neighbors of the current node. For
example, after node $D1$ computes its mapping $[P_1, 1]$ (\ie, one copy of any
packet in $P_1$ will be sent to the correct external ports), it sends this
result to $W$ and $C$, the devices of its upstream neighbors $W1$, $W2$, $W3$
and $C1$. \dduan{This language makes it sound like the nodes in dvnet are
physical things communicating with each other, but it's really still the local
verifiers communicating, right?} \qiao{yes, we need to clarify the connection in
DVNet is logical connection.} Because $C1$ only has one downstream neighbor
$D1$, $C1$ checks the data plane of $C$, computes its mapping also as $[P_1,
1]$, and sends to $W$, the device of its upstream neighbors $W1$ and $W3$. As a
result of this backward counting process, the ingress device $S$ will eventually
receive the
complete counting result as the final verification result. 
}\fi

\para{Counting decomposition and distribution.} 
The centralized counting algorithm in \dvnet{} allows a natural decomposition
into on-device counting tasks to be executed distributively on network devices.
Specifically, for each node $u$ in \dvnet{}, an on-device counting task: (1)
takes as input the data plane of $u.dev$ and the results of on-device counting
tasks of all downstream neighbors of $u$ whose corresponding devices belong to
the set of next-hop devices $u.dev$ forwards packets to; (2) computes the
number of copies that can be delivered from $u$ to the destination along
\dvnet{}, by the network data plane in each universe; and (3) sends the computed
result to devices where its upstream neighbors in \dvnet{} reside in. After the
decomposition, the planner sends the on-device counting task of each $u$, as
well as the lists of $u$'s downstream and upstream neighbors, to device
$u.dev$.

\iffalse{
actual paths from $u$
to the destination provided by the network data plane in each universe; and (3)
it sends the locally computed number to devices where its upstream neighbors in
\dvnet{} reside in. After the decomposition, for each node $u$ in \dvnet{}, the
planner sends its on-device counting task, as well as the lists of $u$'s
downstream and upstream neighbors, to device $u.dev$.
}\fi

\subsubsection{Distributed, Event-Driven Verification using DV Protocol}
\label{sec:overview-dvp}
\iffalse{
After receiving the tasks from the planner, the on-device verifiers on network
devices runs distributively to verify the network data plane by executing their
own tasks and communicating task results along the opposite directions of links
in \dvnet{}. 
}\fi

After receiving the tasks from the planner, on-device verifiers execute them in
a distributed, event-driven way. When events (\eg, rule update, link down and
the arrival of updated results from neighbor devices) happen, on-device
verifiers update the results of their tasks, and send them to neighbors if needed. We
design a DV protocol that specifies how on-device verifiers incrementally update
their on-device tasks, as well as how they communicate task results, efficiently
and correctly.

Consider a scenario in Figure~\ref{fig:workflow-example},
where $B$ updates its data plane to forward $P_1$ to $W$, instead of to $C$. The
changed mappings of different nodes are circled with boxes in
Figure~\ref{fig:workflow-dvnet}.  In this case, device $B$ locally updates
the task results of $B1$ and $B2$ to $[(P_1, 1)]$ and $[(P_1, 0)]$,
respectively, and sends corresponding updates to the devices of their upstream
neighbors, \ie, $[(P_1, 1)]$ sent to $A$ following the opposite of $(A1, B1)$
and $[(P_1, 0)]$ sent to $W$ following the opposite of $(W3, B2)$.  

Upon receiving the update, $W$ does not need to update its mapping for
node $W3$, because $W$ does not forward any packet to $B$. As such, $W$ does
not need to send any update to $A$ along the opposite of $(A1, W3)$.
In contrast, $A$ needs to update its task result for node $A1$ to $[(P_1, 1)]$
because (1) no matter whether $A$ forwards packets in $P_2$ to $B$ or $W$, 1
copy of each packet will be sent to $D$, and (2) $P_2 \cup P_3 = P_1$.  After
updating its local result, $A$ sends the update to $S$ along the
opposite of $(S1, A1)$. Finally, $S$ updates its local result for $S1$ to
$[(P_1, 1)]$, \ie, the requirement is satisfied after the update.

% data plane. 
%$B2$'s upstream neighbor $W3$ 
%$B1$'s upstream neighbor $A1$ will update its mapping to

\iffalse{
to execute an on-device counting task, the on-device verifier
first reads the device's data plane and constructs a table of local equivalence
classes (LECs).  Informally, an LEC at device $X$ is a set of packets whose
actions are identical at $X$. The computation and maintenance of the LEC table can be
realized using existing DPV tools~(\eg, DeltaNet~\cite{deltanet-nsdi17},
VeriFlow~\cite{veriflow}, AP~\cite{ap-icnp13} and APKeep~\cite{apkeep-nsdi20}) 
by considering $X$ the only device in the network. 

Next, the on-device verifier uses the LEC table and the results of other tasks
sent from neighboring devices as input, and counts, across the packet space,
the number of packet copies that can reach the destination node in \dvnet{} from the current node. The results are stored in a counting information
base (CIB) and sent to the on-device verifiers at neighboring devices
incrementally, based on the device-to-device communication instructions
specified by the planner and the DV protocol. 
}\fi

%\input{architecture}
%\input{workflow}

%\newpage

\begin{figure}[t]
\setlength{\abovecaptionskip}{0em}
\setlength{\belowcaptionskip}{0em}
\centering
%\footnotesize
\begin{tabular}{rcll}
%	$prog$ & $::=$ & $[s_1; \ldots; s_n]~[req_1; \ldots;
%	req_n]~\mathbf{verify}~req;$  \\
%	$s$ & $::=$ & $sid: s\_pred |~\{sid\}^+$ \\
%	$s\_pred$ & $::=$ & $packet\_space~|~(packet\_space, ingress\_points)$ \\
%              & $|$ & $(packet\_space, egress\_points)$ \\
%	      & $|$ & $(packet\_space, ingress\_points, egress\_points)$ \\
%	$packet\_space$ & $::=$ &\\
%	$ingress\_location$ & $::=$ &\\
%	$egress\_location$ & $::=$ &\\

  $\var{reqs}$& $::=$ & $\var{req}^*$ \\
$\var{req}$ & $::=$ & $(packet\_space, ingress\_set, behavior)$ &\\
	 $behavior$  & $::=$ & $(match\_op,\hspace{0.5em} path\_exp)$ $|$  $\mathbf{not}$ $behavior$ & \\
	              & $|$ & $behavior$ $\mathbf{or}$ $behavior$ & \\
		      & $|$ & $behavior$ $\mathbf{and}$ $behavior$ & \\
	$path\_exp$ & $::=$ & regular expression over \\ & & the set of devices
	\\
	$match\_op$ & $::=$ & $\mathbf{exist}\hspace{0.5em} count\_exp$ | $\mathbf{equal}$ &\\    
%	$\var{hop}$ & $::=$ & $.$ | $dev$ | $!dev$ | $dev | dev$  & \\
%      & $|$ & . &  \\
%    	& $|$ & * & \\
$exist\_exp$ & $::=$ & $== N$ | $>= N$ | $> N$ | $<= N $ | $< N $ &   \\
%	& $|$ & $\hat{ }$ & (\textsc{Start of Path}) \\
%    	& $|$ & \$ & (\textsc{End of Path}) \\
%         & $|$ & > & (\textsc{Packet Destination}) \\
\end{tabular}
	\caption{The basic abstract syntax of the \system{} requirement specification
	language.}
%\qiao{Three things. First, previously, I made a change to $count\_exp$ before
	%$path\_exp$ because I thought it reads more natural, i.e., subjective
	%verb objective.  What do you think?} \qiao{Second, since we have $ALL$
	%in dataplane model, would using $\mathbf{all}$ be confusing? I used
	%$equal$ and $exist$ previously to map to 2.1. I don't know which one is
	%better. We should discuss about it.} \qiao{Third, I see your point, but
	%that might make $path\_exp$ a little hard to write. So I'm also not
	%sure  about it.}
\label{fig:grammar}
%\vspace{-1.5em}
\end{figure}

\section{Specification Language}\label{sec:lang}
\system{} provides a declarative language for operators to specify
verification requirements %to verify in the network
based on the concepts of traces
and universes. Figure~\ref{fig:grammar} gives its simplified grammar.

%of this language.

\para{Language overview.} 
On a high level, a requirement is specified by a $(packet\_space$ $,
ingress\_set$ $, behavior)$ tuple, with the semantic explained in
\S\ref{sec:overview-spec}.
%, corresponding to the
%definition of requirements in Section~\ref{sec:basic-concepts}. 
%The
%$packet\_space$ and $ingress\_set$ entries must be subsets of the total header
%space and set of devices in the network, respectively.
% The semantic means for every packet
% in $packet\_space$, when it enters the network from any device in
% $ingress\_set$, its packet traces in all universes 
% must satisfy a predicate specified in $req\_pred$. 
To specify behaviors, we use the building block of
$(match\_op,$ $path\_exp)$ entries. The basic syntax provides two
$match\_op$ operators. One is $\mathbf{exist }$ $count\_exp$, 
which requires that in each universe, the number of traces matching $path\_exp$ (a regular
expression over the set of devices) satisfies $count\_exp$. For example,
$\mathbf{exist } >=1$ specifies at least one trace should match $path\_exp$ in
each universe, and can be used to express reachability requirements. 
The other
operator is $\mathbf{equal}$, which
specifies an equivalence behavior: the union of universes for
each $p$ in $pkt\_space$ from each ingress in $ingress\_set$ must be equal to
the set of all
possible paths that match $path\_exp$~\cite{azure}. Finally, behaviors can also be
specified as conjunctions, disjunctions, and negations of these
$(match\_opi,$ $path\_exp)$ pairs. %, to allow compound behaviors. 
%and are evaluated as such.

\iffalse{
For example, for
existence requirements, $count\_exp$ supports expressions such as $>= 1$ (\ie,
to ensure at least one match exists, in the case of reachability), which
corresponds to the $exist\_exp$ nonterminal in the language. For equivalence
requirements, \system{} supports an \textbf{all} token for $count\_exp$, which
denotes the corresponding equivalence behavior (\ie, "the union of universes for
each in $pkt\_space$ from each ingress in $ingress\_set$ must match the set all
possible traces that match $path\_exp$). Finally, behaviors can then be
specified in the language as conjunctions, disjunctions, and negations of these
$(path\_exp, count\_exp)$ pairs (to allow "compound behaviors"), and are
evaluated as such. 
}\fi

These two operators can be used to form a wide range of requirements in data
plane verification. Table~\ref{tab:requirements} provides examples of
requirements that can be specified and verified in \system{}, and the
corresponding specifications in the \system{} language. 
%that can be used to specify each requirement.  
For example, using $\mathbf{exist}$ $count\_exp$, operators can express
simpler requirements such as reachability, waypoint
reachability, and loop-freeness, which are well studied by existing DPV tools~\cite{hsa,
netplumber, veriflow, ap-icnp13, apkeep-nsdi20}, as well as more advanced
requirements, such as multicast, anycast and no-redundant-delivery routing.
Another example is a requirement given in Azure RCDC~\cite{azure}, which
requires that all pairs of ToR devices should reach each other along a shortest
path, and all ToR-to-ToR shortest paths should be available in the data plane.
This can be formulated as an $\mathbf{equal}$ behavior on all
shortest paths across all universes (row 9 in Table~\ref{tab:requirements}).

\iffalse{
a conjunction of an $\mathbf{equal}$ behavior on all
shortest paths across all universes and an $\mathbf{exist}$ $>= 1$ behavior
requiring that at least one shortest path is available in each universe
(row 9 in Table~\ref{tab:requirements}).
}\fi

Note that once a requirement is specified, \system{} checks whether it is
consistently satisfied across all universes. As such, the multipath
consistency~\cite{batfish, nod} is expressed separately as reachability
and isolation requirements. 

\para{Convenience features.}  
\system{} builds and provides operators with a $(device, IP\_prefix)$ mapping
for network devices with external ports (\eg, a ToR switch or a border router),
where each tuple indicates that $IP\_prefix$ can be reached via an external port
of $device$. If a requirement is submitted with inconsistencies between the
destination IPs in $packet\_space$ and the end (destination) devices in its
corresponding $path\_exp$, \system{} will raise an error for operators to update
this requirement.

The language provides syntax sugar to simplify the expression of
requirements. For example, it allows users to specify a device set and
provides device iterators. It provides shortcuts of behaviors, \eg,
$\mathbf{loop\_free}$ for the loop-free requirement. It also provides a third $match\_op$
called $\mathbf{subset}$, which requires for packet $p$ entering the network
from ingress $S$, the set of traces of $p$ in each universe is a non-empty
subset of $path\_exp$. A behavior $\mathbf{subset}\hspace{0.5em}path\_exp$ is a
shortcut of $(\mathbf{match}$ $>=1\hspace{0.5em}path\_exp)$
$\mathbf{and~(match}$ $==0$ $.^*$ $\mathbf{and}$ $(\mathbf{not}$ $path\_exp))$.
We omit their details for the sake of simplicity. 
%\qiao{I'm considering  adding $\mathbf{subset}$ as a $count\_exp$}.

%\dduan{Add a sentence here noting that when you specify a
%behavior using the language, we implicitly use a "consistent" behavior (e.g. it
%must be true in all universes)}

\begin{table}[t]
\setlength{\abovecaptionskip}{0em}
\setlength{\belowcaptionskip}{0em}
    \centering
\footnotesize
	\resizebox{\linewidth}{!}{
	\begin{tabular}{|p{0.4\linewidth}|p{0.6\linewidth}|}
	        \hline
	\textbf{Requirements} & \textbf{\system{} specifications}  \\
        \hline
	\textbf{Reachability} \cite{anteater-sigcomm11, batfish, nod}
		& $(P, [S],$ $(\mathbf{exist} >=1$, $S .^* D))$ \\ \hline
	\textbf{Isolation} \cite{anteater-sigcomm11, batfish, nod}&  
		$(P, [S],$ $(\mathbf{exist} ==0$, $S .^* D))$ \\ \hline
		\textbf{Loop-freeness} \cite{anteater-sigcomm11} &
		$(P, [S],$ $(\mathbf{exist} == 0$, 
		$.^*$ $\mathbf{and}$ $\mathbf{not} ((\mathbf{not} ~X)^*$ 
		 $\mathbf{or~}$ $((\mathbf{not~} X)^* X(\mathbf{not~} X)^*))$
		 $\mathbf{~and~}$ $((\mathbf{not~} Y)^* $
		 $\mathbf{or~} ((\mathbf{not~} Y)^* Y(\mathbf{not~} Y)^*))
		 \ldots, ))$
		\\ \hline
		\textbf{Black hole freeness}\cite{anteater-sigcomm11}
		& $(P, [S],$ $(\mathbf{exist} ==0$, $.^* \mathbf{~and~}$
		$\mathbf{not~} S .^* D))$\\ \hline
		\textbf{Waypoint reachability}~\cite{netplumber} 
		& $(P, [S],$ $(\mathbf{exist} >=1$, $S .^*W.^* D))$\\ \hline
		\textbf{Reachability with limited path length}~\cite{netplumber} 
		& $(P, [S],$ $(\mathbf{exist} >=1$,
		$SD|S.D|S..D))$ \\ \hline
	\textbf{Different-ingress same reachability }
		~\cite{anteater-sigcomm11, veriflow} & $(P, [X, Y]$, $(\mathbf{exist} >=1$,
		$X.^*D | Y.^*D))$% $\mathbf{or} ~(\mathbf{exist} >=1, Y.^*D)$ 
		\\\hline
%	\textbf{Multipath consistent reachability} \cite{batfish, nod} \qiao{same as
%		reachability} & $(``\mathbf{match} >=1"$, $S .^* D)$\\ \hline
%	\textbf{Multipath consistent isolation} \qiao{same as isolation} &
%		$(``\mathbf{match} ==0"$, $S .^* D)$\\ \hline
		\textbf{In-/Cross-pod all-shortest-path reachability} ~\cite{azure} &
		$(P, [S],$  $(\mathbf{equal}$, $S.D))$ 
		%$\mathbf{and}$ $(\mathbf{exist} >=1$, $S.D))$
		/ $(P, [S],$  $(\mathbf{equal}$, $S...D))$ 
		%$\mathbf{and}$ $(\mathbf{exist} >=1$, $S...D))$
		\\ \hline
%new things done by us
		\textbf{Non-redundant reachability} [\system{}]&
		$(P, [S],$ $(\mathbf{exist} ==1$, $S .^* D))$ \\ \hline
		%\textbf{Redundant reachability} \qiao{?} [\system{}]&
		%$(``\mathbf{match} ==2"$, $S .^* D)$ \\ \hline
		\textbf{Mulicast} [\system{}]& $(P, [S],$ $((\mathbf{exist} >=1, S .^* D)$ 
		$\mathbf{and} ~(exist >=1, S .^* E)))$\\
	    \hline
		\textbf{Anycast} [\system{}]& $(P, [S],$ $((\mathbf{exist} >=1, S .^*
		D)$ $\mathbf{and} ~(\mathbf{exist}
		==0, S .^* E))$ $\mathbf{or~}$ $((\mathbf{exist} ==0, S .^*
		D)$ $\mathbf{and} ~(\mathbf{exist} ==1, S .^* E)))$\\
	    \hline
	% \textbf{1+1 routing} & \\ \hline
    \end{tabular}
}
	\caption{Selected requirements and their \system{} specifications.}
    \label{tab:requirements}
\end{table}

\para{Expressiveness and limitation.} 
Our language is expressive in that it can specify all verification
requirements studied in DPV literature, except for middlebox
traversal symmetry~\cite{nod} (\ie, $S$-$D$ and $D$-$S$ must pass the
same middlebox). In addition, although not studied in DPV
literature, path node- / link-disjointness cannot be expressed using our
language, either. To be precise, our language can express all "single-path"
requirements that compare the packet traces of one packet space with a
regular-expression $path\_exp$, but cannot specify "multi-path" requirements
that compare the packet traces of two packet spaces. We discuss how to extend
\system{} to specify and verify such requirements in \S\ref{sec:ext}.

\begin{figure}[t]
\setlength{\abovecaptionskip}{0.1cm}
\setlength{\belowcaptionskip}{-0.cm}
\centering
\includegraphics[width=0.5\columnwidth]{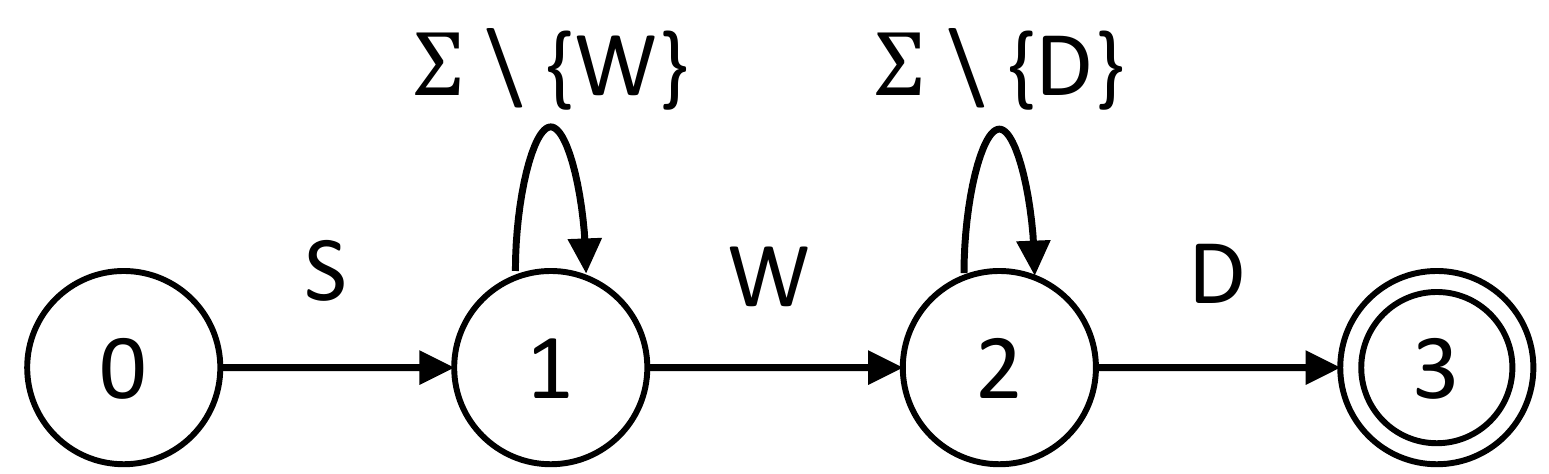}
\caption{The finite automata of $S.^*W.^*D$ with an alphabet $\Sigma=\{S, W, A,
B, C, D\}$.}
\label{fig:automata-example}
%\reducespace
%\vspace{-1em}
\end{figure}

\section{Verification Planner}\label{sec:dvnet}
We now present the design details of the planner.  For ease of presentation, we
first introduce \dvnet{} and how to use it for 
verification decomposition assuming a requirement has only one regular
expression. We then describe how to handle requirements with more than one
regular expressions.

\subsection{\dvnet{}}
Given a regular expression $path\_exp$ and a network, \dvnet{} is a DAG
compactly representing all paths in the network that matches $path\_exp$. There
are different ways to construct a \dvnet{} (\eg, graph dual variables). In
\system{}, we are inspired by network synthesis~\cite{merlin, propane,
contra}, and leverage the automata theory~\cite{lewis1998elements} for \dvnet{}
construction.

%For presentation simplicity, we
%first describe this transformation process assuming the requirement has only
%one regular expression $path\_set$. 

Specifically, given a $path\_exp$, we first convert it into a finite automata $(\Sigma, Q, F,
q_0, \delta)$. $\Sigma$ is the alphabet whose symbols are network device
identifiers.
%of a network device in the network. 
$Q$ is the set of states. %in the automata.
$q_0$ is the initial state. $F$ is the set of accepting states.  $\delta: Q \times
\Sigma \rightarrow Q$ is the state transition function. For example, for 
regular expression $S.^*W.^*D$ with a network of devices $S, W, A, B, C, D$, its
corresponding finite automata is in Figure~\ref{fig:automata-example}.

After converting $path\_exp$ to a finite automata, the planner multiplies it with
the topology. The multiplication yields a \textit{product
graph} $G'=(V', E')$.  Each node $u \in V'$ has an attribute $dev$ representing the
identifier of a device in the network and attribute $state$ representing its
state in the finite automata of $path\_exp$. Given two nodes $u, v \in V'$,
there exists a directed link $(u, v) \in E'$ if (1) $(u.dev, v.dev)$ is a link in
the network, and (2) $\delta(u.state, v.dev) = v.state$.

Finally, the planner performs state minimization on $G'$ to
remove redundant nodes,
and assigns each remaining node $u$ a unique identifier (a concatenation of
$u.dev$ and an integer), to get the \dvnet{}. An example of \dvnet{} was
given in Figure~\ref{fig:workflow-example}.

\iffalse{
Figure~\ref{fig:workflow-dvnet} gives the \dvnet{} computed from the loop-free,
waypointing reachability requirement and the network topology in
Figure~\ref{fig:workflow-topo}.
}\fi

%\subsection{From Verification to Distributed Counting}\label{sec:dv-detail}
\subsection{Verification Decomposition}\label{sec:dv-detail}
The core insight of \system{} is that we can transform DPV
into a counting problem on \dvnet{}, which can be naturally decomposed to
small on-device counting tasks running distributively. To be concrete, first
consider a requirement on a packet $p$ in the form of $(\mathbf{exist}~
count\_exp,$ $path\_exp)$. We can verify it by counting whether the network can
deliver a satisfactory number of copies of $p$ to the destination along paths in
the \dvnet{} in each universe. Such a problem can be solved by a traversal of
nodes in \dvnet{} in reverse topological order (Algorithm~\ref{alg:counting}),
during which each node $u$ counts the number of copies of $p$ in all $p$'s
universes that can reach the destination nodes of \dvnet{} from $u$.

\iffalse{
For requirement $(packet\_space,$ $ingress\_set,$ $path\_set,$ $req\_context)$, 
after its \dvnet{} is constructed, the problem of verifying whether for any
packet $p$ in $pacekt\_space$ entering the network from $ingress\_set$, the
number of traces satisfying $path\_set$ in all its universes satisfies the
condition in $req\_context$ is equivalent to counting whether the number of
copies of $p$ delivered to the destination in \dvnet{} in all $p$'s universes
satisfies the condition in $req\_context$. A strawman counting approach is to
collect the data planes of all devices and perform a backward depth-first
traversal in \dvnet{}, however, this method is still centralized and not
scalable.

\para{Basic idea.} 
In \system{}, the planner makes the counting process distributed by decomposing
the backward depth-first traversal in \dvnet{} on a node-level. In particular,
in \dvnet{}, each node $u$ counts for each packet $p \in packet\_space$, the
number of copies in all $p$'s universes that can reach the destination nodes of
\dvnet{} from $u$.  $u$ then sends this counting result backwards to all its
upstream neighbors in \dvnet{}. 
}\fi

\begin{algorithm}[t]
\setlength{\abovecaptionskip}{0.1cm}
\setlength{\belowcaptionskip}{-0.cm}
	\footnotesize
	Sort nodes in \dvnet{} in reverse topological order: $u_1, \ldots, u_n$\;
	\ForEach{$u_i$, $i = 1, \ldots, n$}{
		\If{$u_i$ is a destination}{$\mathbf{c}_i \leftarrow 1$}\Else{
			\ForEach{$v_j \in N_{d}(u_i)$}{
				\If{$v_j.dev \in u_i.dev.fwd(p)$}{$b_{ij} \leftarrow 1$\;}
			}
			\If{$u_i.dev.fwd(p).type == ALL$}{
				Update $\mathbf{c}_u$ with Equation~(\ref{eqn:count-all})\;
			}
			\Else{
				Update $\mathbf{c}_u$ with Equation~(\ref{eqn:count-any})\;
			}
		}
	}
	\Return $\mathbf{c}_n$\;
	\caption{{\sc Count}$(\mbox{\dvnet{}}, p)$.}
	\label{alg:counting}
%\vspace{-1.5em}
\end{algorithm}

\para{Counting at nodes.}
Each $u_i$ only keeps unique counting results of different
universes to avoid information explosion. If $u_i$ is a destination node in
\dvnet{}, its count is 1.
Denote the downstream neighbors of $u_i$ in
\dvnet{} as $N_d(u_i)=\{v_j\}_j$, and their counting results as sets
$\{\mathbf{c}_{v_j}\}_j$. Let $b_{ij}=1$ if the group of next-hops for
$p$ on $u_i.dev$ includes $v_j.dev$, and 0 otherwise. Define
$\otimes$ as the \textit{cross-product sum} operator for sets, \ie,
$\mathbf{c}_1 \otimes \mathbf{c}_{2} = (a+b | a \in \mathbf{c}_1, b \in
\mathbf{c}_{2})$. If $u_i.dev$'s forwarding action for $p$ is of type $ALL$, \ie, a
copy of $p$ will be forwarded to each next-hop in the rule, the
count of $p$ at $u_i$ is, %computed  as,
%\vspace{-0.5em}
\begin{equation}\label{eqn:count-all}
	\mathbf{c}_{u_i} = \otimes_{j: b_{ij}=1} (\mathbf{c}_{v_j}).
%\vspace{-0.8em}
\end{equation}
For example, in Figure~\ref{fig:workflow-dvnet}, for packets in
$P_1$, the count at $W1$ is $[1]$, the result of $C1$, because $W$
only forwards $P_1$ to $C$, not other devices.

Next, define $\oplus$ as the \textit{union} operator for sets. Let $\delta=1$ if $u_i.dev$ forwards $p$ to at least one device that does not have a corresponding node in $N_d(u_i)$ in \dvnet{}, and 0 otherwise. If $u_i$'s forwarding action for $p$ is of type $ANY$, \ie, $p$ will be forwarded
to one of the next-hops in the rule, the count of $p$
at $u_i$ is, %calculated as,
%\vspace{-0.5em}
\begin{equation}\label{eqn:count-any}
	\mathbf{c}_{u_i} = 
	\begin{cases}
	\oplus_{j: b_{ij}=1} (\mathbf{c}_{v_j}), & \text{if $\delta=0$}, \\
	(\oplus_{j: b_{ij}=1} (\mathbf{c}_{v_j}))\oplus \mathbf{0}, & \text{if $\delta=1$}.
	\end{cases}
%\vspace{-0.8em}
\end{equation}
Still in Figure~\ref{fig:workflow-dvnet}, for packets in $P_2$, the count at
$A1$ is $[0, 1]$, the union of $[0]$ from $B1$ and $[1]$ from $W3$ because
$A1$'s device $A$ forwards packets in $P_2$ to either $B$ or $W$.  We give a
proof sketch of Algorithm~\ref{alg:counting}'s correctness in
Appendix~\ref{sec:counting-correctness}.

\para{Distributed counting.} 
Algorithm~\ref{alg:counting} can be naturally decomposed to small counting
tasks, one for each node $u$ in \dvnet{}, to enable distributed counting. The
planner sends the task of $u$ to $u.dev$, with the lists of downstream and
upstream neighbors of $u$. $u.dev$ receives the counts from $v_j.dev$,
where $v_j \in N_d(u)$, computes  
$\mathbf{c}_{u}$ using Equations~(\ref{eqn:count-all})(\ref{eqn:count-any}),
%$u.dev$ takes as input its own data plane and the
%received counting task results from $v_j.dev$ where $v_j \in N_d(u)$, computes
%$\mathbf{c}_{u}$ using Equation~(\ref{eqn:count-all})(\ref{eqn:count-any}), 
and sends $\mathbf{c}_{u}$ to the corresponding devices of all $u$'s upstream
neighbors in \dvnet{}. In the end, the counts at source node of
\dvnet{} (\eg, $\mathbf{c}_{S1}$ at $S1$ in Figure~\ref{fig:workflow-dvnet}) is
the numbers of copies of $p$ delivered to the destination of \dvnet{} in all
$p$'s universes. The device of the source node can then easily verify the
requirement.  %with a simple comparison.

\para{Optimizing counting result propagation.} 
%The counting results at node $u$ are sent to the devices of all
%$u$'s upstream neighbors in \dvnet, with some exceptions for
When there are a huge number of paths in \dvnet{}, the counting result set
$\mathbf{c}_u$ can be large due to $ANY$-type actions at devices (\eg, 
a chained diamond topology). Letting
$u.dev$ send the complete $\mathbf{c}_u$ to the devices of $u$'s upstream
neighbors may result in large communication and computation overhead at devices.
Given a requirement, we define the \textit{minimal counting
information} of each node $u$ as the minimal set of elements in $\mathbf{c}_u$
that needs sending to its upstream nodes so that the source node in \dvnet{}
can correctly verify the requirement, assuming arbitrary data
planes at devices. We design optimizations to find such minimal counting information for
requirements with $\mathbf{exist}$ $count\_exp$ and $\mathbf{equal}$ operations,
respectively.

\iffalse{
We design two optimizations to improve the efficiency of
counting propagation, while guaranteeing the correctness of the final
verification result. 
}\fi

%New universes are created due to $ANY$-type actions. Cross
%\qiao{stop here, continue tomorrow}

For $\mathbf{exist} ~count\_exp$ operation,
our optimization is based on the \textit{monotonicity} of 
$\otimes$.
%\begin{property}[Counting Monotonicity]
Suppose two sets $\mathbf{c}_1$, $\mathbf{c}_2$ whose elements are all
non-negative. For any $x \in \mathbf{c}_1$ and $y \in
\mathbf{c}_2$, $a = x+y \in \mathbf{c}_1 \otimes \mathbf{c}_2$ satisfies
$a \geq x$ and $a \geq y$. We then have:

\iffalse{
For $\mathbf{exist} ~count\_exp$ operation,
our optimization is based on the \textit{monotonicity} property of 
$\otimes$ and $\oplus$.
%\begin{property}[Counting Monotonicity]
Suppose two sets $\mathbf{c}_1$, $\mathbf{c}_2$ whose elements are all
non-negative. For any $x \in \mathbf{c}_1$ and $y \in
\mathbf{c}_2$, we have $a = x+y \in \mathbf{c}_1 \otimes \mathbf{c}_2$,
$a \geq x$ and $a \geq y$; and $b = x\in \mathbf{c}_1 \oplus
\mathbf{c}_2$ and $b \geq x$. As such, we have
}\fi
%\end{property}

\iffalse{
\begin{property}[Counting Monotonicity]
Suppose two vectors $\mathbf{c}_1$, $\mathbf{c}_2$ whose elements are all
non-negative. Given any elements $x \in \mathbf{c}_1$ and $y \in
\mathbf{c}_2$ and the corresponding element $z = y+z \in \mathbf{c}_1 \otimes
\mathbf{c}_2$, $z \geq y$ and $z\geq z$. Given any element $x \in \mathbf{c}_1$
and the corresponding element $z = x\in \mathbf{c}_1 \oplus \mathbf{c}_2$, $z
\geq x$.
\end{property}
}\fi

\begin{proposition}\label{prop:mic-exist}
%[Minimal counting information for $\mathbf{exist}$ $count\_exp$]
Given a requirement with $\mathbf{exist}$ $count\_exp$ operation, the minimal
counting information of node $u$ is $min(\mathbf{c}_u)$ ($max(\mathbf{c}_u)$) if
$count\_exp$ is $\geq N$ or $> N$ ($\leq N$ or $< N$), and the first
$min(|\mathbf{c}_u|, 2)$ smallest elements in $\mathbf{c}_u$ if $count\_exp$
is $== N$. The proof is in Appendix~\ref{sec:proof-mic-exist}.
\end{proposition}

%We prove such sets of elements are the minimal
%counting information of node $u$ with a requirement with an $exist ~count\_exp$
%context.

\iffalse{
\begin{theorem}[Minimal Counting Information]
	Given a requirement with context $exist ~count\_exp$, for each $u$ in the
	\dvnet{}, the minimal counting information it needs to send to its
	upstream nodes for is  $min(\mathbf{c}_u)$ /
	$max(\mathbf{c}_u)$ / the first $min(|\mathbf{c}_u|, 2)$ smallest
	elements of $\mathbf{c}_u$ for
	different $count\_exp$s: $\geq N | > N$ /  $\leq N | < N$ / $== N$.
\end{theorem}
}\fi

%This is the  provide
%the minimal counting information needed to verify whether $exist count\_exp$

\iffalse{
The first optimization is based on the \textit{monotonicity} property of
distributed counting on \dvnet{} of \textit{one} universe. Specifically,  
given a count $a$ at node $u$ and a count $b$ at $v$, 
\begin{theorem}[Counting Monotonicity]
	Assume two nodes $u$ and $v$, such that $u \rightarrow v$ is a link in
	\dvnet{}, and for packet $p$, $u.dev$ forwards to $v.dev$. Then $\forall
	a \in \mathbf{c}_{v}$ and $\forall b \in \mathbf{c}_{u}$ that takes $a$ in
	computation, $b \geq a$.	
\end{theorem}
The correctness of this theorem follows the formula to compute $\mathbf{c}_{u}$,
and is omitted.
}\fi

For a requirement with an $\mathbf{equal}$ operator, we prove that the minimal
counting information of any $u$ is $\emptyset$, making distributed verification
become local verification. Specifically, no node $u$ even needs to compute
$\mathbf{c}_u$. Instead, it only needs to check if $u.dev$ forwards any packet
specified in the requirement to all the devices corresponding to the downstream
neighbors of $u$ in \dvnet{}. If not, a network error is identified, and $u.dev$
can immediately report it to the operators.  This optimization achieves local
verification on generic equivalence requirements, making the local contracts
on all-shortest-path availability in Azure RCDC~\cite{azure} a special
case.

\para{Computing consistent counting results.} 
Counting tasks are event-driven. Given a task for $u$, when an event (\eg, a
rule update or a physical port down at $u.dev$, or a count update
received from the device of a downstream neighbor of $u$), $u.dev$ updates the
counting result for $u$, and sends it to the devices of $u$'s upstream neighbors
if the result changes. As such, assuming the network becomes stable at some point, the
device of source node of \dvnet{} will eventually update its count result to
be consistent with the network data plane.

\para{Updating tasks.}
The process of the planner decomposing requirement to count tasks and sending to
devices is similar to configuring routing protocols. After the planner sends the
tasks to devices, the on-device verifiers run independently without the need of
the planner. Only when the planner receives a new requirement, or the 
topology is changed by administrator (\eg, adding a link), the planner will update the tasks by
regenerating \dvnet{}, and send them to the devices.

\subsection{Compound requirements}
%The presentation above focuses on a requirement with one regular expression. 
We now describe how the
planner decides the on-device counting tasks
for a requirement with multiple $path\_exp$s.  We focus on
requirements with a logic combination of multiple $(\mathbf{exist}$
$count\_exp,$ $path\_exp)$ pairs, because $\mathbf{equal}$ operator can
be verified locally. 
In particular, the case where a compound requirement with $path\_exp$s
with different sources can be handled by adding a virtual source device
connecting to all the sources. As such, we divide compound requirements
based on the destinations of regular expressions.

\begin{figure}[t]
\setlength{\abovecaptionskip}{0.cm}
\setlength{\belowcaptionskip}{-0.cm}
    \centering
	\begin{subfigure}[t]{0.46\linewidth}
\setlength{\abovecaptionskip}{0.cm}
\setlength{\belowcaptionskip}{-0.cm}
	    \centering\includegraphics[width=0.95\linewidth]{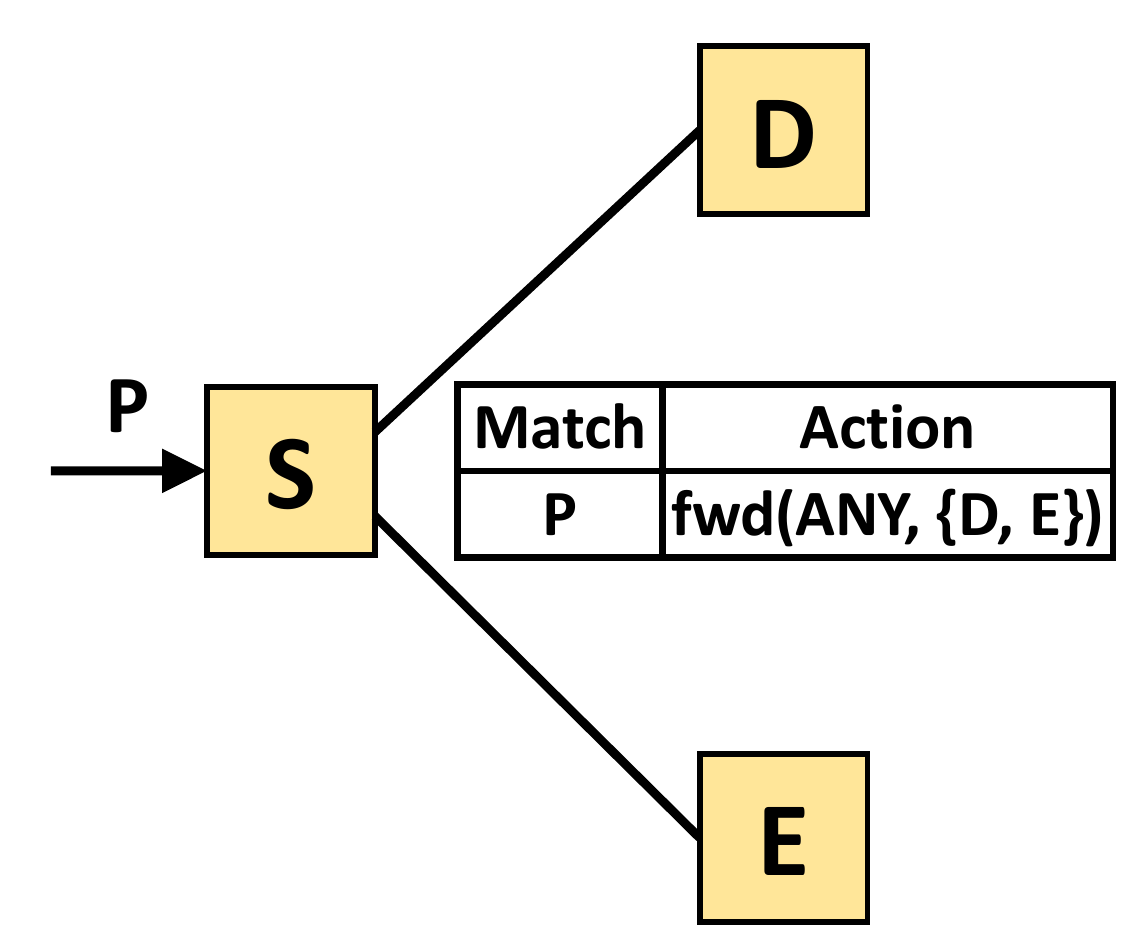}
		\caption{\label{fig:anycast-topo} \footnotesize A network for 
		anycast.}
	\end{subfigure}
\hfill
	\begin{subfigure}[t]{0.5\linewidth}
\setlength{\abovecaptionskip}{0.cm}
\setlength{\belowcaptionskip}{-0.cm}
	    \centering\includegraphics[width=0.95\linewidth]{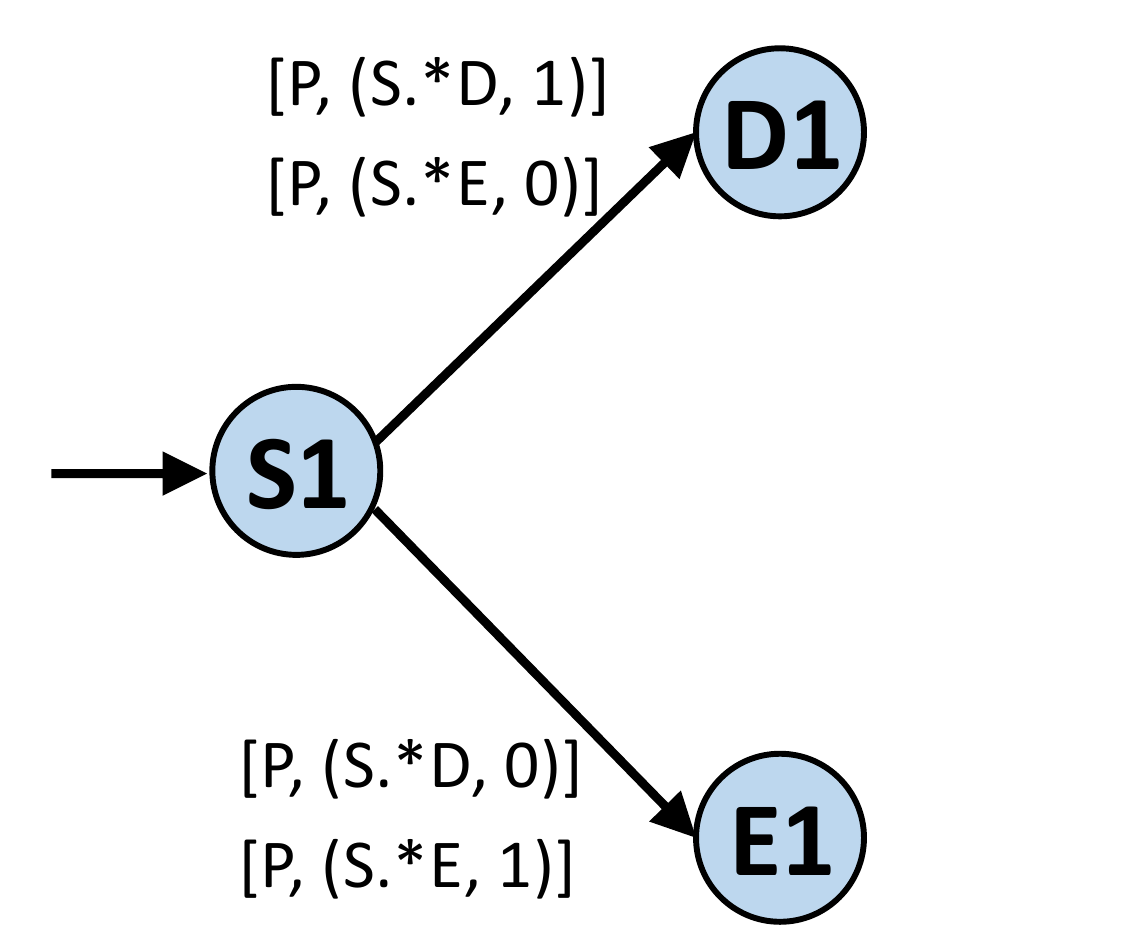}
		\caption{\label{fig:anycast-dvnet} \footnotesize The correct \dvnet{}
		and counting.}
	\end{subfigure}
    \caption{Verifying an anycast, a requirement with multiple $path\_exp$ with
	different destinations.}
    \label{fig:anycast-example}
\end{figure}

\para{Regular expressions with different destinations.}
One may think a natural solution is to build a \dvnet{} for each $path\_exp$, 
let devices count along all \dvnet{}s, and cross-produce the results
at the source. However, this is incorrect.  Consider an anycast requirement for
$S$ to reach $D$ or $E$, but not both (Figure~\ref{fig:anycast-topo}).  This
requirement is satisfied in the network. However, if we build two \dvnet{}s, one
for each destination, we get two chains $S1 \rightarrow D1$, and $S2 \rightarrow
E1$. After counting on both \dvnet{}s, $S1$ gets a set $[0, 1]$ for
reaching $D1$, and $S2$ gets $[0, 1]$ for reaching $E1$. The
cross-product computed by device $S$ would be $[(0, 0), (0, 1), (0, 1), (1,
1)]$, raising a false positive of network error.
%, a false positive
%that the network does not satisfy the anycast
%requirement.

To address this issue, for a requirement with multiple $(\mathbf{exist}$
$count\_exp,$ $path\_exp)$ pairs where $path\_exp$s have different destinations,
we let the planner construct a single \dvnet{} representing all paths in the
network that match at least one $path\_exp$, by multiplying the union of all
regular expressions with the topology, and specify one counting task for one
regular expression, at all nodes in \dvnet{}, including all destination nodes.
Consider the same anycast example.  The planner computes one \dvnet{} in
Figure~\ref{fig:anycast-dvnet}.  Each node counts the number of packets reaching
both $D$ and $E$. The count of $D1$ is $[(S.^*D, 1), (S.^*E,
0)]$ and $E1$ is $[(S.^*D, 0), (S.^*E, 1)]$. Such results are sent to $S1$.
After $S1$ processes it using Equation~(\ref{eqn:count-any}), it determines that
in each universe, a packet is sent to $D$ or $E$, but not
both, \ie, the requirement is satisfied.

\para{Regular expressions with the same destination.}
Following the design for the case of different destinations, one may be tempted
to handle this case by also constructing a single \dvnet{} for the union of such
$path\_exp$s. However, because these $path\_exp$s have the same
destination, the counting along \dvnet{} cannot differentiate the counts for
different $path\_set$s, unless the information of paths are collected and sent
along with the counting results. This would lead to large communication and
computation overhead across devices. 

%The key challenge to handle such a requirement is that constructing a single
%\dvnet{} for the union of such regular expressions cannot different which regular
%expressions a given path is matched to. 

\begin{figure}[t]
\setlength{\abovecaptionskip}{0.cm}
\setlength{\belowcaptionskip}{-0.cm}
    \centering
	\begin{subfigure}[t]{0.55\linewidth}
\setlength{\abovecaptionskip}{0.cm}
\setlength{\belowcaptionskip}{-0.cm}
		\centering\includegraphics[width=0.95\linewidth]{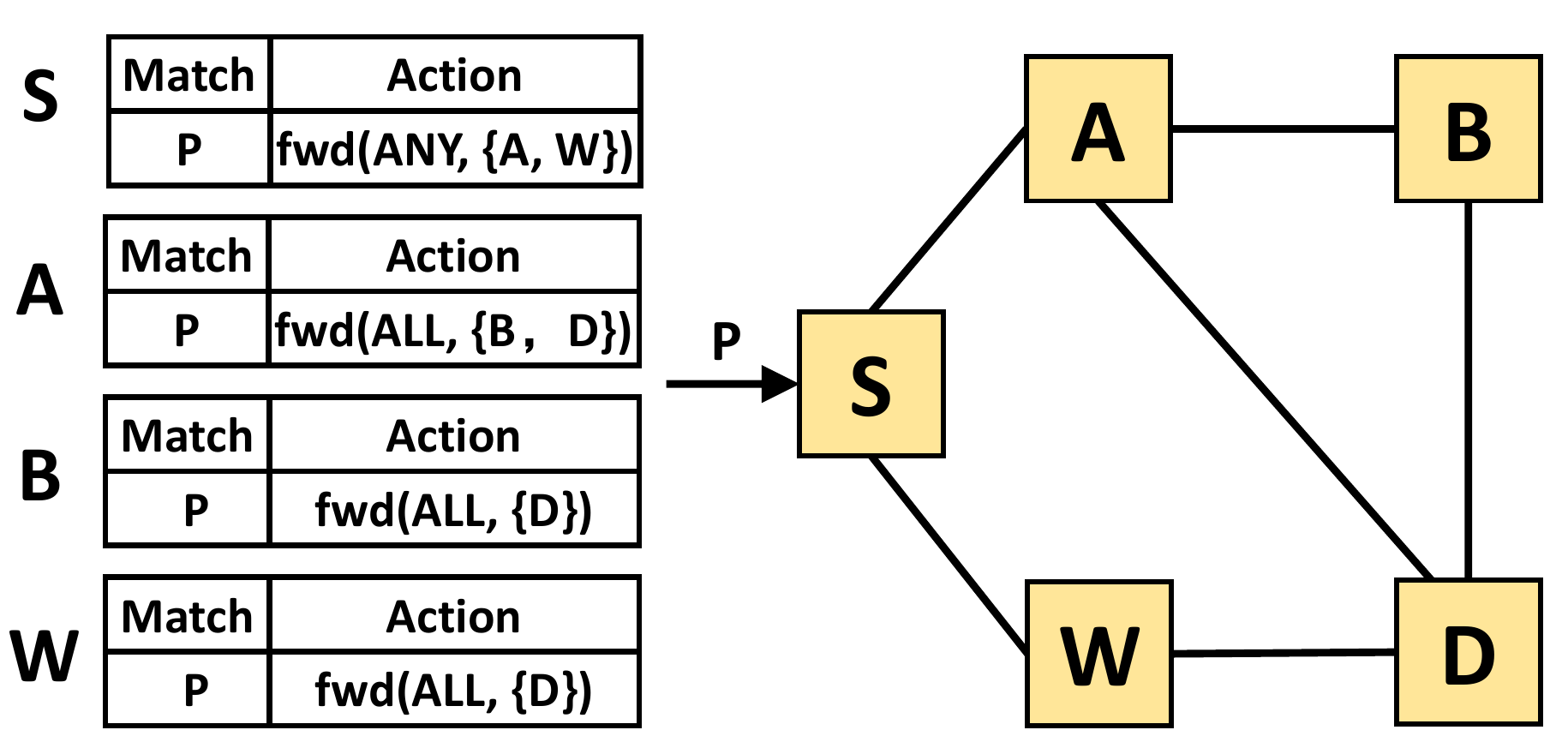}
		\caption{\label{fig:waypoint-and-reachability-topo}
		\footnotesize A network and its data plane.}
	\end{subfigure}
\hfill
	\begin{subfigure}[t]{0.42\linewidth}
\setlength{\abovecaptionskip}{0.cm}
\setlength{\belowcaptionskip}{-0.cm}
		\centering\includegraphics[width=0.95\linewidth]{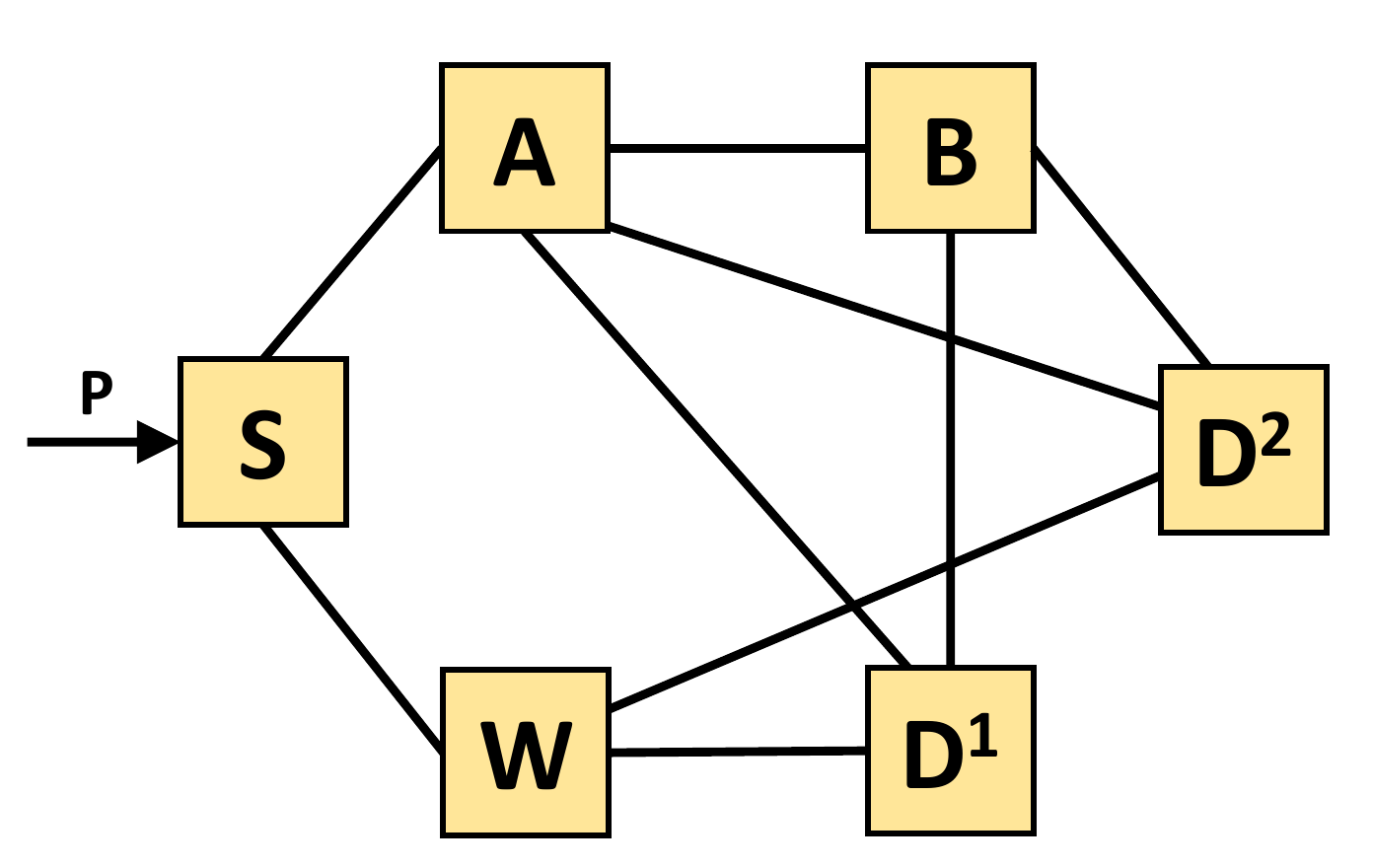}
		\caption{\label{fig:waypoint-virtual-topo}\footnotesize The updated topology
		with virtual destinations.}
	\end{subfigure}
	\caption{Verifying Equation~(\ref{eqn:waypoint}),
	a requirement with multiple $path\_exp$s with the
	same destination.}
    \label{fig:waypoint-example}
\end{figure}

Another strawman is to construct one \dvnet{} for one regular expression, count
separately and aggregate the result at the source via cross-producing. However,
false positives again can arise. Consider 
Figure~\ref{fig:waypoint-and-reachability-topo} and the requirement 
\begin{equation}\label{eqn:waypoint}
%\vspace{-0.5em}
\footnotesize
\begin{aligned}
	(P, [S], & (\mathbf{exist}
	>=2, (S .^* D~\mathbf{and~loop\_free}) \\
	& \mathbf{or}~(\mathbf{exist} >=1, S .^*W.^* D~\mathbf{and~loop\_free}))),
\end{aligned}
%\vspace{-0.5em}
\end{equation}
which specifies at least two copies of each packet in $P$ should be sent to $D$
along a loop-free path, or at least one copy should be sent to $D$ along a loop-free path
passing $W$. We observe that the data plane satisfies this requirement.
However, suppose we construct a \dvnet{} for each $path\_exp$,
and perform counting separately. $S$ will receive a counting result $[1, 2]$
for reaching $D$ with a simple path, and a counting result $[0, 1]$ for reaching
$D$ with a simple path passing $W$. The cross-product results $[(1, 0), (1,
1), (2, 0), (2, 1)]$ indicate that a phantom violation is found.

To address issue, we handle this requirement by adding virtual destination
devices.  Suppose a requirement has $m$ $(\mathbf{exist}$ $count\_exp_i,$
$path\_exp_i)$ pairs where $path\_exp_i$s have the same destination $D$. The
planner changes $D$ to $D^1$ and adds $m-1$ virtual devices $D^{i}$ ($i=2,
\ldots, m$). Each $D^i$ has the same set of neighbors as $D$ does, in the
network topology.  It then rewrites the destination of $path\_exp_i$ to $D^i$
($i=1, \ldots, m$).  Figure~\ref{fig:waypoint-virtual-topo} gives the updated
topology of Figure~\ref{fig:waypoint-and-reachability-topo} to handle the
requirement in Equation~(\ref{eqn:waypoint}).

Afterwards, the planner takes the union of all $path\_exp$s, intersects it with
an auxiliary $path\_exp$ specifying no any two $D^i, D^j$ should co-exist
in a path. It then multiplies the resulting regular expression with the new topology
to generate one single \dvnet{}. Counting can then proceed as the case for
regular expressions with different destinations, by letting each device treat
all its actions forwarding to $D$ as forwarding to \textit{all} $D^i$s, and
adjust Equations~(\ref{eqn:count-all})(\ref{eqn:count-any}) accordingly.

\begin{figure*}[t]
\setlength{\abovecaptionskip}{0cm}
\setlength{\belowcaptionskip}{-0.cm}
	\centering
	\begin{subfigure}[t]{0.44\linewidth}
	\setlength{\abovecaptionskip}{0cm}	
	    \centering\includegraphics[width=0.95\linewidth]{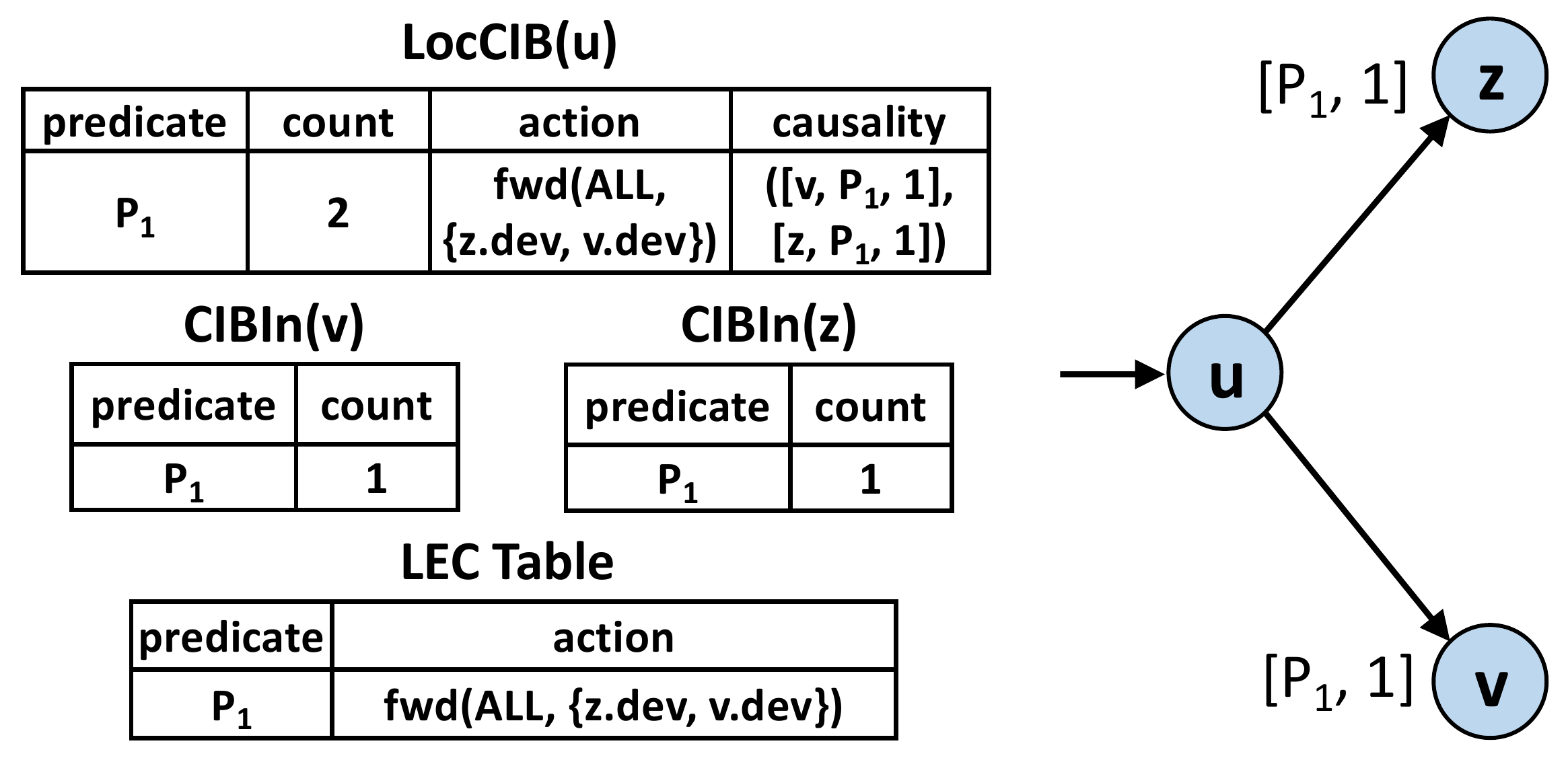}
		\caption{\label{fig:dvp-loccib}\footnotesize A \dvnet{} with
		LEC table of $u.dev$, $CIBIn$ and $LocCIB$ of $u$.}
	\end{subfigure}
\hfill
	\begin{subfigure}[t]{0.55\linewidth}
	\setlength{\abovecaptionskip}{0cm}
	    \centering\includegraphics[width=0.95\linewidth]{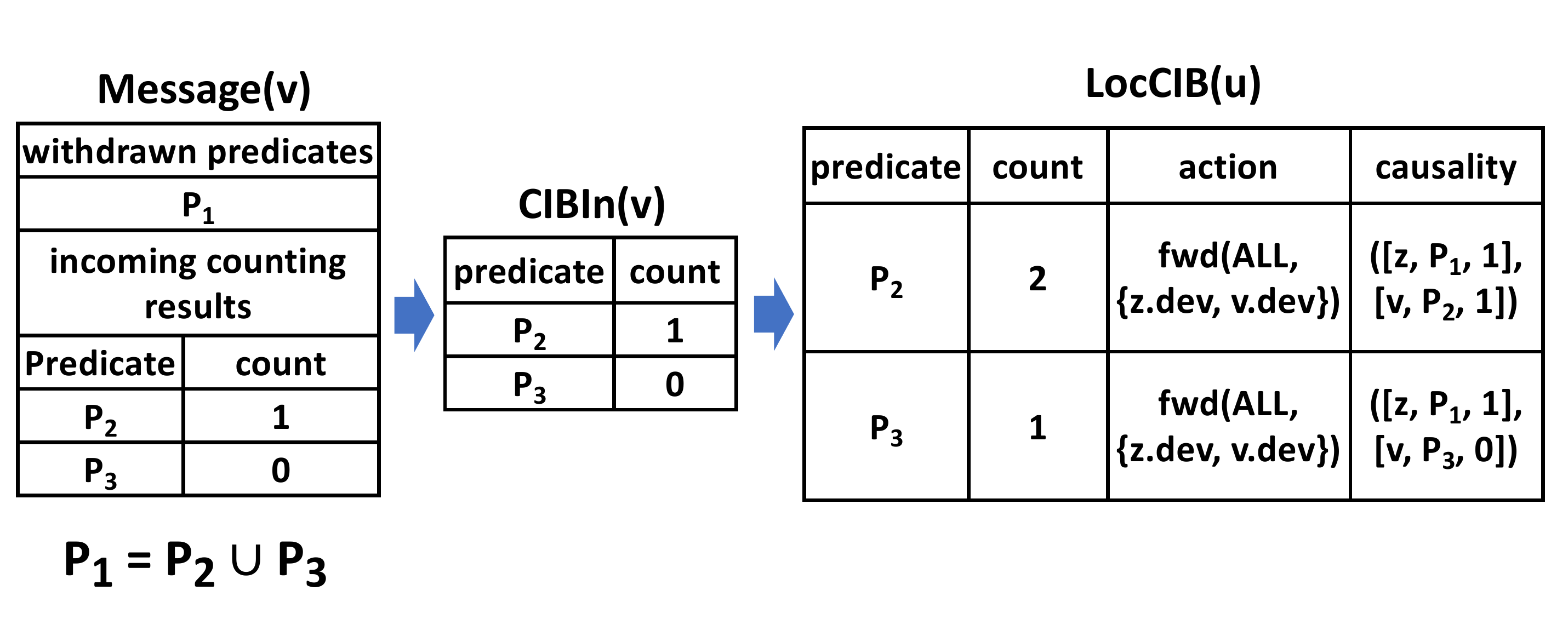}
		\caption{\label{fig:dvp-update}\footnotesize $u.dev$ handles an UPDATE from
		$v.dev$ to update $CIBIn(v)$ and $LocCIB(u)$.}
	\end{subfigure}
    \caption{An illustration example to demonstrate the key data structure and
	process of the DV protocol.}
    \label{fig:dvp-example}
\vspace{-1.5em}
\end{figure*}
\section{DV Protocol}\label{sec:dvp}
This section presents the details of the DV protocol. While the 
planner specifies the on-device tasks,
the DV protocol specifies how on-device verifiers share their counting results
with neighbors in an efficient, correct way, to collaboratively verify a
requirement. For ease of presentation, we introduce the protocol assuming a
single destination in \dvnet{}. 
%constructs the \dvnet{} and 

Given link $(u, v)$ in \dvnet{}, the DV protocol specifies the format and order
of messages $v.dev$ needs to send to $u.dev$, and the actions $u.dev$ needs to
take when receiving the messages. The DV protocol design is inspired by
vector-based routing protocols (\eg, RIP~\cite{RIP} and BGP~\cite{BGP}). One
distinction is that it needs no loop-prevention mechanism.  This is because the
messages are sent along the reverse direction in the DAG \dvnet{}. As a result,
no message loop will be formed. 

\subsection{Information Storage}
Each device stores two types of information: LEC (local equivalence class) and
CIB (counting information base). Given a device $X$, an LEC is a set of packets
whose actions are identical at $X$. $X$ stores its LECs in a table of
$(packet\_space, action)$ mapping called the LEC table.  Multiple existing DPV
tools can be used to compute and maintain the LEC table.  In our implementation,
we choose to encode the packet sets as \textit{predicates} using Binary Decision
Diagram (BDD~\cite{bryant1986graph}), and use BDD-based DPV
tools~\cite{ap-icnp13, apkeep-nsdi20} to maintain a table of a minimal number of
LECs at devices. This is because the DV protocol requires devices to perform
packet set operations (\eg, intersection and union), which can be realized
efficiently using logical operations on BDD. 
%Following the convention of these
%tools, we use \textit{predicate} to refer to a set of packets.

Given a device $X$, CIB stores for each $X.node$ in \dvnet{} (\ie, nodes with a
device ID $X$), for different packet sets, the number of packet copies that can reach
from $X.node$ to the destination node in \dvnet{}. Specifically, for each
$X.node$, $X$ stores three distinct types of CIB:
\begin{itemize}[leftmargin=*]
		\setlength\itemsep{0em}
	\item $CIBIn(v)$ for each of $X.node$'s downstream neighbors $v$: 
		it stores the latest, unprocessed counting results received from $v$ in a $(predicate, count)$ mapping;
	\item $LocCIB(X.node)$: it stores for different predicates, the latest
		number of packet copies
		that can reach from $X.node$ to the destination node in 
		$(predicate, count, action, causality)$ tuples, where the
		$causality$ field records the input to get the $count$ field
		(\ie, the right hand side of
		Equations~(\ref{eqn:count-all})(\ref{eqn:count-any}));
	\item $CIBOut(X.node)$: it stores the counting results to be sent to the upstream
		nodes of $X.node$ in $(predicate, count)$ tuples.
\end{itemize}
Figure~\ref{fig:dvp-loccib} gives an example \dvnet{}, with the counts of 
node $v$, $z$, the LEC table of $u.dev$, and $CIBIn(v)$, $CIBIn(z)$ and
$LocCIB(u)$ at node $u$. Specifically, the $causality$ field is $([v, P_1,$ $1],$
$[z, P_1, 1])$ because the $count$ 2 of predicate $P_1$ is computed via the
results of both $v$ and $z$ (\ie, $2=1+1$).

\iffalse{
In summary, for $X.node$, its $CIBIn(v)$s contain received, unprocessed counting results from
its downstream nodes $(v)$s; its Loc-CIB stores its latest counting results by applying
Equations~(\ref{eqn:count-all})(\ref{eqn:count-any}); its CIB-Out reads the counting
results to be sent to its upstream nodes from Loc-CIB, strips out the $action$
attribute and groups the tuples by $count$.\qiao{need to introduce a new
name CLEC?}
}\fi

\subsection{Message Format and Handling}
Messages in the DV protocol are sent over TCP connections. 
%A message is processed only after it is entirely received. 
The protocol defines control messages
(\eg, OPEN and KEEPALIVE) to manage the connections between devices.
We focus on the
UPDATE message, which is used to transfer counting results from the device of a
node to the devices of its upstream neighbors in \dvnet{}.

\para{UPDATE message format.} 
An UPDATE message includes three fields: (1) intended link: a tuple indicating
along which link in \dvnet{} the counting result is propagated oppositely;
%the message is intended for the counting result propagation along which link in
%\dvnet{};  
(2) withdrawn predicates: a list of predicates whose counting results
are obsolete; and (3) incoming counting results: a list of predicates with their
latest counts. The intended link is to differentiate links in \dvnet{} with the
same pair of devices (\eg, $(W1, C1)$ and $(W3, C1)$ in
Figure~\ref{fig:workflow-dvnet}).

\para{UPDATE message invariant.}
For the withdrawn predicates and incoming counting results, the DV protocol
maintains an important invariant: for each UPDATE message, the union of the
withdrawn predicates equals the union of the predicates in the incoming
counting results. This ensures that a node always receives the latest, complete
counting results from its downstream neighbors, guaranteeing the eventual
consistency between the verification result at the device of source node of
\dvnet{} and a stable data plane.
%This invariant ensures that there is no packet whose count is 

\para{UPDATE message handling.} 
Consider link $(u, v)$ in \dvnet{}. Suppose
$u.dev$ receives from $v.dev$ an UPDATE message whose intended link is
$(u, v)$. $u.dev$ handles it in three steps.

\para{Step 1: updating $CIBIn(v)$.}
$u.dev$ updates $CIBIn(v)$ by removing entries whose predicates belong to
withdrawn predicates and inserting all entries in incoming counting
results.

\para{Step 2: updating $LocCIB(u)$.}
To update $LocCIB(u)$, $u.dev$ first finds all affected entries, \ie, the
ones that need to be updated. To be concrete, an entry in $LocCIB(u)$ needs to
be updated if its $causality$ field has one predicate from $v$ and belongs to
the withdrawn predicates of this message. It then updates the counting results
of all affected entries one by one. Specifically, for each pair of an affected
entry $r$ and an entry $r'$ from the incoming counting results, $u.dev$ computes
the intersection of their predicates. If the intersection is not empty, a new
entry $r^{new}$ is created in $LocCIB(u)$ for predicate $r.pred \cap r'.pred$.
The $count$ of $r^{new}$ is computed in two steps: (1) perform an inverse
operation of $\otimes$ or $\oplus$ between $r.count$ and $v$'s previous
counting result in $r.causality$, to remove the impact of the latter; and (2)
perform $\otimes$ or $\oplus$ between the result from the last step and $r'.count$
to get the latest counting result. The $action$ field is the same as $r$. The
$causality$ of this entry inherits from that of $r$, with tuple $(v, r')$
replacing $v$'s previous record. After computing and inserting all new entries, 
all affected entries are removed from $LocCIB(u)$.

Figure~\ref{fig:dvp-update} shows how $u$ in
Figure~\ref{fig:dvp-loccib} processes an UPDATE message from $v.dev$ to update its
$CIBIn(v)$ and $LocCIB(u)$.

\para{Step 3: updating $CIBOut(u)$.}
$u.dev$ puts the predicates of all
entries removed from $LocCIB(u)$ in the withdrawn predicates. For all
inserted entries of $LocCIB(u)$, it strips the $action$ and $causality$
fields, merges entries with the same $count$ value, and puts the results 
in the incoming counting results. 

After processing the UPDATE message, for each upstream neighbor
$w$ of $u$, $u.dev$ sends an UPDATE messaging consisting of an intended link
$(w, u)$ and $CIBOut(u)$.

%\para{Aggregating $LocalCIB(u)$.} 

\para{Internal event handling.} 
If $u.dev$ has an internal event (\eg, rule update or
link down), we handle it similar to handling an UPDATE message.
For example, if a link is down, we consider
predicates forwarded to that link update their counts to 0.
Specifically, the predicates whose forwarding actions are changed by the event
are considered as withdrawn predicates and the predicates in incoming
count results of an UPDATE message.  
Different from handling regular UPDATE messages, no $CIBIn(v)$ needs
updating. The counts of newly inserted entries in $LocCIB(u)$ are computed
by inverting $\otimes$/$\oplus$ and reading related entries in different $CIBIn(v)$s. 
Only predicates with new counts are included as withdrawn predicates and
incoming counting results in $CIBOut(u)$.

\para{Outbound UPDATE message suppression.} 
Multiple rule updates may occur in a short time during a network event
(\eg, a configuration update). Verifying the transient DP 
may not be necessary, and waste computation and communication resources.
As such, the DV protocol provides an optional dampening mechanism:
%inspired by BGP: 
after $u.dev$ finishes processing an UPDATE message,
before sending any UPDATE messages, it first checks if it still has
unprocessed UPDATE messages with an intended link from $u$ or internal events. 
If so, it continues processing them until no one is left unprocessed, and 
sends the latest $CIBOut(u)$ in one UPDATE message.

%To avoid the dampeing, A device may receive multiple UPDATE messages
%in a short time interval. Processing  

%\vspace{-1em}
\section{Extensions}\label{sec:ext}
\para{Packet transformations.} 
%\system{} verifies 
For data planes with packet transformations, \system{} uses BDD to
encode such actions~\cite{scalable-ap-ton17}, and extends the
CIB and the DV UPDATE message to record and share the count results of packet
transformation actions. 

\para{Large networks with a huge number of valid paths.}
One concern is that \dvnet{} may be too large to generate in
large networks with a huge number of valid paths. First, our survey and private
conversations with operators suggest that they usually want the network
to use paths with limited hops, if not the shortest one. The number of
such paths is small even in large networks. Second, if a network wants to
verify requirements with a huge number of valid paths, \system{} is 
inspired by BGP to verify them via 
divide-and-conquer: divide the network into partitions abstracted as
one-big-switches, construct \dvnet{} on this abstract network, and perform
intra-/inter-partition distributed verifications.

\para{Incremental deployment.} 
%\para{Q: [\system{} Deployment Flexibility]} 
%Does \system{} have to be distributively deployed at network devices? 
%verify requirements related to packet transformation?
\system{} can be deployed incrementally in two ways.  The first is to assign
an off-device instance (\eg, VM) for each device
without an on-device verifier, who plays as a verifier
to collect the data plane from the device and exchange messages with other
verifiers based on \dvnet{}. This is a generalization of the deployment of
RCDC, whose local verifiers are deployed in off-device instances. The
second is the divide-and-conquer approach above. We deploy
one verifier on one server for each partition. The verifier collects the data
planes of devices in its partition to perform intra-partition verification, and
exchanges the results with verifiers of other partitions for inter-partition
verification. The two approaches are not exclusive.

\para{Multi-path comparison.}
%What are the limitations of \system{}? In other words, what is difficult for
%\system{} to achieve?
The \system{} specification language (\S\ref{sec:lang}) currently
does not support specifying "multi-path" requirements that compare the packet
traces of two packet spaces (\eg, route symmetry and node-disjointness).  To
address this issue, one may extend the syntax
with an $id$ keyword to refer to different packet spaces, and allow
users to define trace comparison operators using predicate logic. To verify
them, one may construct the reachability \dvnet{} for each
packet space, let on-device verifiers collect the actual downstream paths 
and send them to their upstream neighbors, and eventually perform the user-defined
comparison operation with the complete paths of the two packet spaces
as input. We leave its full investigation as future work.

\iffalse{
\para{A:} First, \system{} currently only supports ``single-path" requirements,
\ie, requirements specifying each trace should satisfy certain patterns. It
cannot verify ``multi-path" requirements, \eg, node-disjoint / link-disjoint
paths. Verifying these requirements may require the help of requirement-related
algorithms, \eg, the Suurballe's algorithm~\cite{suurballe1974disjoint}. Second,
the performance of \system{} highly depends on the representation of predicates.
Compared with BDD, novel data structures (\eg, ddNF~\cite{bjorner2016ddnf} and
\#PEC~\cite{horn2019precise}) may help further improve \system{}'s efficiency.
Third, \system{} does not support verifying stateful (\eg, middlebox) or
programmable (\eg, P4\cite{bosshart2014p4}) data planes. We leave the
investigation of these limitations as future work.
}\fi

%\vspace{-0.5em}
\section{Performance Evaluation}~\label{sec:eval}
%\vspace{-0.3em}
We implement a prototype of \system{} in Java with \textasciitilde 8K LoC 
(Appendix~\ref{sec:impl}) and conduct extensive evaluations.
\iffalse{, to
demonstrate the benefits of \system{} for achieving scalable,
on-device data plane checking in diverse topologies under various scenarios.}\fi 
We focus on answering the
following questions: 
(1) What is the capability of \system{} in verifying a wide range of
requirements? (\S\ref{sec:eval-function}) 
(2) What is the performance of \system{} in a testbed environment with different
types of network devices, mimicking a real-world wide area network (WAN)?
(\S\ref{sec:eval-testbed})
(3) What is the performance of \system{} in various real-world, large-scale
networks (WAN/LAN/DC) under various DPV scenarios?
(\S\ref{sec:eval-simulation})
(4) What is the overhead of running \system{} on commodity network
devices?  (\S\ref{sec:eval-micro})

\iffalse{
a real-world,
large-scale network with a large number of data plane updates in a short
time? (\S~\ref{sec:eval-green-start}) (4) What is the performance of
\system{} on incrementally verifying various networks?
(\S~\ref{sec:eval-incremental}) (5) What is the overhead \system{} bring to
network devices (\eg, CPU load and memory)?  (\S~\ref{sec:eval-micro})
}\fi

\iffalse{
We conduct extensive evaluations on real data plane datasets to answer the following questions:
\begin{enumerate}
    \item What is the performance of \system{} in the green-start verification of a large-scale network with large FIB size?(Sec. \ref{sec:eval-green-start})
    \item What is the performance of \system{} for incremental verification?(Sec. \ref{sec:eval-incremental})
    \item What is the overhead and resource consumption of \system?(Sec. \ref{sec:eval-micro})
\end{enumerate}
}\fi

%\input{implementation}
\begin{table}[t]
    \centering
    \resizebox{\linewidth}{!}
    {
    \begin{tabular}{|c|c|c|c|}
            \hline
        \textbf{Models} & \textbf{CPU} & \textbf{Cores} \\
        \hline
        Mellanox SN2700 \cite{mellanox} & Intel(R) Celeron(R) CPU 1047UE @ 1.40GHz & 2 \\
        \hline
        Edgecore Wedge32-100X \cite{edgecore} & Intel(R) Pentium(R) CPU D1517 @ 1.60GHz & 4\\
        \hline
        Barefoot S9180-32X \cite{barefoot} & Intel(R) Xeon(R) CPU D-1527 @ 2.20GHz & 8 \\
        \hline
    \end{tabular}
    }
    \caption{Devices in the testbed.}
    \label{tab:testbed}
\vspace{-1.5em}
\end{table}

\subsection{Functionality Demonstrations}\label{sec:eval-function}
To demonstrate the capability of \system{} in verifying a wide range of DPV
requirements, we assemble a network of six switches: 
4 Mellanox SN2700 switches, 1 Edgecore whitebox
switch and 1 Barefoot Tofino switch
(Table~\ref{tab:testbed}). The first two models are installed
SONiC~\cite{sonic}, and the third is installed ONL~\cite{onl}. The topology is
the same as in Figure~\ref{fig:workflow-topo}. For each device, we
deploy a \system{} on-device verifier.

We run experiments to verify (1) loop-free, waypoint
reachability from $S$ to $D$ in Figure~\ref{fig:workflow-topo}, (2)
loop-free, multicast from $S$ to $C$ and $D$, (3) loop-free,
anycast from $S$ to $B$ and $D$, (4) different-ingress consistent
loop-free reachability from $S$ and $B$ to $D$, and (5) all-shortest-path
availability from $S$ to $C$. After the planner sends the on-device tasks to
switches, we disconnect it from the switches. We then configure devices with a
correct data plane satisfying these requirements and run the experiment. As
expected, no data plane error is reported by on-device verifiers. Next, we
iteratively reconfigure devices with an erroneous data plane that violates one
of the requirements above, and rerun the experiments. Each time, the on-device
verifier at $S$ successfully reports the error, except for the final
experiment, where we configure device $B$ with an incorrect data plane to violate the
all-shortest-path reachability from $S$ to $C$. This time, the verifier on $B$ locally
detects and reports this error without propagating any message to neighbors.
This shows that \system{} can verify the all-shortest-path availability locally
as RCDC does, making it a special case of \system{}.  Details of these
demos can be found at~\cite{demo}.

\subsection{Testbed Experiments}\label{sec:eval-testbed}
We extend our testbed with 3 Barefoot switches, to mimic the 9-device Internet2
WAN~\cite{internet2}. We install the forwarding rules of different devices to
corresponding switches in the testbed, and inject latencies between switches,
based on the propagation latencies between the locations of Internet2
devices~\cite{latency}.  We verify the conjunction of loop-freeness,
blackhole-freeness and all-pair reachability between switches along paths with
($\leq$x+2) hops, where $x$ is the smallest-hop-count for each pair of switches.

\begin{figure*}[h!tbp]
\setlength{\abovecaptionskip}{0cm}
\setlength{\belowcaptionskip}{-0.cm}
    \centering
    \begin{minipage}[b]{0.7\linewidth}
	\centering
        \begin{subfigure}{\linewidth}
    	   \setlength{\abovecaptionskip}{0cm}
            \setlength{\belowcaptionskip}{-0.cm}
            \centering\includegraphics[width=\linewidth]{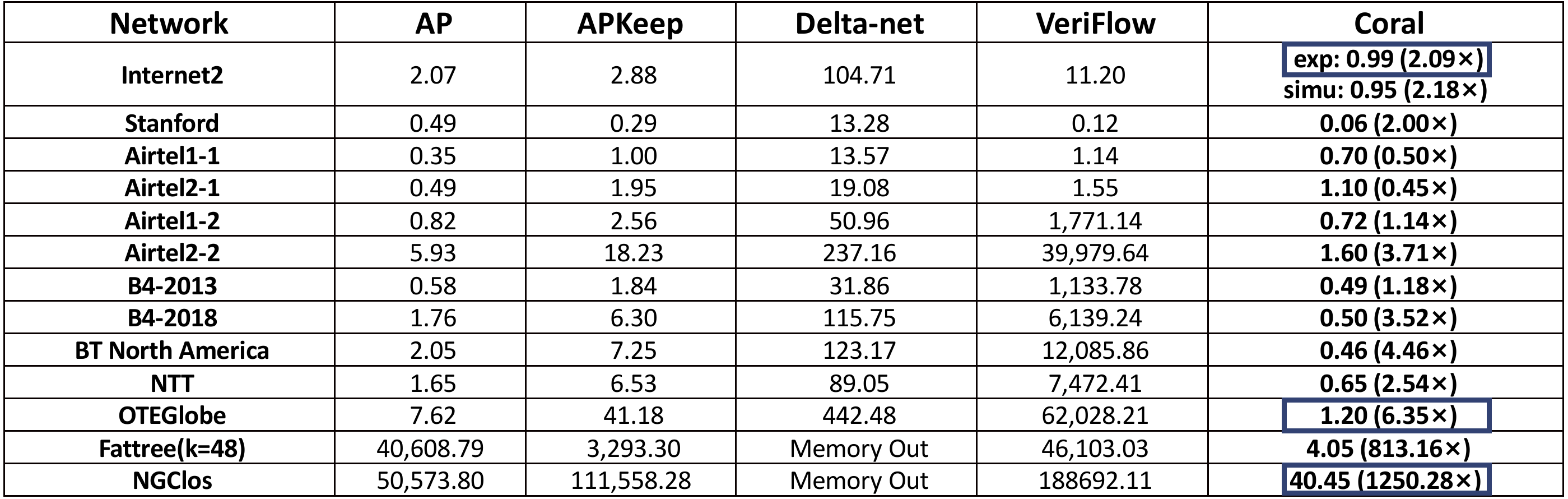}
	       \caption{\label{fig:green-start2} \footnotesize Verification time of burst update (seconds).}
    	\end{subfigure}
    	\begin{subfigure}{\linewidth}
    	   \setlength{\abovecaptionskip}{0cm}
            \setlength{\belowcaptionskip}{-0.cm}
		\centering\includegraphics[width=\linewidth]{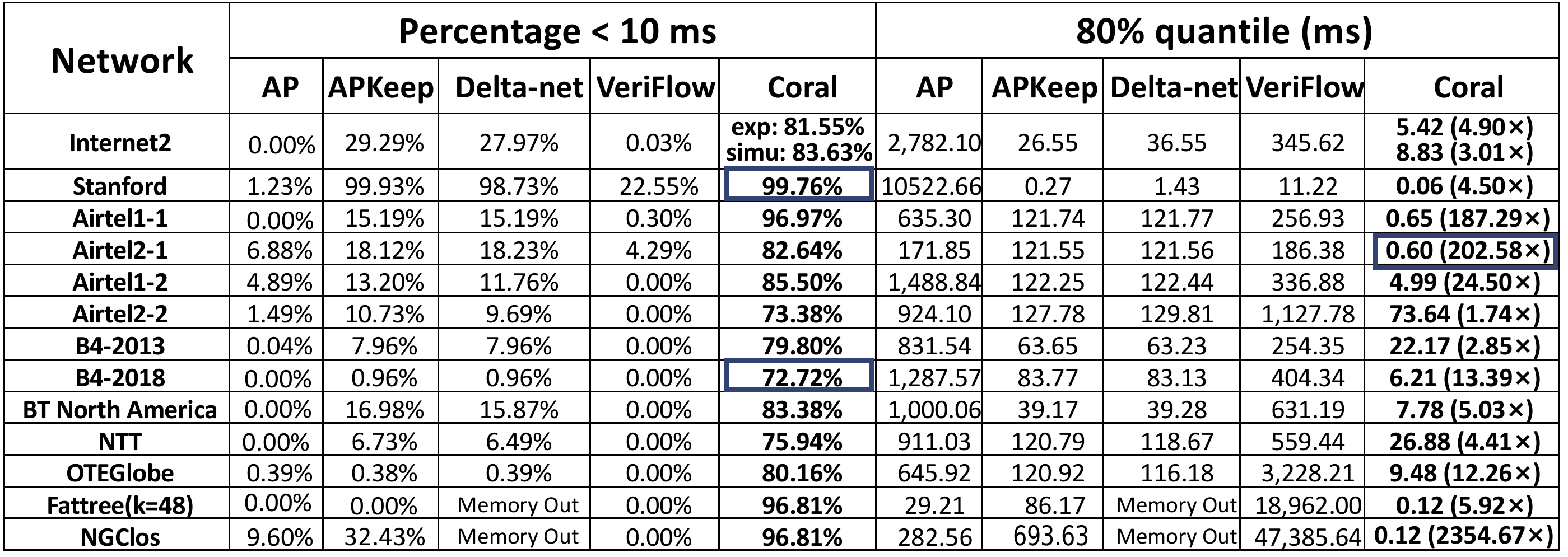}
        	\caption{\label{fig:incremental}\footnotesize Verification time of incremental update.}
    	\end{subfigure}
    	\caption{Verification time of experiments and large-scale simulations.}
    \end{minipage}
    \begin{minipage}[b]{0.29\linewidth}
        \centering
        \begin{subfigure}{\linewidth}
    	   \setlength{\abovecaptionskip}{0cm}
            \setlength{\belowcaptionskip}{-0.cm}
            \centering\includegraphics[width=.98\linewidth]{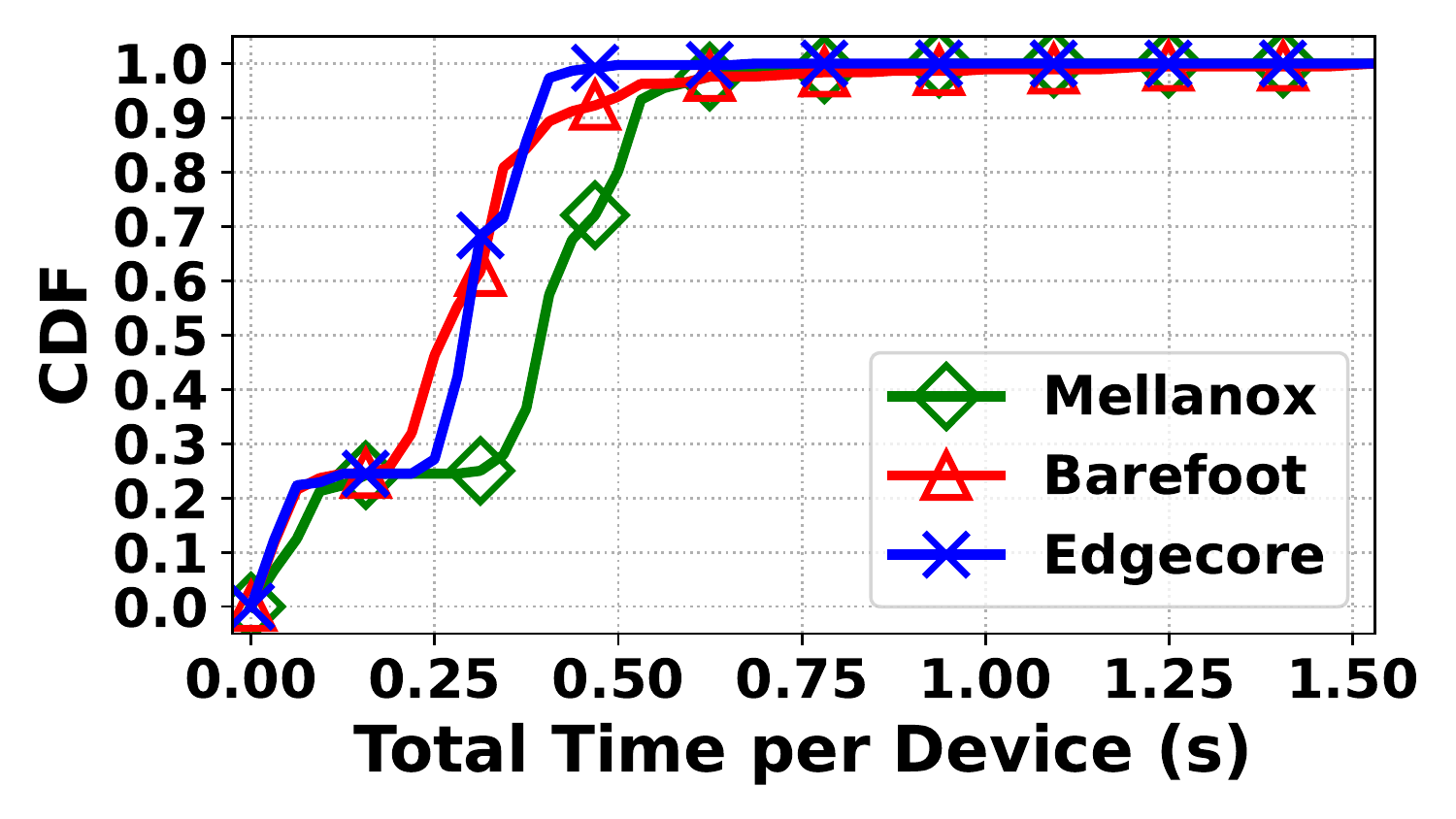}
	 %       \caption{\label{fig:init-total-time} \footnotesize Total time.}
    	\end{subfigure}
    	\begin{subfigure}{\linewidth}
    	   \setlength{\abovecaptionskip}{0cm}
            \setlength{\belowcaptionskip}{-0.cm}
		\centering\includegraphics[width=.98\linewidth]{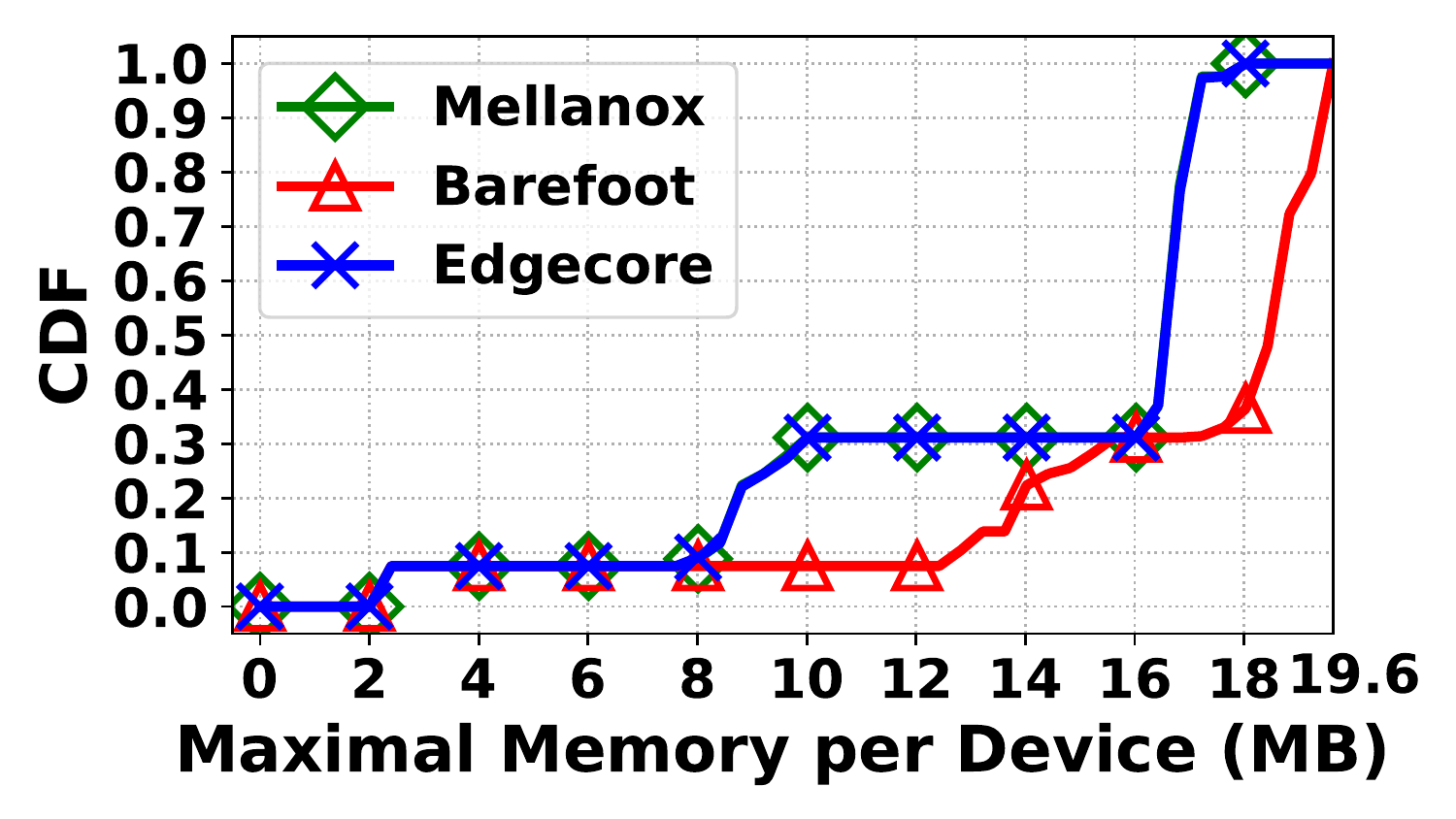}
        %	\caption{\label{fig:init-memory}\footnotesize Maximal memory.}
    	\end{subfigure}
    	\begin{subfigure}{\linewidth}
            \setlength{\abovecaptionskip}{0cm}
            \setlength{\belowcaptionskip}{-0.cm}
            \centering\includegraphics[width=.98\linewidth]{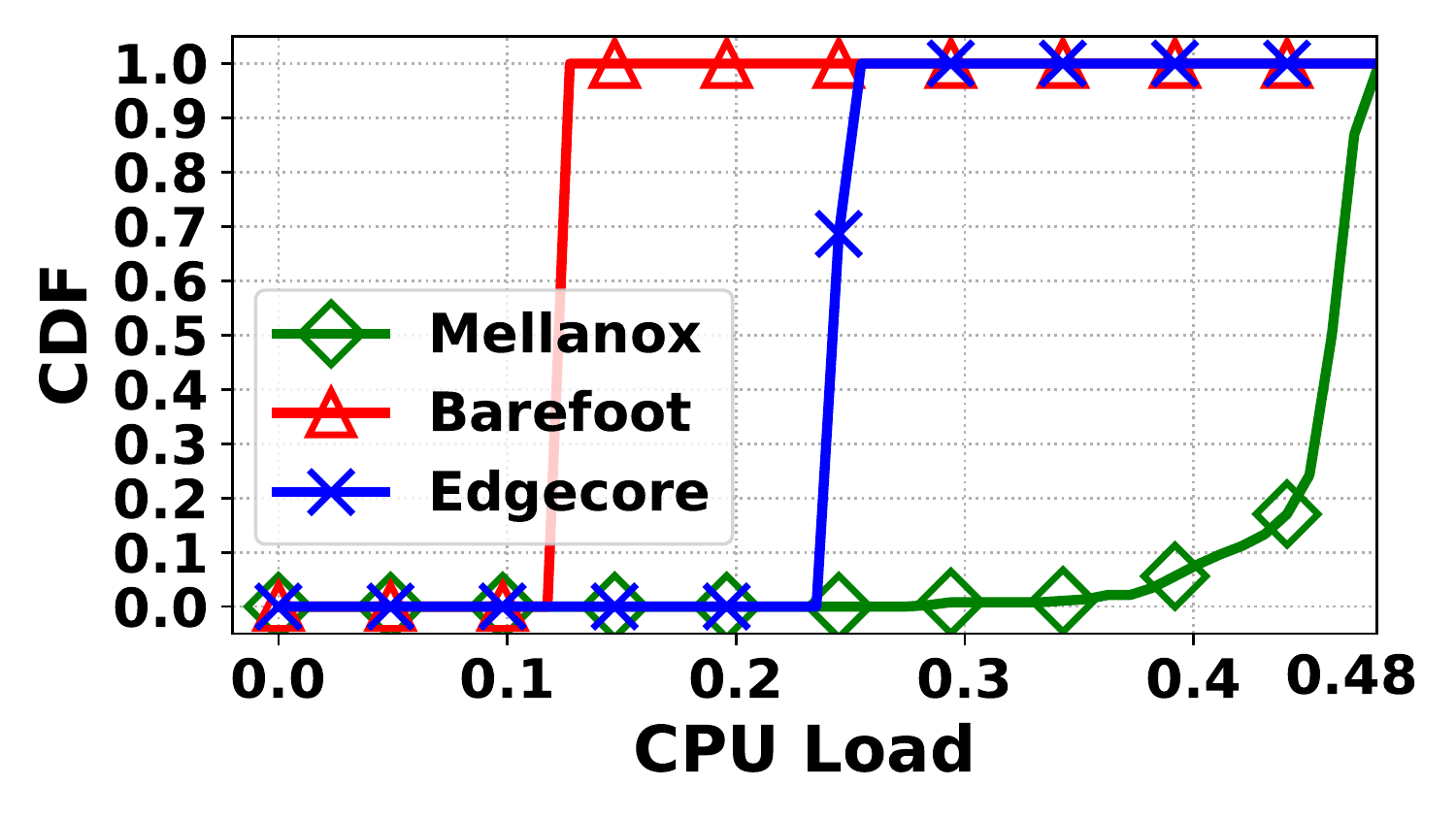}
    	    %\caption{\label{fig:init-cpu-load} \footnotesize CPU load.}
    	\end{subfigure}
	    \caption{Initialization overhead.}
%\caption{\footnotesize{Initialization overhead.}}
    	\label{fig:eval-init}
    \end{minipage}
\vspace{-1em}
\end{figure*}

%As such, we mimic Internet2 in the testbed.
\para{Experiment 1: burst update.} 
We first evaluate \system{} in the scenario of burst update, \ie, all forwarding
rules are installed to corresponding switches all at once. \system{} finishes
the verification in 0.99 seconds, outperforming the best centralized DPV in
comparison by 2.09$\times$ (Figure~\ref{fig:green-start2}). 
%takes over 2 seconds on a powerful
%server .

%This shows the
%efficacy and efficiency of \system{} in performing on-device data plane
%checking. 

%\hcy{0.403+0.582}
\para{Experiment 2: incremental update.}
We start from the snapshot after the burst update, randomly
generate $10K$ rule updates distributed evenly across devices and apply them
one by one. After each update, we incrementally verify the network.
For 80\% of the updates, \system{} finishes the incremental verification 
$\leq 5.42ms$, outperforming the best centralized DPV in comparison
by 4.90$\times$. This is because in \system{}, when a rule update happens,
only devices whose on-device task results are affected need to incrementally
update their results, and only these changed results are sent to neighbors
incrementally. For most rule updates, the number of these affected devices
is small (shown in Appendix~\ref{sec:update-msg}). 
%\todo{further clarify?}

For both experiments, we also measure the overhead of running \system{}
on-device verifiers on switches. 
%CPU load and memory
%consumption of each switch running the \system{} on-device verifier, 
We present the results in a more
comprehensive way in \S\ref{sec:eval-micro}.

\subsection{Large-Scale Simulations}\label{sec:eval-simulation}
We implement an event-driven simulator to evaluate 
\system{} in various real-world networks, 
on a server with 2 Intel Xeon Silver 4210R CPUs and 128 GB
memory.

%(10 cores each)

\subsubsection{Simulation Setup} We first introduce the settings.

\para{Datasets.} We use 13 datasets in 
Table~\ref{tab:datasets}. The first four
are public datasets and the others are synthesized with public topologies~\cite{knight2011internet, spring2002measuring,
jain2013b4, hong2018b4}. 
Fattree is a $48$-ary fattree~\cite{fattree}. \lnet{} is a real, large,
Clos-based DC.  For WAN, we assign link latencies based on the device locations
in the datasets~\cite{latency}. For LAN and DC, we assign 10$\mu$s latency for
each link. 

\iffalse{
Both use BGP with routing policies 
in~\cite{rfc7938, azure}.
}\fi
%~\qiao{too high? 1 microseconds?}.
%We collect their forwarding rules by simulating the network. 

\iffalse{
The Facebook dataset is from one of the Facebook's production
data centers, the topology follows the Fabric~\cite{fabric} architecture and
contains 112 Pods. The Fattree dataset contains 48 Pods. 
}\fi

\para{Comparison methods.} 
We compare \system{} with four state-of-the-art centralized DPV tools:
AP~\cite{ap-icnp13}, APKeep~\cite{apkeep-nsdi20},
Delta-net~\cite{deltanet-nsdi17} and Veriflow~\cite{veriflow}. We reproduce APKeep and
Delta-net, and use the
open-sourced version of AP and Veriflow.

%, and based on their pseudocode.

\para{Requirements.}
We verify the all-pair loop-free, blackhole-free, ($\leq$x+2)-hop reachability
in \S\ref{sec:eval-testbed} for WAN/LAN and the all-ToR-pair shortest path
reachability for DC. We also use \system{} to verify the local contracts of 
all-shortest-path availability of DC, as RCDC does, in Appendix~\ref{sec:rcdc}.

%For a fair comparison, we evaluate common DPV requirements including 
%loop-freeness, blackhole freeness, and all-pair reachability.  

\para{Metric.} 
We study the verification time. It is computed as the 
period from the arrival of data plane updates at devices to the time when all
requirements are verified, including the propagation delays. For
centralized DPV, we randomly assign a device as the location of the
server, and let all devices send data planes to the server
along lowest-latency paths. We also measure \system{}'s message
overhead in Appendix~\ref{sec:update-msg}.

\iffalse{
We believe this is fair because although \system{} needs to
transmit counting results among devices, centralized DPV tools also need to
collect data plane updates from devices. To demonstrate that \system{} can be
deployed on commodity network devices, we also measure its resource overhead
(\ie, memory, and DV UPDATE message size).
}\fi

\iffalse{
According to \cite{apkeep-nsdi20}, the changes of
forwarding behavior for each device are locally performed at each device, we
consider the predicate serialization/transmission/deserialization time in our
evaluations. Meanwhile, we capture the resource overhead (i.e., memory
consumption, number of entires in CIB and the announcement message size between
devices) of \system{} in the green-start verification process.
}\fi

\iffalse{
\para{Methodology.} We conduct two types of evaluations on the datasets.
\begin{enumerate}
    \item \textbf{Green-start verification.} Start the verification from scratch and verifies the all-pair reachability.
    \item \textbf{Incremental verification.} Based on the green-start verification, we randomly update the data plane by modifying/deleting a forwarding rule, then trigger the incremental verification for all-pair reachability.
\end{enumerate}
}\fi

\iffalse{
\para{Environment.} All evaluations are performed using event-driven simulation
on a server with 2 Intel Xeon Silver 4210R CPUs (10 cores each) and a 128 GB
memory.
}\fi

\iffalse{
All  evaluations are conducted on cloud Ubuntu servers with
8 vCPUs (2.5GHz) and 32GB memory.
}\fi

\begin{table}[t]
    \centering
%\footnotesize
\resizebox{0.7\linewidth}{!}{
    \begin{tabular}{|c|c|c|c|c|}
	        \hline
        \textbf{Network} & \textbf{\#Devices} & \textbf{\#Links} & \textbf{\#Rules} & \textbf{Type}\\
        \hline
	    Internet2~\cite{internet2} & 9 & 28 & $7.74 \times 10^4$ & WAN \\
	    \hline
	    Stanford~\cite{hsa} & 16 & 74 & $3.84 \times 10^3$ & LAN \\
	    \hline
	    Airtel1-1~\cite{deltanet-nsdi17} & 16 & 26 & $2.83 \times 10^4$ & WAN \\
        \hline
        Airtel2-1~\cite{deltanet-nsdi17} & 68 & 158 & $3.81 \times 10^4$ & WAN \\
        \hline
	Airtel1-2 & 16 & 26 & $9.60 \times 10^4$ & WAN \\
        \hline
        Airtel2-2 & 68 & 158 & $4.56 \times 10^5$ & WAN \\
        \hline
        B4-2013 & 12 & 18 & $7.92 \times 10^4$ & WAN \\
        \hline
        B4-2018 & 33 & 56 & $2.11 \times 10^5$ & WAN \\
        \hline
        BT North America & 36 & 76 & $2.52 \times 10^5$ & WAN \\
        \hline
        NTT & 47 & 63 & $1.98 \times 10^5$ & WAN \\
        \hline
        OTEGlobe & 93 & 103 & $7.22 \times 10^5$ & WAN \\
        \hline
	    Fattree ($k=48$) & 2,880 & 55,296 & $3.31 \times 10^6$ & DC \\
        \hline
	    \lnet{} & 6,016 & 43,008 & $3.23 \times 10^7$ & DC \\
	    \hline
    \end{tabular}
}
    \caption{Datasets statistics.}
    \label{tab:datasets}
 \vspace{-1.7em}
\end{table}

\iffalse{
\begin{figure*}[t]
    \centering
	\begin{subfigure}[t]{0.24\linewidth}
	    \centering\includegraphics[width=1\linewidth]{figures/eval/inc_fb.png}
		\caption{\label{fig:eval-inc-fb} \lnet.}
	\end{subfigure}
\hfill
	\begin{subfigure}[t]{0.24\linewidth}
	    \centering\includegraphics[width=1\linewidth]{figures/eval/inc_ft.png}
		\caption{\label{fig:eval-inc-ft} Fattree.}
	\end{subfigure}
\hfill
	\begin{subfigure}[t]{0.24\linewidth}
	    \centering\includegraphics[width=1\linewidth]{figures/eval/inc_i2.png}
		\caption{\label{fig:eval-inc-i2} Internet2.}
	\end{subfigure}
\hfill
	\begin{subfigure}[t]{0.24\linewidth}
	    \centering\includegraphics[width=1\linewidth]{figures/eval/inc_st.png}
		\caption{\label{fig:eval-inc-st} Stanford.}
	\end{subfigure}
    \caption{The incremental verification time (microseconds).}
    \label{fig:eval-incremental}
\end{figure*}
}\fi

\subsubsection{Results: Burst Update}\label{sec:eval-green-start}
%We first evaluate \system{}'s performance in burst updates. 
%Examples  include
%the green start of a data center~\cite{sung2016robotron, jupiter} and the update of large regional networks. To fully test its
%scalability, we consider the green start of all four topologies. 
%(2 data centers, 1 wide area network and 1 campus network).
Figure~\ref{fig:green-start2} gives the results. For WAN/LAN, 
\system{} completes the verification in $\leq 1.60s$ and
achieves an up to 6.35$\times$ speedup than the fastest centralized DPV. For DC, 
this speedup is up to 1250.28$\times$. This is because \system{}
decomposes verification into lightweight on-device tasks, which have
a dependency chain roughly linear to the network diameter. A DC
has a small diameter (\eg, 4 hops).  As such, on-device verifiers achieve a
very high level of parallelization, enabling high scalability of \system{}.

\iffalse{
We observe that in datacenters,
the verification time of \system{} is up to $2,758\times$ faster than APKeep.  Even for
smaller networks (\ie, Internet 2 and Stanford), the verification time of
\system{} is still at least $5\times$ faster than APKeep. 
}\fi

We note that \system{} is slower than AP in Airtel1-1
and Airtel1-2, but faster in Airtel1-2 and Airtel2-2 whose
topologies are the same pairwise. This is because the latter two
have a much higher number of rules (3.39$\times$ and
11.97$\times$). The bottleneck of AP is to transform rules to equivalence
classes (ECs), 
whose time increases linearly with the number of rules, leading to a
linear increase of total time. In contrast, \system{} only computes
LEC on devices in parallel, and is not a
bottleneck (Appendix~\ref{sec:green-break}). 
As such, with more rules, \system{} becomes faster than AP.

\subsubsection{Results: Incremental Update}\label{sec:eval-incremental}

\iffalse{
For each topology in Table~\ref{tab:datasets}, we use the data plane after the burst update experiment as the
starting point, randomly generate 1,000 data plane rule updates distributed
evenly across devices and applies them one by one. After each update, we incrementally verify the network
data plane. 
}\fi

% \input{incremental-evaluation}

We evaluate \system{} for incremental verification using the
same methodology as in \S\ref{sec:eval-testbed}. Figure~\ref{fig:incremental}
gives the results.
%compares the incremental verification time of \system{} with the other methods.
The 80\% quantile verification time of \system{} is up to 202.58$\times$ faster
than the fastest centralized DPV.  Among all datasets, \system{} finishes
verifying at least 72.72\% rule updates in less than 10$ms$, while this lower
bound of other tools is less than 1\%. This is for the same reason as in 
experiments (\S\ref{sec:eval-testbed}) and shown in Appendix~\ref{sec:update-msg}, 
and demonstrates that \system{} consistently enables scalable data
plane checkups under various networks and DPV scenarios.

\subsection{On-Device Microbenchmarks}\label{sec:eval-micro}
We run extensive microbenchmarks to measure 
%the initialization and message processing 
the overhead of \system{}
on-device verifiers. 
\iffalse{, both initializing LEC and CIB and processing DV protocol
messages, on commodity network devices.}\fi 
We also measure the latency of computing
\dvnet{} and on-device tasks in Appendix~\ref{sec:dvnet-overhead}.

\begin{figure*}[!htbp]
\setlength{\abovecaptionskip}{0cm}
\setlength{\belowcaptionskip}{-0.cm}
    \centering
	\begin{subfigure}{0.24\linewidth}
\setlength{\abovecaptionskip}{0cm}
\setlength{\belowcaptionskip}{-0.cm}
	    \centering\includegraphics[width=1\linewidth]{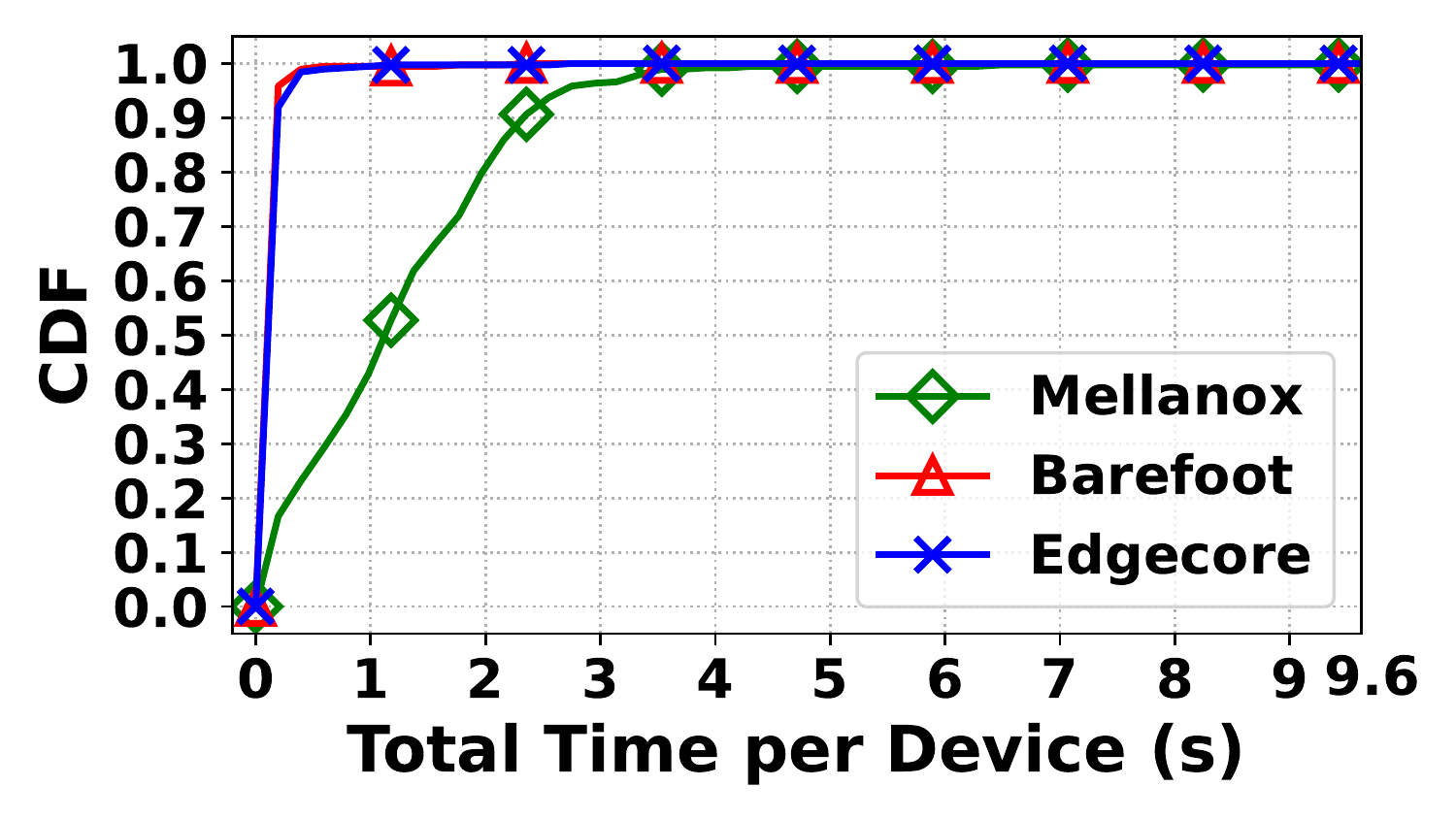}
%		\caption{\label{fig:trace-total-time} Total time.}
	\end{subfigure}
\hfill
	\begin{subfigure}{0.24\linewidth}
\setlength{\abovecaptionskip}{0cm}
\setlength{\belowcaptionskip}{-0.cm}
	    \centering\includegraphics[width=1\linewidth]{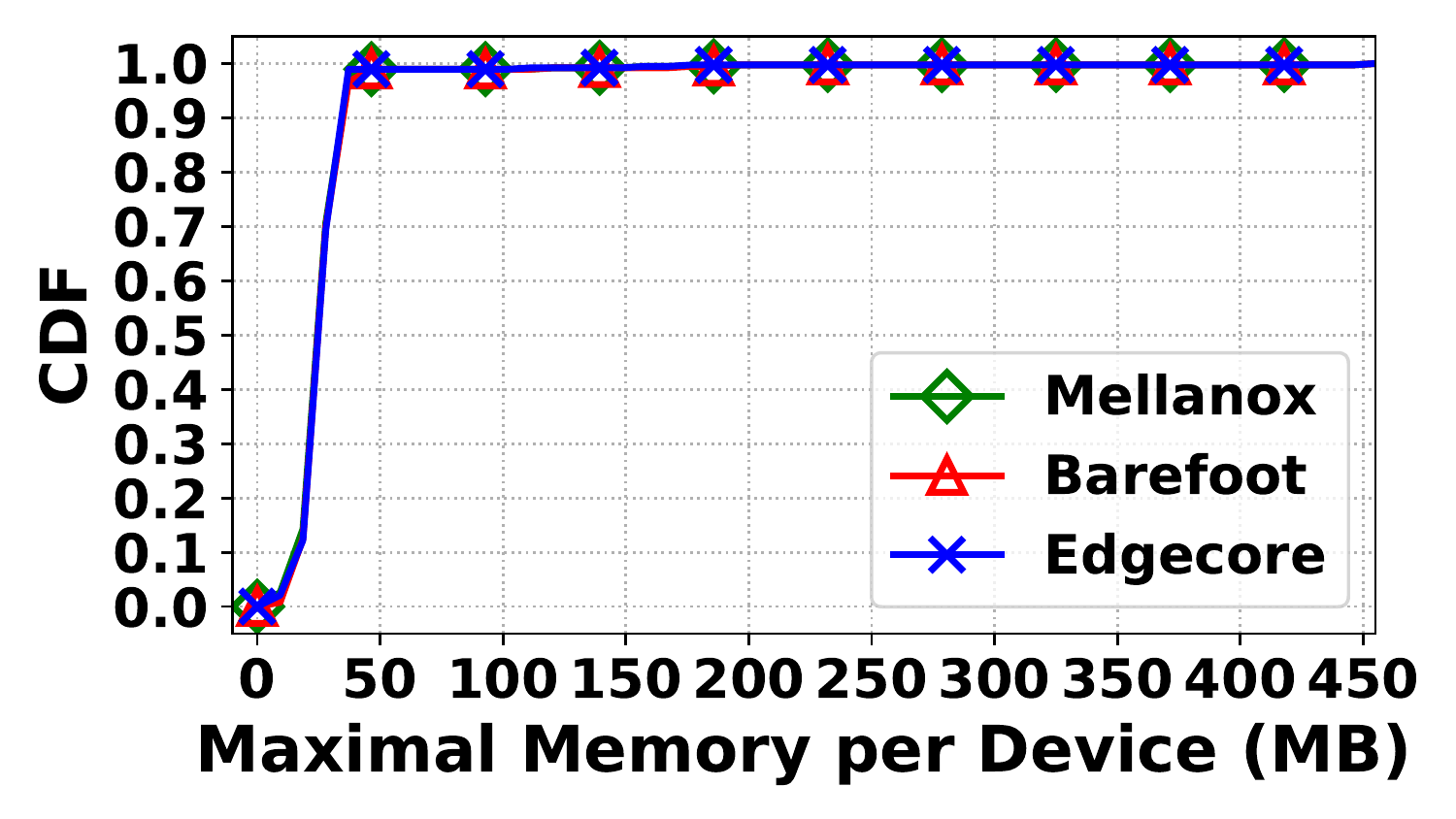}
%		\caption{\label{fig:trace-memory} Memory.}
	\end{subfigure}
\hfill
    \begin{subfigure}{0.24\linewidth}
\setlength{\abovecaptionskip}{0cm}
\setlength{\belowcaptionskip}{-0.cm}
        \centering\includegraphics[width=1\linewidth]{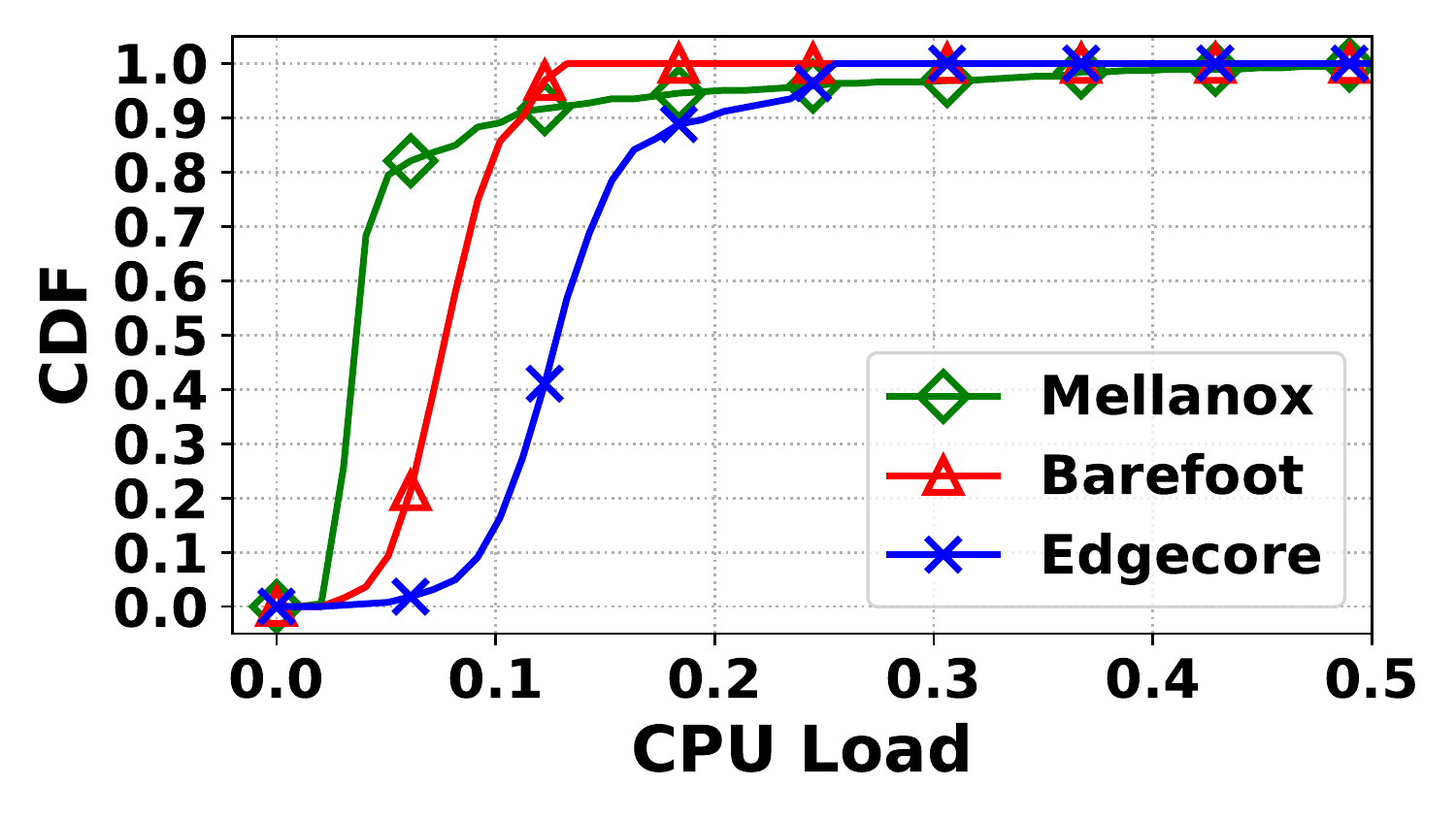}
%        \caption{\label{fig:trace-cpu-load} CPU load.}
    \end{subfigure}
\hfill
	\begin{subfigure}{0.24\linewidth}
\setlength{\abovecaptionskip}{0cm}
\setlength{\belowcaptionskip}{-0.cm}
	    \centering\includegraphics[width=1\linewidth]{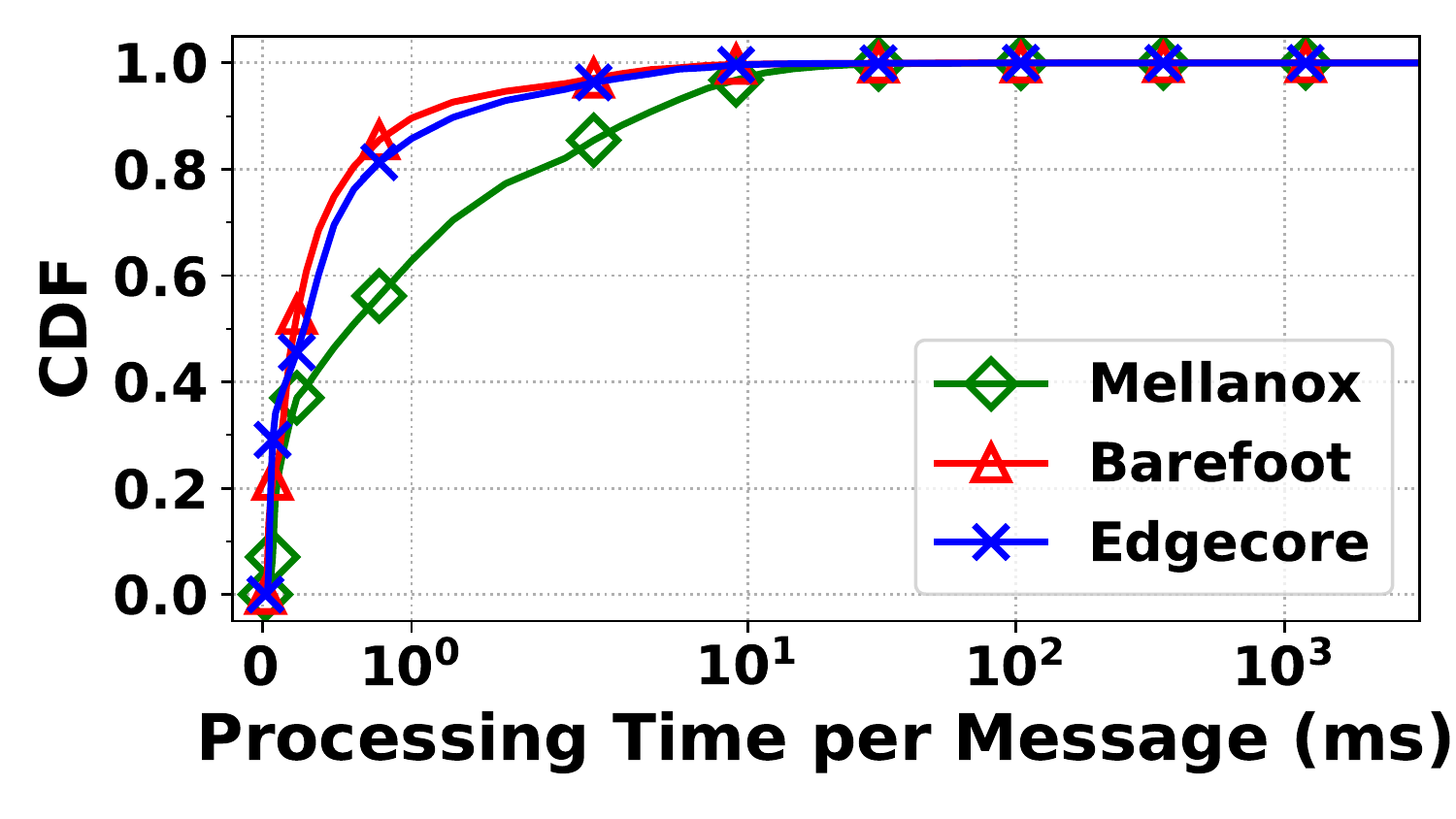}
%		\caption{\label{fig:trace-message} Time for each message.}
	\end{subfigure}
	\caption{DV UPDATE message processing overhead.}
	\label{fig:trace}
\vspace{-1.7em}
\end{figure*}

\para{Initialization overhead.} 
For each of 414 devices from WAN / LAN and 6 devices from \lnet{}/Fattree
(one edge, aggregation and core switch, respectively), we measure the overhead
of its initialization phase in burst update (\ie, computing the initial LEC and CIB), 
in terms of total time, maximal memory
and CPU load, on the three switch models in our testbed. The CPU load is
computed as $CPU~time$ $/(total~time$ $\times$ $number~of~cores)$.
Figure~\ref{fig:eval-init} plots their CDFs. On 
all three switches, all devices in the datasets complete initialization in $\leq
1.5s$, with a CPU load $\leq0.48$, and a maximal memory
$\leq 19.6MB$. 
\iffalse{As such, we conclude that the initialization overhead of
\system{} on-device verifier on commodity switches is reasonably low.}\fi

\para{DV UPDATE message processing overhead.}
For each of the same set of devices in the datasets, we collect the trace of
their received DV UPDATE messages during burst update and incremental update
experiments, replay the traces consecutively on each of the three switches, and
measure the message processing overhead in terms of
total time, maximal memory, CPU load and per message processing time. 
Figure~\ref{fig:trace} shows their
CDFs.  For 90\% of devices, all three switches process all UPDATE messages
in $\leq 2.29s$, with a maximal memory $\leq 32.08MB$, and a CPU load $\leq
0.20$. And for 90\% of all $835.2k$ UPDATE messages, the switches can process it
in $\leq 4ms$. 

To summarize, the initialization and messaging processing overhead
microbenchmarks show that \system{} on-device verifiers can be deployed on
commodity network devices with little overhead.
%The total number of DV UPDATE messages recorded and replayed is 835187. 

\iffalse{
As such, we conclude that the DV UPDATE
message processing overhead of \system{} on-device verifier on commodity
switches is also reasonably low.

From these microbenchmarks, we conclude that \system{} on-device verifiers can
be deployed on commodity network devices with a low overhead.
}\fi

\iffalse{
\subsection{Micro Benchmark}\label{sec:eval-micro}
To further understand the overhead of \system{}, we capture the resource consumption of \system{} including memory consumption, number of entries of CIB and the size of the announcement messages in the process of green-start verification. 

\para{Memory overhead.} Figure~\ref{fig:eval-micro-mem} shows the distribution of the memory consumption for each device, we see that 90\% are under ??, which is a reasonable overhead for modern network devices (e.g., the edge-core AS7800 series switches~\cite{edgecoreas7800} have up to 16GB memory). 

\para{CIB size.} Figure~\ref{fig:eval-micro-cib} shows the distribution of the number of entries in CIB of each device, we see that even for a large-scale network, the size of the CIBs are not large.

\para{Message size.} Figure~\ref{fig:eval-micro-size} shows the distribution of the message size, 90\% are smaller than ?? bytes, which is reasonable for modern networks with over 10Gbps links.

In conclusion, \system{} can achieve up to ??x faster green-start verification compared with the state-of-the-art, and compatible performance for incremental verification with a reasonable overhead.
}\fi

%\input{discussion}
\section{Related Work}\label{sec:related-work}
Network verification includes control plane
verification (CPV) that finds errors in configurations~\cite{bagpipe, batfish, minesweeper,
ali-sigcomm20, era-osdi16, bonsai, shapeshifter, gember2016fast, arc-sosp17,
tiramisu, plankton-nsdi20, netdice-sigcomm20, nv-pldi20, spider, groot,
rcc-nsdi05, fsr, ball2014vericon}; and DPV that checks the
correctness of the data plane. \system{} is a DPV tool, and can help 
simulation-based CPV~\cite{batfish,
liu2017crystalnet, lopes2019fast} verify the
simulated DP.

\para{Centralized DPV.}
Existing DPV tools \cite{xie2005static, flowchecker, anteater, hsa,
nod, netplumber, veriflow, deltanet-nsdi17, ap-icnp13, ap-ton16,
practical-ap-conext15, practical-ap-ton17, scalable-ap-ton17, apkeep-nsdi20,
symmetry, libra, azure} use a centralized server to collect and analyze the data planes of 
devices. Despite substantial efforts on performance
optimization, the centralized design makes them unscalable in nature due to the
need for reliable server-network connections and the server being a bottleneck and
single PoF.
%, in that (1) it requires reliable connections between the
%server and network devices, and (2) the server becomes both a performance
%bottleneck and a single point of failure. 
%In contrast, \system{} adopts a
%distributed design, which systematically decomposes the job of DPV into smaller
%tasks executed at devices, achieving scalable data plane correctness checkups. 
%\para{Scaling up DPV.}
%Several designs explore how to scale DPV~\cite{symmetry, libra, azure}. 
To scale up DPV, Plotkin
\textit{et al.}~\cite{symmetry} exploit the symmetry and
surgery in topology to aggregate the network to a smaller one.
%, but requires data
%planes at devices to be isomorphic. 
Libra~\cite{libra} parallelizes DPV by partitioning the data plane into
subnets. RCDC~\cite{azure} parallelizes verifying all-shortest-path
availability by partitioning the data plane by device. However, they are still
centralized designs with the limitations above.  In contrast, \system{} adopts a
distributed design, to systematically decompose DPV into small
tasks executed on network devices, achieving scalable data plane checkups on a wide
range of requirements.

\iffalse{
Its packet
space decomposition is orthogonal to the topology decomposition of
\system{}. Two can be integrated to further scale up DPV. The only known DPV tool
decomposing verification to local tasks is RCDC in  
Azure~\cite{azure}. It lets devices to locally verify its
data plane, but only for the availability
and fault-tolerance of shortest paths. \system{} is a more  generic
framework that can verify a wide range of 
requirements distributively,
including RCDC as a special case.
}\fi

\para{Verification of stateful/programmable DP.} 
%Other than verifying the data plane generated by routing protocols, 
Some studies investigate the verification of stateful DP (\eg, middleboxes) 
~\cite{panda2017verifying, zhang2020automated, zaostrovnykh2017formally, netsmc, chen2019performance} and
programmable DP (\eg, P4~\cite{bosshart2014p4})~\cite{liu2018p4v,
dumitrescu2020bf4}. Studying how to extend \system{} to verify
 stateful and programmable DP would be an interesting future
work.

\para{Network synthesis.}
Synthesis~\cite{merlin, contra, propane, tian2019safely,
el2018netcomplete} is complementary to
verification.
% compute  device configurations (\eg, routing protocol
%configurations and programmable data plane) that will generate data planes
%satisfying the specified intents. 
\system{} is inspired by some of
them~\cite{merlin, propane, contra} to use automata theory to
generate \dvnet{}. %, a compact representation of all valid paths.
%in the network allowed by requirement.

\para{Predicate representation.}
\system{} chooses BDD~\cite{bryant1986graph} to represent packets
%because %state-of-the-art DPV tools show that 
for its efficiency.
%it is efficient for predicate
%operations. 
Recent data structures (\eg, ddNF~\cite{bjorner2016ddnf}
and \#PEC~\cite{horn2019precise}) may have better
performance and benefit \system{}. We leave this as future work.
%\vspace{-0.3em}

%\vspace{-0.3em}
\section{Conclusion}\label{sec:conclusion}
%\vspace{-0.5em}
%We tackle the scalability challenge of DPV by
We design \system{}, a distributed DPV
framework to achieve scalable DPV by decomposing verification to lightweight 
on-device counting tasks. Extensive experiments demonstrate the benefits of
\system{} to achieve scalable DPV. This work does not raise any ethical issues.

\iffalse{
A DV protocol is
designed for on-device verifiers to communicate the
counting results. 
}\fi

%\end{spacing}

%\input{resa-srg}
%\end{spacing}
%\newpage
%{\small
%\bibliographystyle{abbrv}
\newpage
\bibliographystyle{ACM-Reference-Format}
\bibliography{all}  
%\appendix
%\input{tmp/icd}
%\input{backup}

\appendix
\newpage
\section{Proofs of \dvnet{} Backward Counting}
\subsection{Proof Sketch of the Correctness of
Algorithm~\ref{alg:counting}}\label{sec:counting-correctness}

Given a packet $p$ and a \dvnet{}, the goal of Algorithm~\ref{alg:counting} is
to compute the number of copies of $p$ that can be delivered by the network to
the destination of \dvnet{} along paths in the \dvnet{} in each universe.
Suppose Algorithm~\ref{alg:counting} is incorrect. There could be three cases:
(1) there exists a path in \dvnet{} that is provided by the network data plane,
but is not counted by Algorithm~\ref{alg:counting}; (2) There exists a path in
\dvnet{} that is not provided by the network data plane, but is counted by
Algorithm~\ref{alg:counting}; (3) Algorithm~\ref{alg:counting} counts a path 
out of \dvnet{}.  None of these cases could happen because at each node
$u$, Equations~(\ref{eqn:count-all}) (\ref{eqn:count-any}) only counts
$\mathbf{c}_{v_j}$ of $v_j$ with $b_{ij}=1$, \ie, the downstream neighbors of
$u$ whose devices are in the next-hops of $u.dev$ forwarding $p$ to. As such,
Algorithm~\ref{alg:counting} is correct.

\subsection{Proof of Proposition~\ref{prop:mic-exist}}\label{sec:proof-mic-exist}
Consider $\mathbf{c}_u$ of packet $p$ at $u$, and an upstream neighbor of $u$, denoted as $w$.
Suppose $u.dev$ is in the group of next-hops where $w.dev$ forwards $p$. Because
of the monotonicity of $\otimes$, in each universe that $w.dev$ forwards $p$ to
$u.dev$, the number of copies of $p$ that can be sent from $w$ to the
destination in \dvnet{} is greater than or equal to the number of copies of $p$
that can be sent from $u$ to the destination in \dvnet{}. As such, 
\begin{itemize}[leftmargin=*]
		\setlength\itemsep{0em}
\item When $count\_exp$ is $\geq N$ or $> N$, each $u$ only sends
$min(\mathbf{c}_u)$ to its upstream neighbors. With such information, in the
end, the source node of \dvnet{} can compute the lower bound of the number of
copies of $p$ delivered in all universes. If this lower bound satisfies
$count\_exp$, then all universes satisfy it. If this lower bound does  not
satisfy $count\_exp$, a network error is found. 
\item When $count\_exp$ is $\leq N$ or $< N$, each $u$ only sends
$max(\mathbf{c}_u)$ to its upstream neighbors. The analysis is similar, with the
source node computing the upper bound.
\item When $count\_exp$ is $==N$, if $\mathbf{c}_u$ has more than 1 count, it
means any action to forward $p$ to $u$ would mean a network error. In this case,
 $u$ only needs to send its upstream neighbors any 2 counts in $\mathbf{c}_u$ to
let them know that. If $\mathbf{c}_u$ has only 1
count, $u$ sends it to $u$'s upstream neighbors for further counting. 
Summarizing these two sub-cases, $u$ only needs to send the first
$min(|\mathbf{c}_u|, 2)$ smallest elements in $\mathbf{c}_u$ to its upstream
neighbors. 
\end{itemize}

With this analysis, we complete the proof of
Proposition~\ref{prop:mic-exist}.

%\vspace{-1em}

\begin{figure}
\setlength{\abovecaptionskip}{0cm}
\setlength{\belowcaptionskip}{-0.cm}
\centering
\includegraphics[width=0.9\columnwidth]{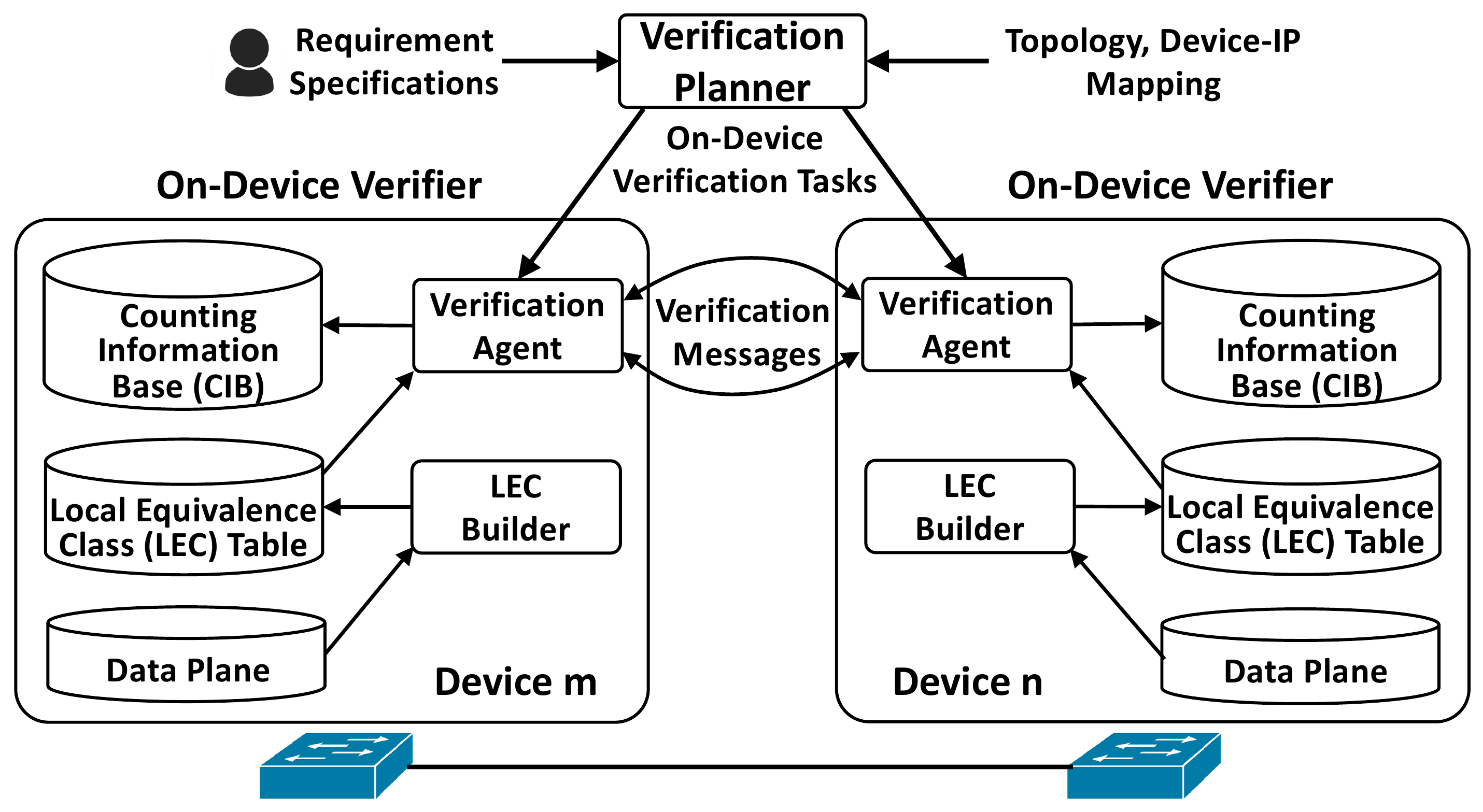}
\caption{The implementation of \system{}.}
\label{fig:impl}
%\reducespace
%\vspace{-1em}
\end{figure}

\section{Implementation}\label{sec:impl}
Our \system{}  prototype has \textasciitilde 8K lines of Java code, including a
verification planner and on-device verifiers. Figure~\ref{fig:impl} shows the
implementation structure. The
planner computes the \dvnet{} based on the requirement and the topology, and
decides the 
counting task of on-device verifiers.

An on-device verifier has (1) an LEC builder that reads the
data plane of the device to maintain an LEC table of a minimal number of LECs, and
(2) a verification agent that maintains TCP connections with the verifiers of
neighbor devices, takes in the LEC table and the DV protocol UPDATE messages
from neighbor devices to update the on-devices CIBs, and sends out UPDATE
messages with latest counting results to neighbor devices, based on counting tasks. 
For the verification agent, we
use a thread pool implementation, where a thread is assigned for a node in a
\dvnet{}. To avoid creating too many threads and hurting the system performance,
we design an opportunistic algorithm to merge threads with similar
responsibilities (\eg, requirements with different source IP prefixes but same
destination IP prefixes) into a single thread. A dispatcher thread receives
events (\eg, an LEC table update or a DV protocol UPDATE message), and
dispatches events to the corresponding thread. An LEC table update is sent to
all threads whose requirements overlap with the update, and an UPDATE message is
dispatched based on the intended link field of the UPDATE message. For predicate
operation and transmission, we adapt and modify the JDD~\cite{jdd} library to
support the serialization and deserialization between BDD and the Protobuf data
encoding~\cite{protobuf}, so that BDDs can be efficiently transmitted between devices
in UPDATE messages.

\iffalse{
We fully implement \system{} by ?? lines of Java code. \system{} works as agent
programs running at each device operating system. The device agent mainly
contains of 4 components: 1) a requirement parser that translates the regular
expression requirement to \dvnet. 2) a south-bound interface for listening for
the FIB changes, we provide both command line interfaces for
insert/delete/modify FIB and programmable interfaces as a library to adapt to
different vendors. 3) a set of north-bound TCP channels for transmitting the
counting information between neighbors, the counting messages are encoded by the
Protobuf~\cite{protobuf} data structures. 4) a local verification engine that
handles FIB change events from south-bound or counting messages from
north-bound, and emits update messages if the local CIB is changed.
For predicate operation and transmission, we modify the JDD~\cite{jdd} library to support the BDD serialization/deserialization to the Protobuf structures.
In our evaluation, we simulate the devices by threads and run the device agents in the thread scopes.
}\fi

\section{Verification Time Breakdown of Burst Update}\label{sec:green-break}

\begin{figure*}[t]
\setlength{\abovecaptionskip}{0cm}
\setlength{\belowcaptionskip}{-0.cm}
\centering
	\begin{subfigure}{0.33\linewidth}
\setlength{\abovecaptionskip}{0cm}
\setlength{\belowcaptionskip}{-0.cm}
	    \centering\includegraphics[width=1\linewidth]{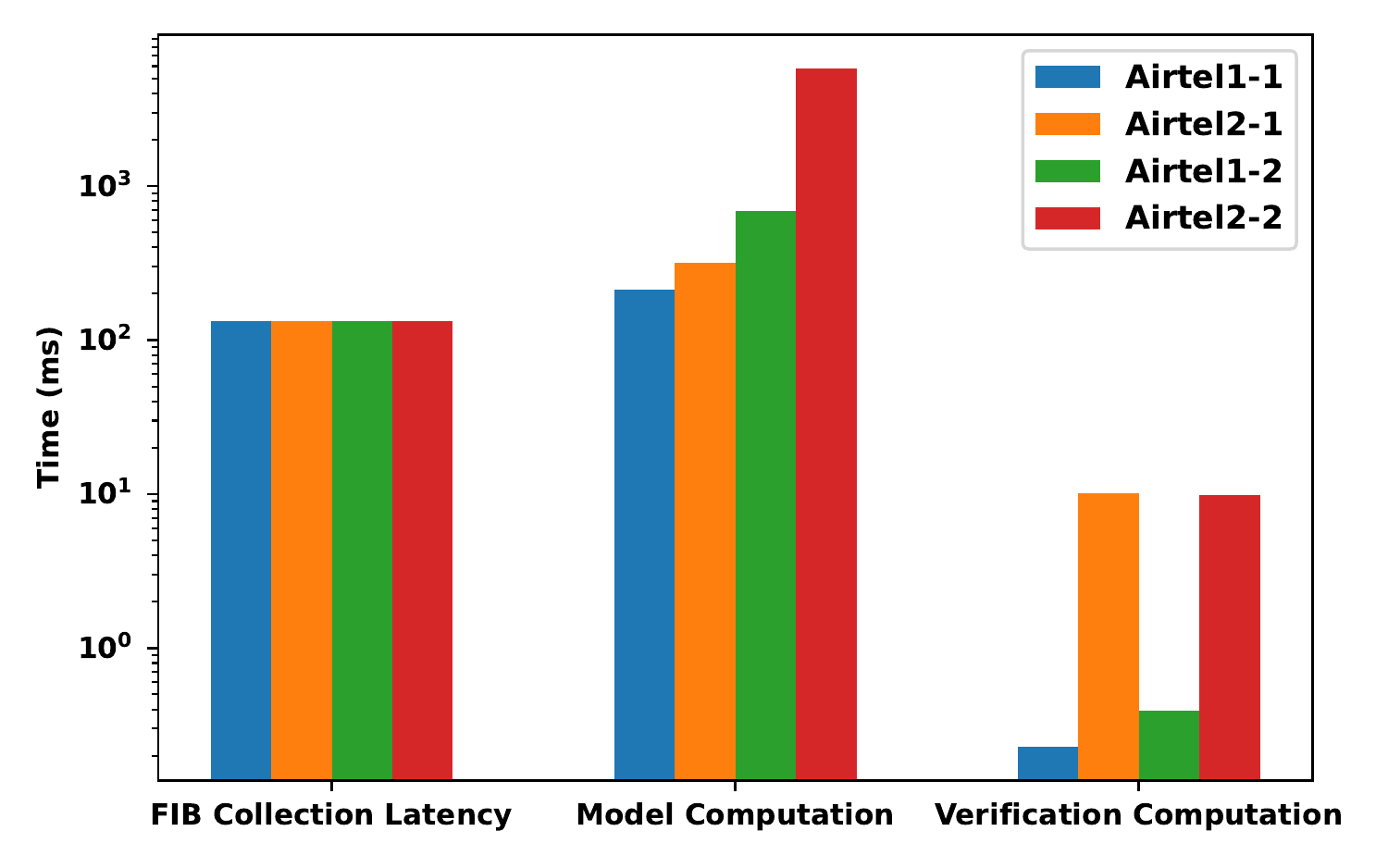}
		\caption{\label{fig:AP-time} AP.}
	\end{subfigure}
\hfill
	\begin{subfigure}{0.33\linewidth}
\setlength{\abovecaptionskip}{0cm}
\setlength{\belowcaptionskip}{-0.cm}
		\centering\includegraphics[width=1\linewidth]{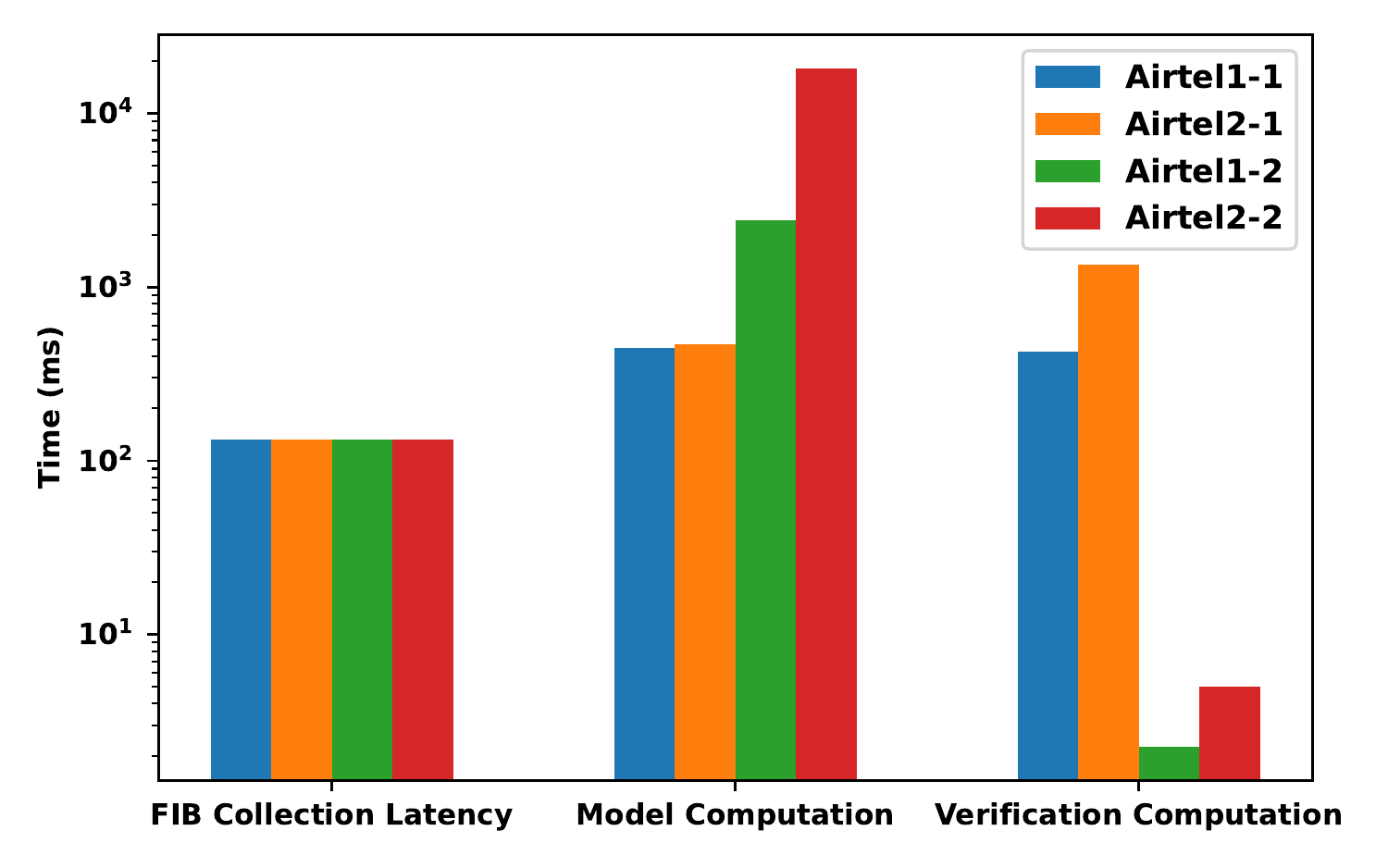}
		\caption{\label{fig:dc-memory} APKeep.}
	\end{subfigure}
\hfill
	\begin{subfigure}{0.33\linewidth}
\setlength{\abovecaptionskip}{0cm}
\setlength{\belowcaptionskip}{-0.cm}
		\centering\includegraphics[width=1\linewidth]{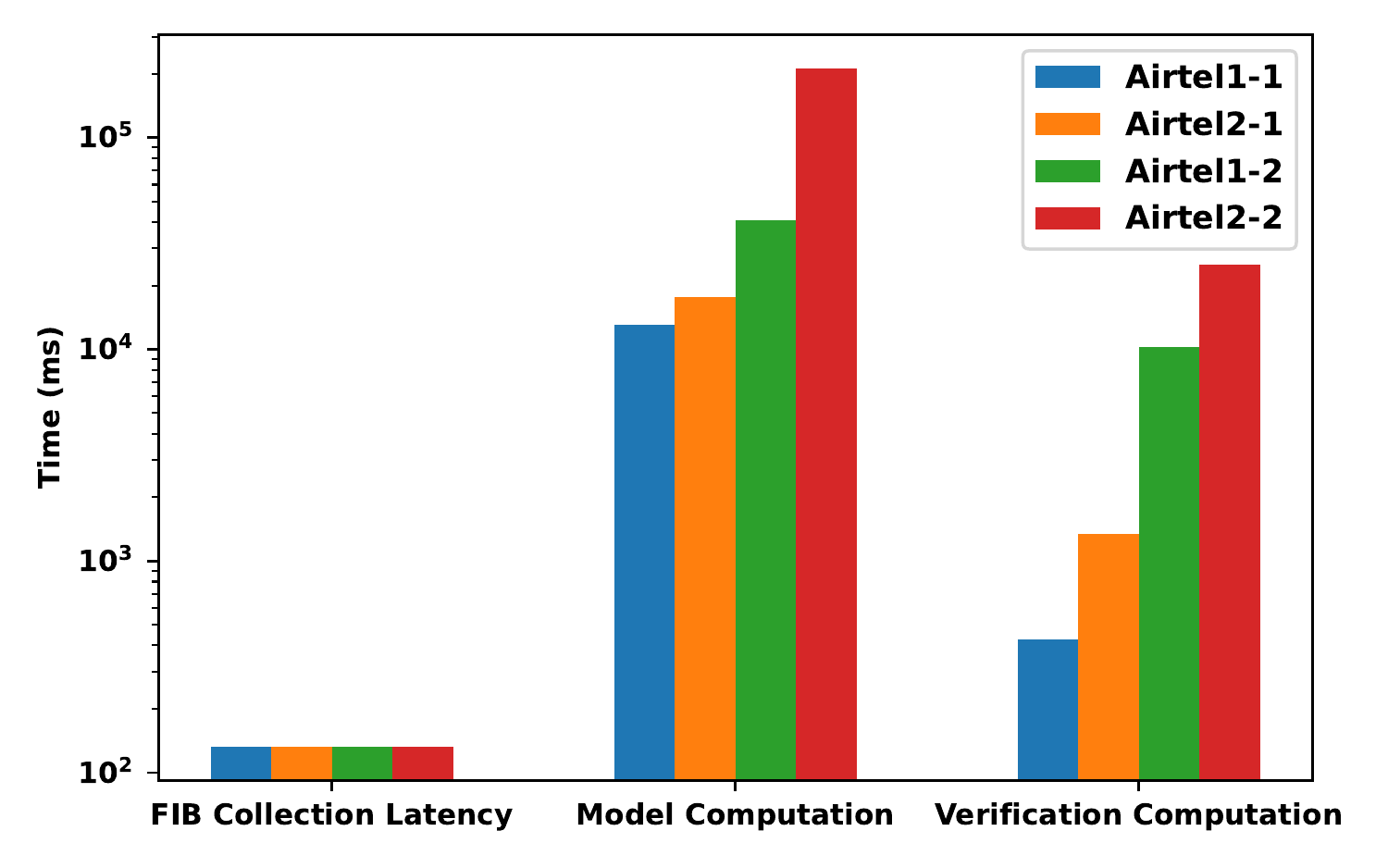}
		\caption{\label{fig:Delte-net-Airtel} Delta-net.}
	\end{subfigure}
\hfill
	\begin{subfigure}{0.33\linewidth}
\setlength{\abovecaptionskip}{0cm}
\setlength{\belowcaptionskip}{-0.cm}
		\centering\includegraphics[width=1\linewidth]{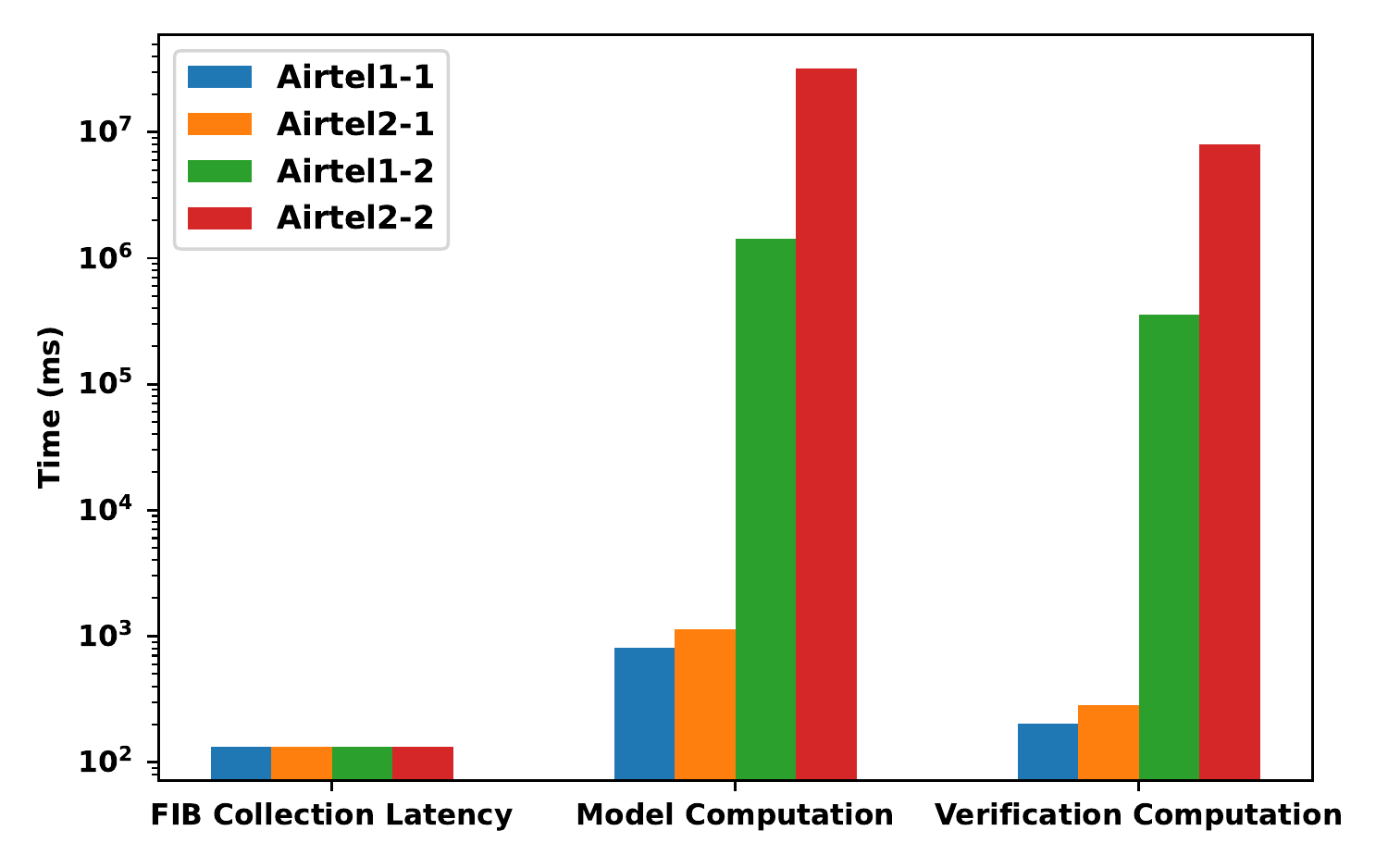}
		\caption{\label{fig:VeriFlow-Airtel} VeriFlow.}
	\end{subfigure}
%\hfill
	\begin{subfigure}{0.33\linewidth}
\setlength{\abovecaptionskip}{0cm}
\setlength{\belowcaptionskip}{-0.cm}
		\centering\includegraphics[width=1\linewidth]{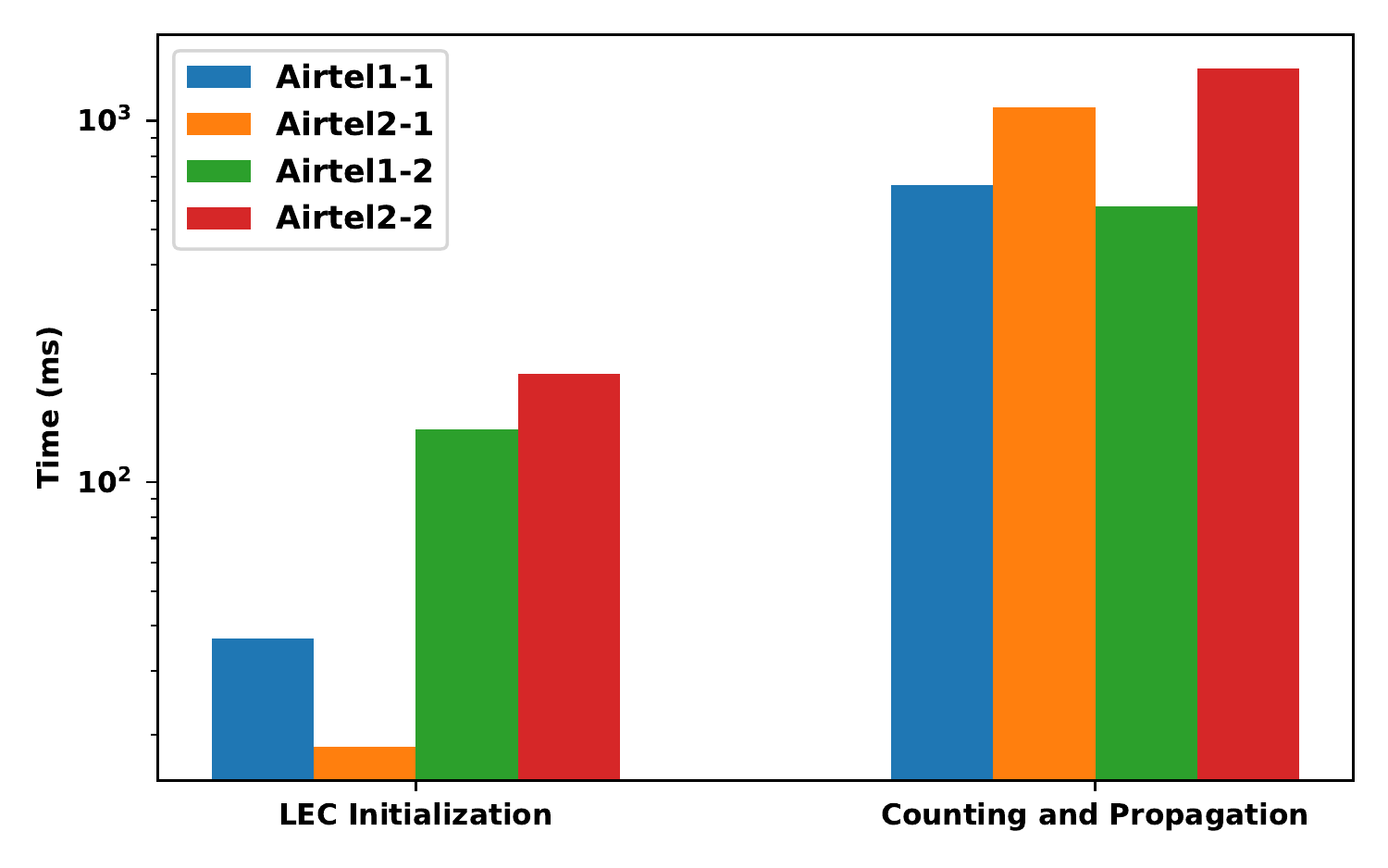}
		\caption{\label{fig:DDPV-Airtel} \system{}.}
	\end{subfigure}
	\caption{Verification time breakdown of different tools in burst update.}
    \label{fig:Green-Start-Airtel}
\end{figure*}

% \begin{figure*}
%     \includegraphics[width=0.5\textwidth,height=0.4\textwidth]{figures/eval/Green_start.pdf}
% \caption{Verification time of burst update (seconds).}
% \label{fig:Green-Start}
% \end{figure*}

This section studies why \system{} is slower than AP in burst update
verification for Airtel1-1 and Airtel2-1, but faster for Airtel1-2 and
Airtel2-2. Figure~\ref{fig:Green-Start-Airtel} provides a verification time breakdown of
burst update of different DVP tools for these four datasets. 
Airtel1-1 and Airtel1-2 have the same topology but
different numbers of rules. So are Airtel2-1 and Airtel2-2. In particular,
centralized DPV tools work in three phases: (1) FIB collection that collects data
planes of all devices to the server, (2) model computation that takes as input
the data planes of all devices and computes the equivalence classes, and (3)
verification computation that takes as input the computed ECs to verify
requirements.  We observe that model computation is the bottleneck of all four
centralized DPV tools (except for VeriFlow, whose bottleneck is both model
computation and verification computation). The model computation time is
proportional to the total number of rules in the data plane. As such, with the
number of rules increasing from Airtel1-1/Airtel2-1 to Airtel1-2/Airtel2-2 (3.39$\times$ and
11.97$\times$), the performance of centralized DPV tools degrades approximately
with the same ratio.

In contrast, to verify burst update, \system{} operates in two phases: (1) LEC
initialization at devices and (2) counting and propagation among devices.
Because LEC initialization is performed by each device in parallel, its time is
only proportional to the number of rules at each device. As such, it is not the
performance bottleneck of \system{}, and has only a small impact on the total
verification time of \system{} when the total number of rules increases (\eg,
from Airtel1-1/Airtel2-1 to Airtel1-2/Airtel2-2). As such, with more rules,
\system{} becomes faster than AP. 
As a result, we conclude that \system{} achieves better scalability
than centralized DVP tools such as AP when the total number of rules increases.

\begin{figure*}[t]
    \centering
%	\begin{subfigure}[t]{0.48\linewidth}
%	    \centering\includegraphics[width=1\linewidth]{figures/eval/memory.png}
%		\caption{\label{fig:eval-micro-mem} Memory consumption.}
%	\end{subfigure}
%\hfill
%	\begin{subfigure}[t]{0.24\linewidth}
%	    \centering\includegraphics[width=1\linewidth]{figures/eval/cib.png}
%		\caption{\label{fig:eval-micro-cib} Number of entries in CIB}
%	\end{subfigure}
%\hfill
\setlength{\abovecaptionskip}{0cm}
\setlength{\belowcaptionskip}{-0.cm}
	\begin{subfigure}[t]{0.43\linewidth}
\setlength{\abovecaptionskip}{0cm}
\setlength{\belowcaptionskip}{-0.cm}
		\centering\includegraphics[width=1\linewidth]{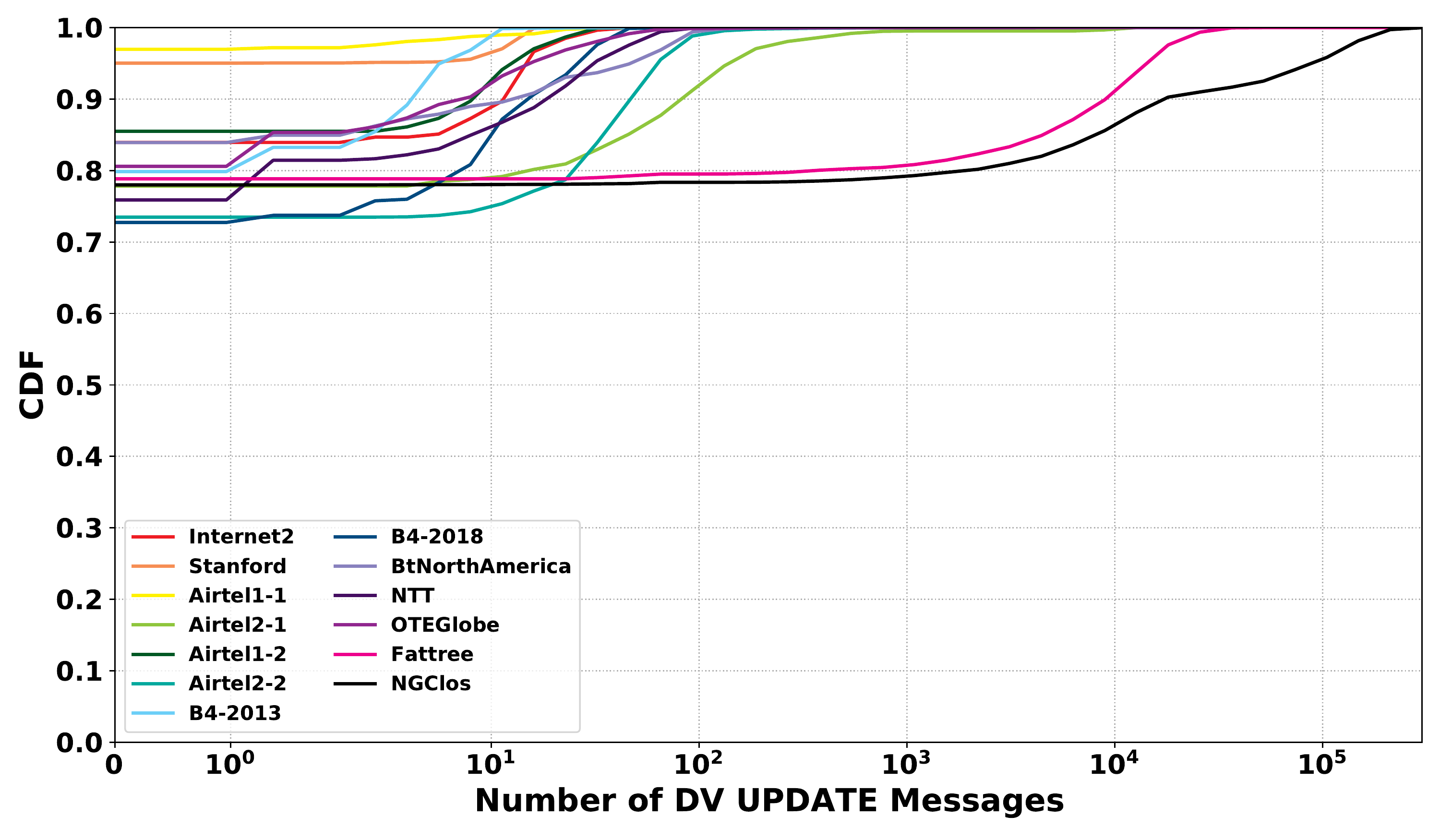}
		\caption{\label{fig:eval-micro-cib-message} Number of DV UPDATE
		messages per rule update.}
	\end{subfigure}
%	\hfill
		\begin{subfigure}[t]{0.43\linewidth}
\setlength{\abovecaptionskip}{0cm}
\setlength{\belowcaptionskip}{-0.cm}
		\centering\includegraphics[width=1\linewidth]{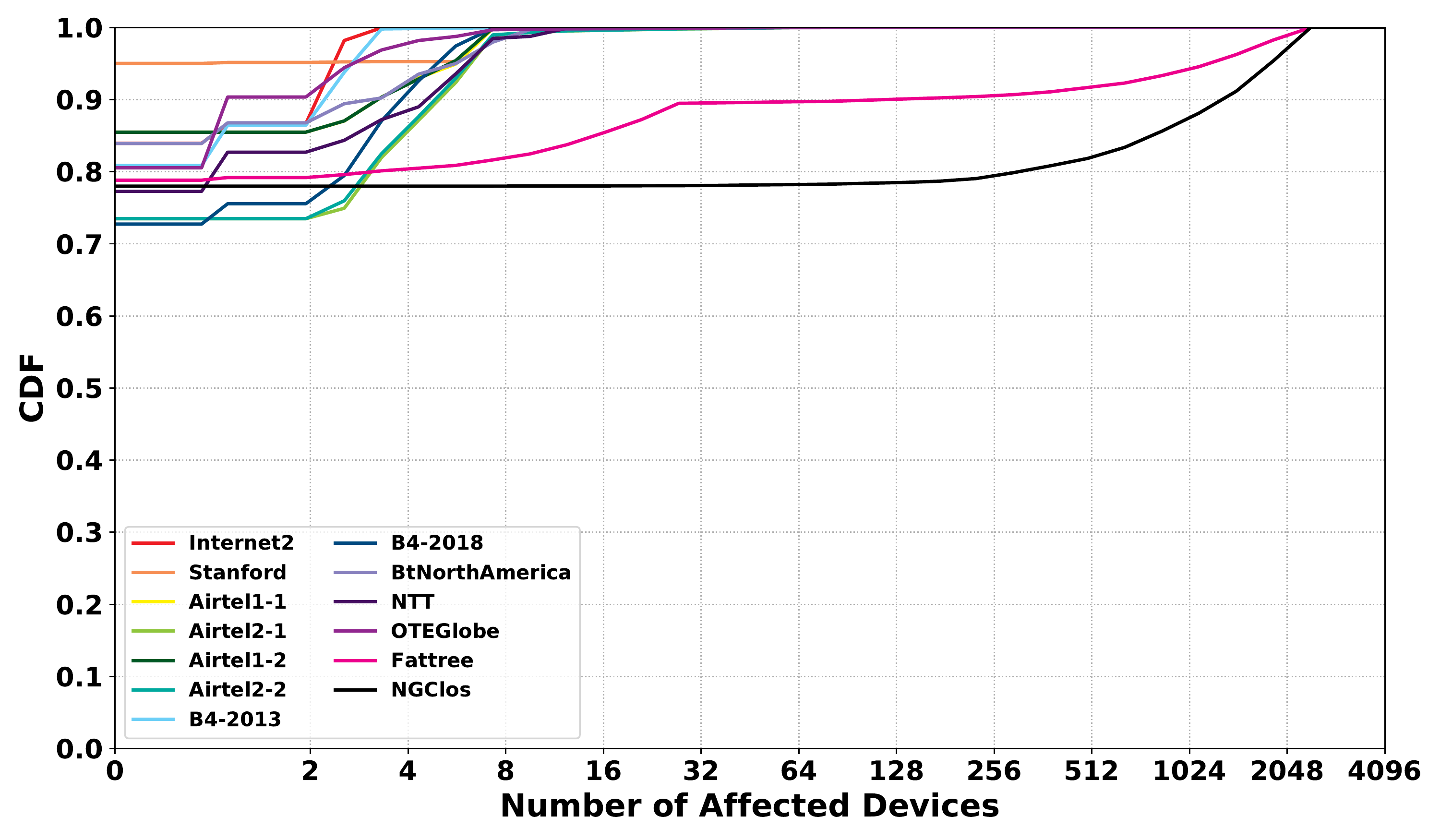}
		\caption{\label{fig:eval-micro-device-number} Number of devices with changed counting results per rule update.}
	\end{subfigure}
	\hfill
	\begin{subfigure}[t]{0.43\linewidth}
\setlength{\abovecaptionskip}{0cm}
\setlength{\belowcaptionskip}{-0.cm}
	\centering\includegraphics[width=1\linewidth]{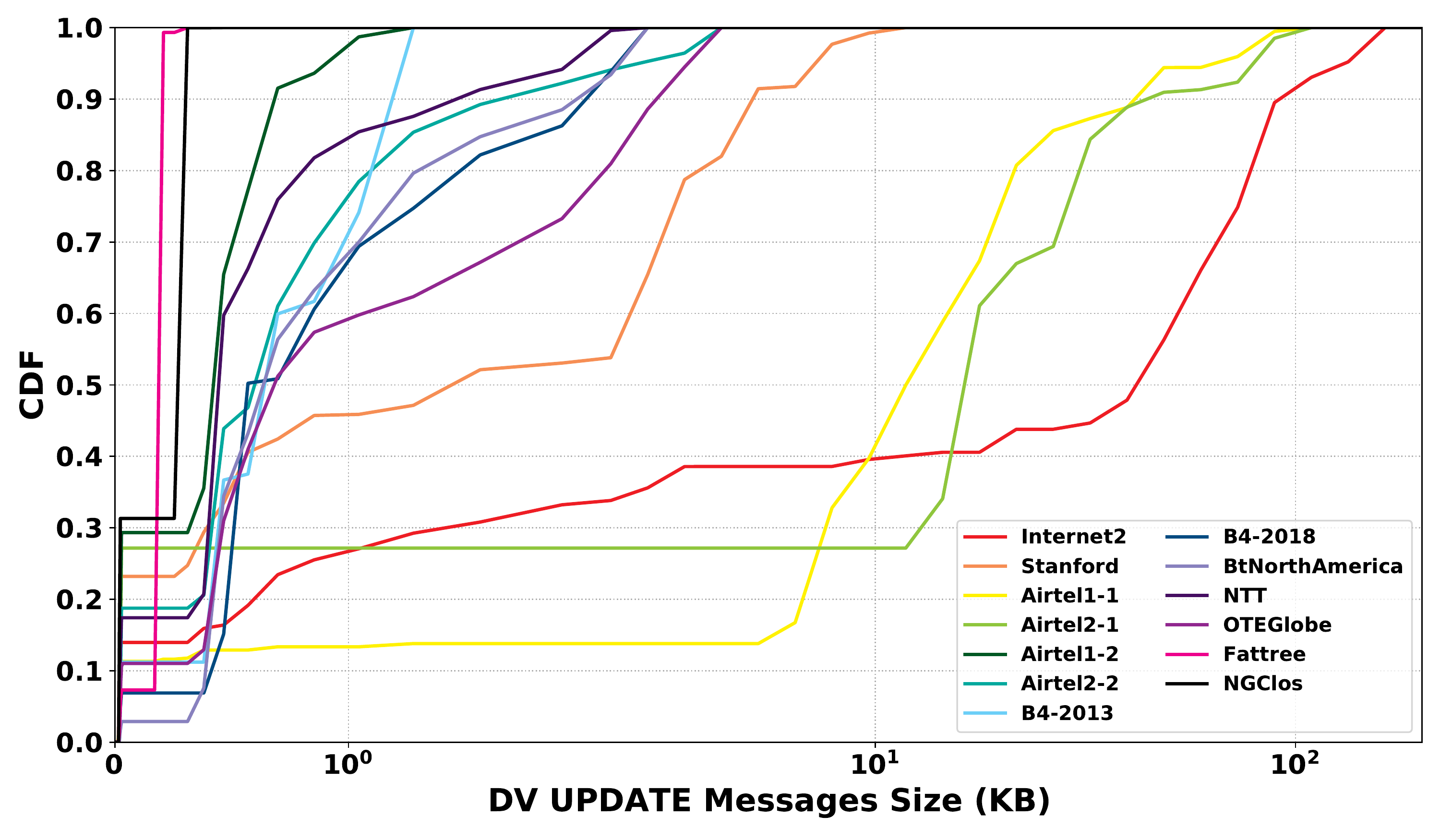}
		\caption{\label{fig:eval-micro-size} Size of DV UPDATE message.}
	\end{subfigure}
%	\hfill
	\begin{subfigure}[t]{0.43\linewidth}
\setlength{\abovecaptionskip}{0cm}
\setlength{\belowcaptionskip}{-0.cm}
	\centering\includegraphics[width=1\linewidth]{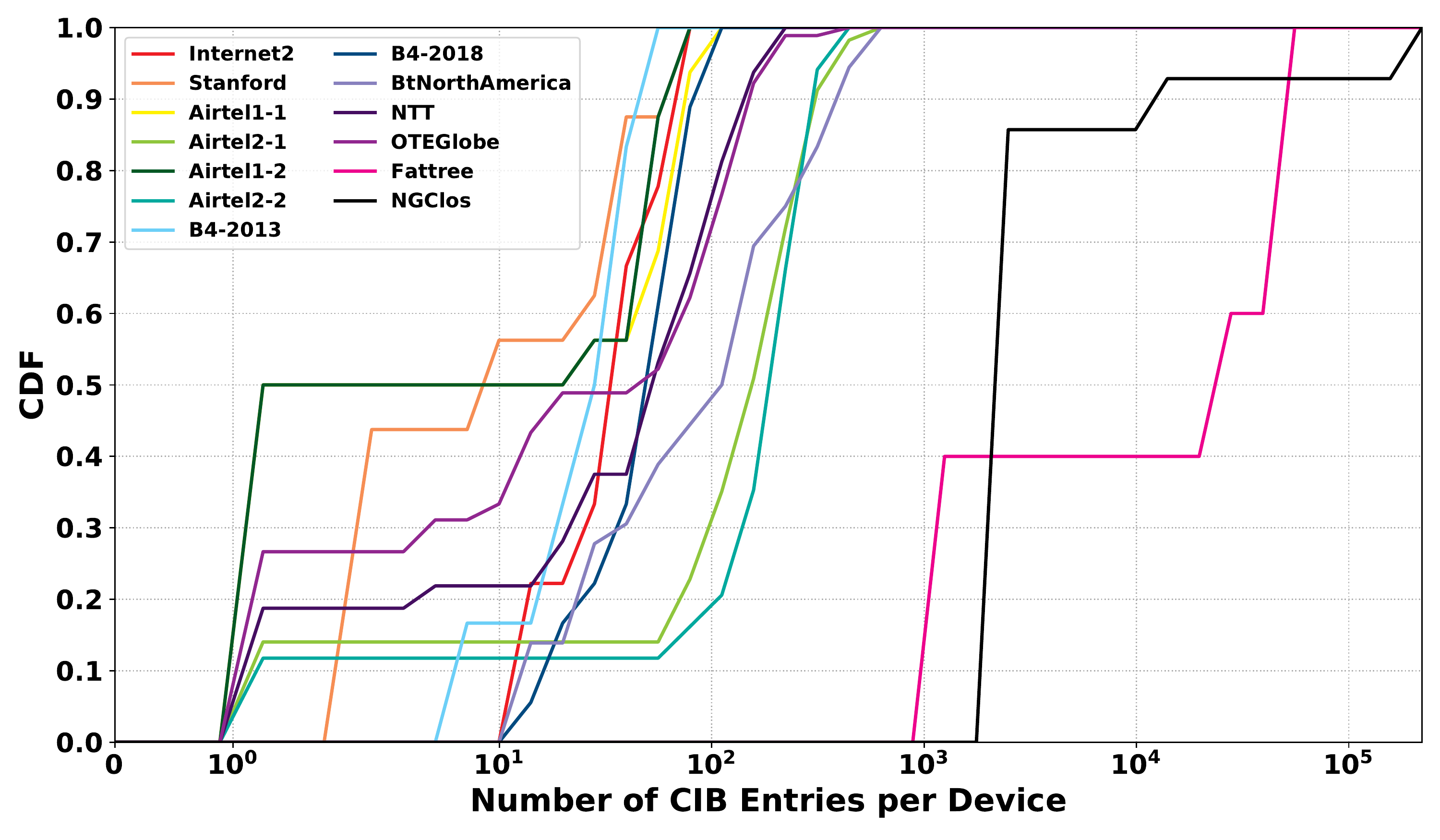}
		\caption{\label{fig:eval-micro-cib} Number of CIB entries per
		device after all rule updates.}
	\end{subfigure}
		\hfill
    \caption{The messaging overhead of \system{} in incremental verification.}
    \label{fig:eval-micro-cdf}
\end{figure*}

\section{Messaging Overhead of \system{} in Incremental Verification}
\label{sec:update-msg}
For all datasets in our experiments and simulations, we first plot CDFs of the
number of DV UPDATE messages sent in the network per rule update
(Figure~\ref{fig:eval-micro-cib-message}) and the
number of devices whose counting results change per rule update
(Figure~\ref{fig:eval-micro-device-number}).

Figure~\ref{fig:eval-micro-cib-message} shows that for each dataset, at least
70\% of rule updates do not incur any DV UPDATE message in \system{}. 
Figure~\ref{fig:eval-micro-device-number}
further shows that for at least 75\% of rule updates, the number of devices whose
counting results change is no more than two. This shows that by decomposing
verification to on-device counting tasks, a large portion of incremental
verifications become local verification on a single network device, or only
require sharing counting results among a small number of network devices. As
such, the \system{} achieves substantial scaling up on incremental verification. 

We next plot the size of DV UPDATE message incurred across 10,000 rule updates
(Figure~\ref{fig:eval-micro-size}). We observe that all UPDATE messages are
smaller than 150KB, in particular, for \lnet{} and Fatree, their
UPDATE messages are all smaller than 396 bytes. This indicates that
the bandwidth overhead of \system{} is very low. 

In the end, we plot the number of CIB entries of each device after 10,000 rule
updates (Figure~\ref{fig:eval-micro-cib}), in supplementary to the maximal
memory microbenchmark results in Figure~\ref{fig:trace}, which shows that the
\system{} on-device verifiers only consume a small amount of memory on commodity
network switches.

%Specifically, in large data center networks Fattree and
%\lnet{}, the 90\% quantile of incremental verification tiem of \system{} is
%13.47$\times$ and 5$\times$, respectively.

\begin{figure*}[]
\setlength{\abovecaptionskip}{0cm}
\setlength{\belowcaptionskip}{-0.cm}
    \centering
	\begin{subfigure}{0.24\linewidth}
\setlength{\abovecaptionskip}{0cm}
\setlength{\belowcaptionskip}{-0.cm}
	    \centering\includegraphics[width=1\linewidth]{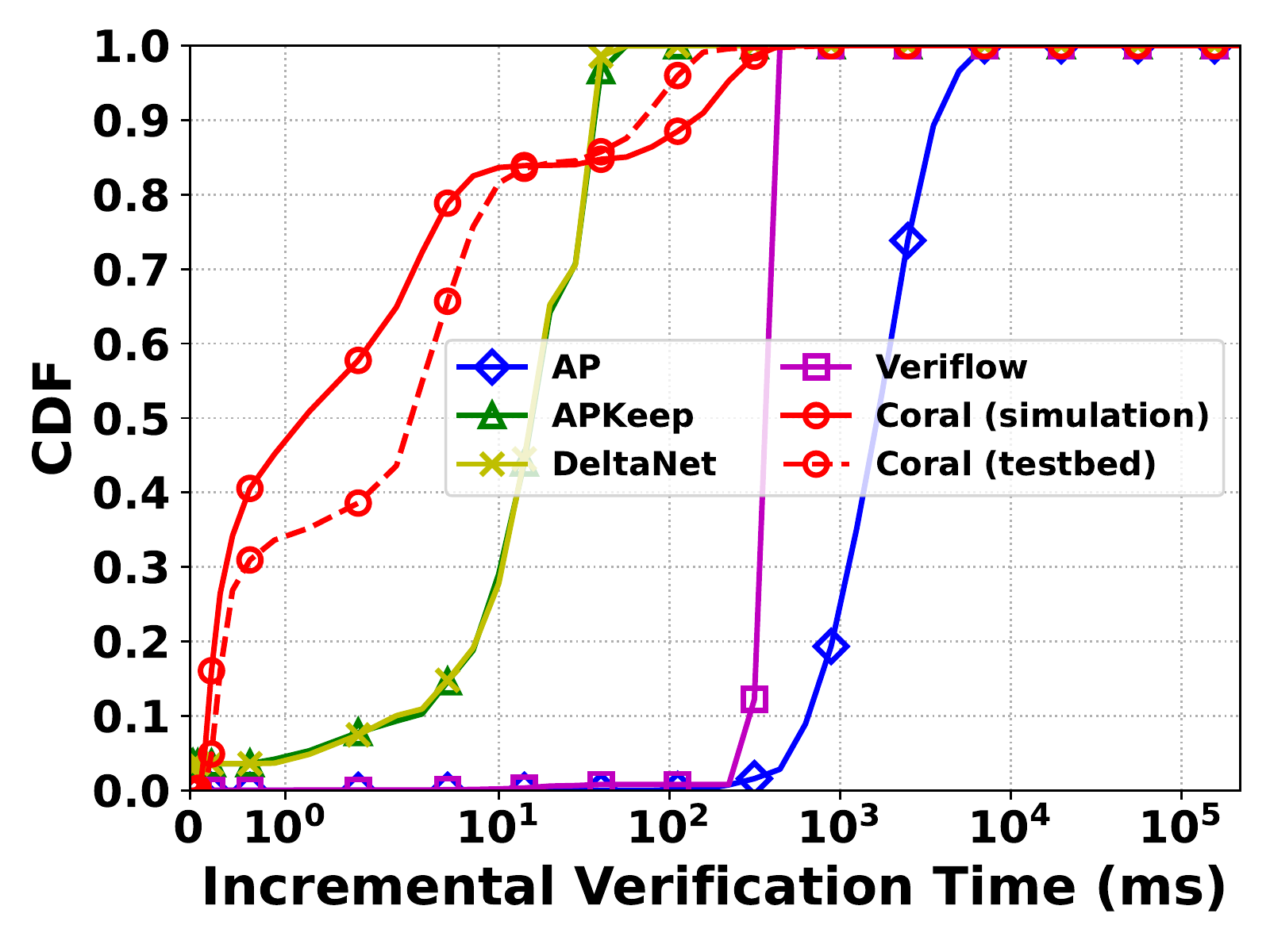}
		\caption{\label{fig:incremental-i2} Internet2.}
	\end{subfigure}
\hfill
	\begin{subfigure}{0.24\linewidth}
\setlength{\abovecaptionskip}{0cm}
\setlength{\belowcaptionskip}{-0.cm}
	\centering\includegraphics[width=1\linewidth]{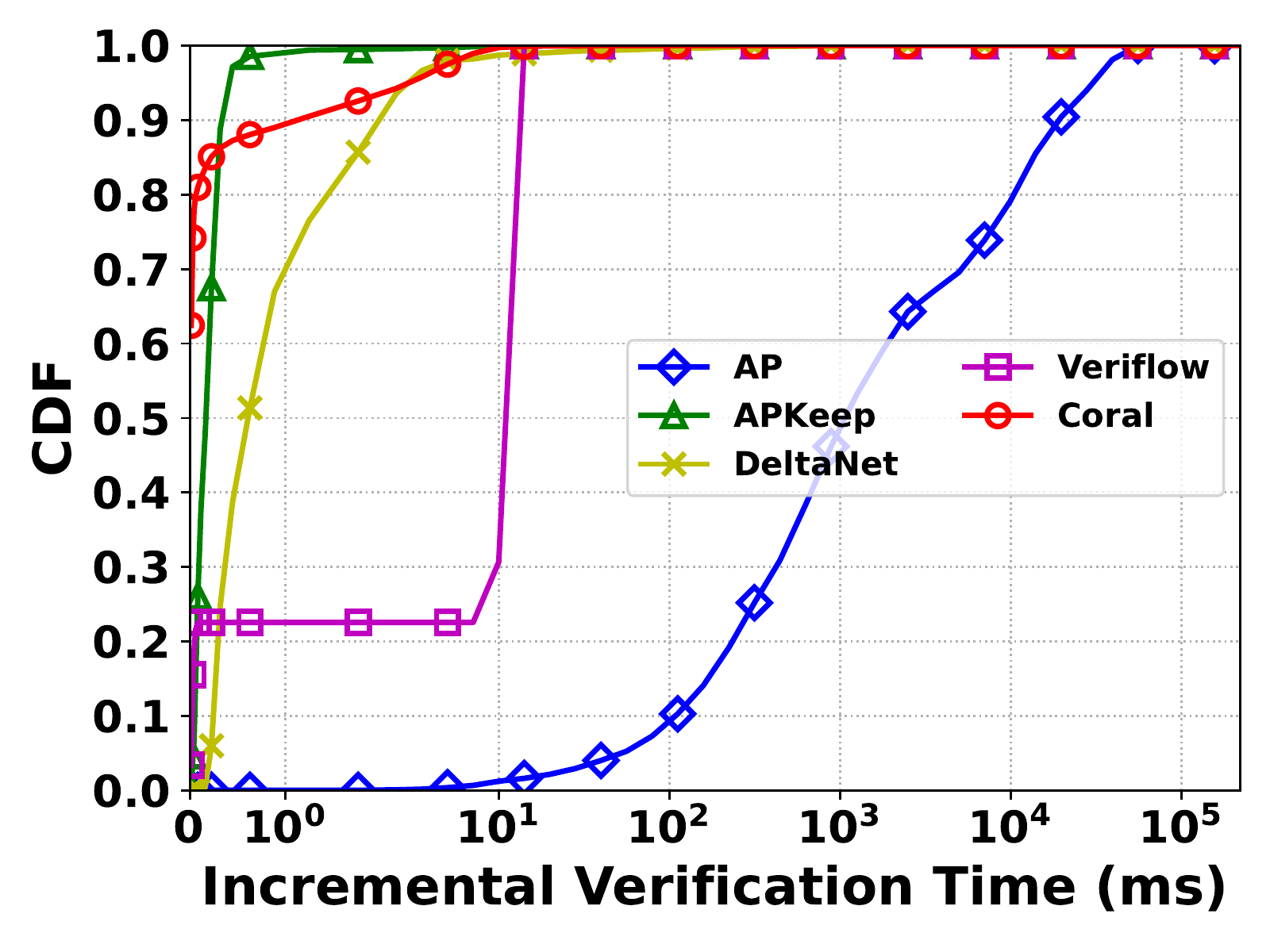}
		\caption{\label{fig:incremental-st} Stanford.}
	\end{subfigure}
\hfill
	\begin{subfigure}{0.24\linewidth}
\setlength{\abovecaptionskip}{0cm}
\setlength{\belowcaptionskip}{-0.cm}
	\centering\includegraphics[width=1\linewidth]{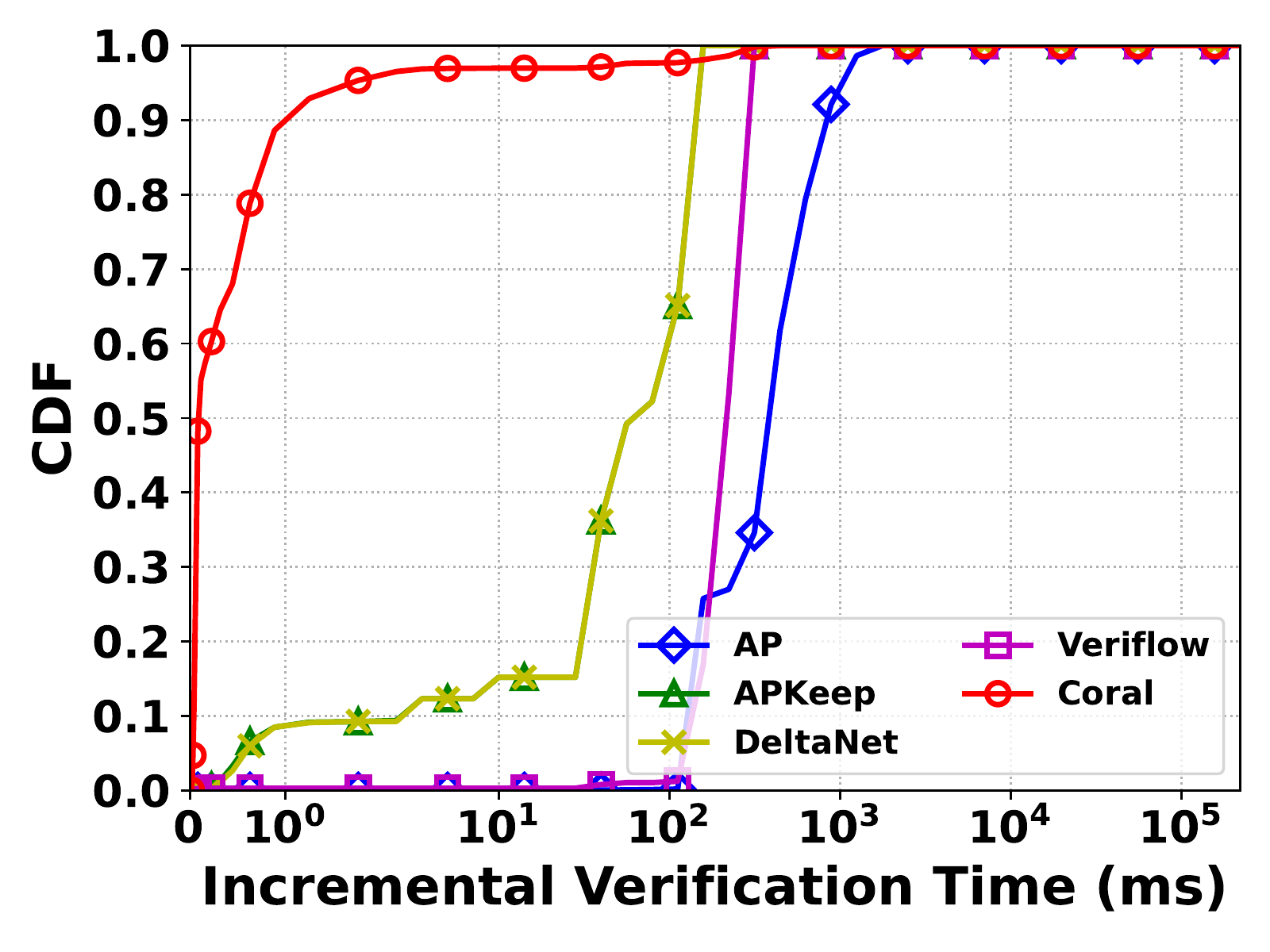}
		\caption{\label{fig:incremental-Airtel1-1} Airtel1-1.}
	\end{subfigure}
\hfill
    \begin{subfigure}{0.24\linewidth}
\setlength{\abovecaptionskip}{0cm}
\setlength{\belowcaptionskip}{-0.cm}
	 \centering\includegraphics[width=1\linewidth]{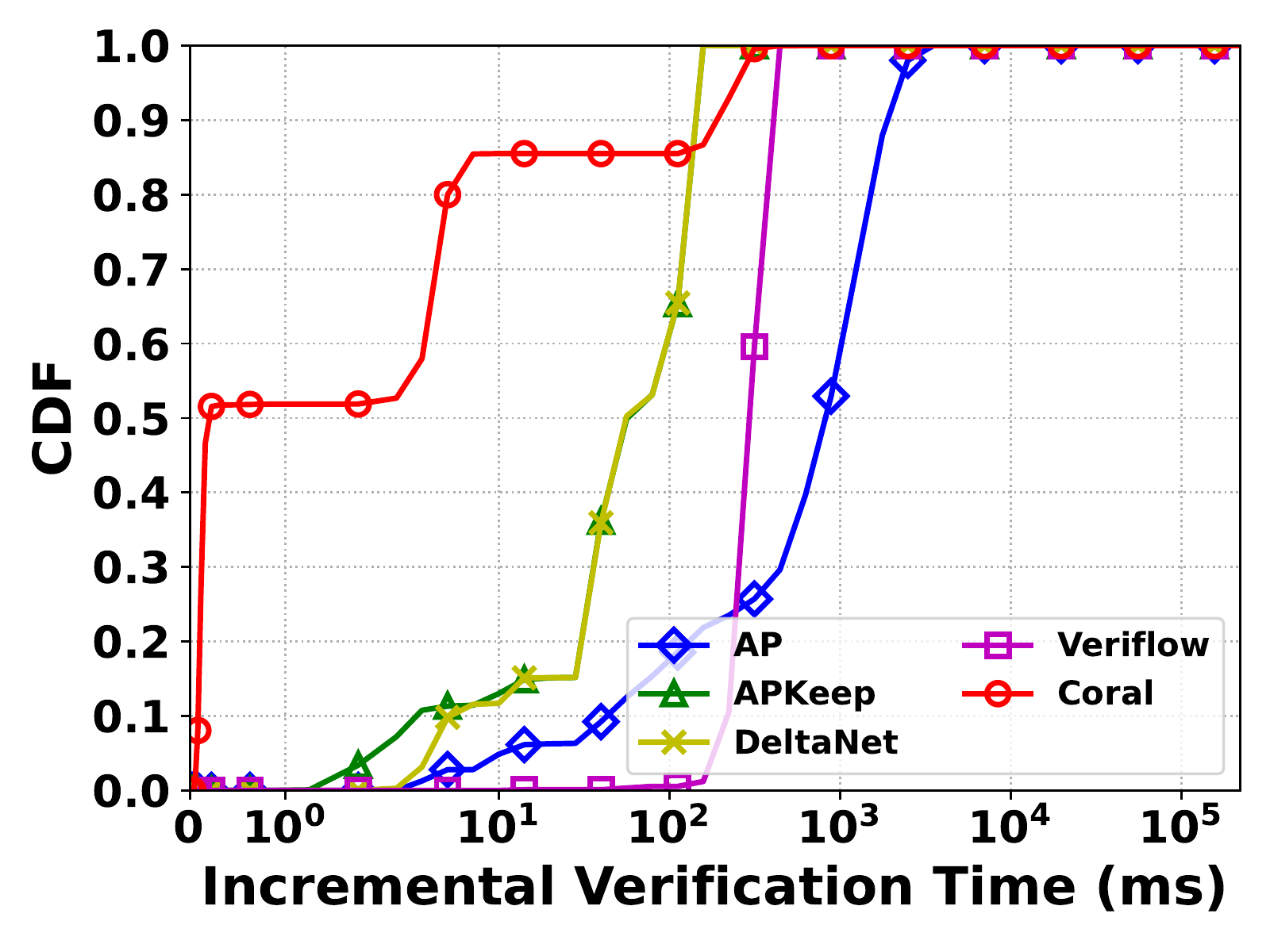}
        \caption{\label{incremental-Airtel1-2} Airtel1-2.}
    \end{subfigure}
\hfill
    \begin{subfigure}{0.24\linewidth}
\setlength{\abovecaptionskip}{0cm}
\setlength{\belowcaptionskip}{-0.cm}
        \centering\includegraphics[width=1\linewidth]{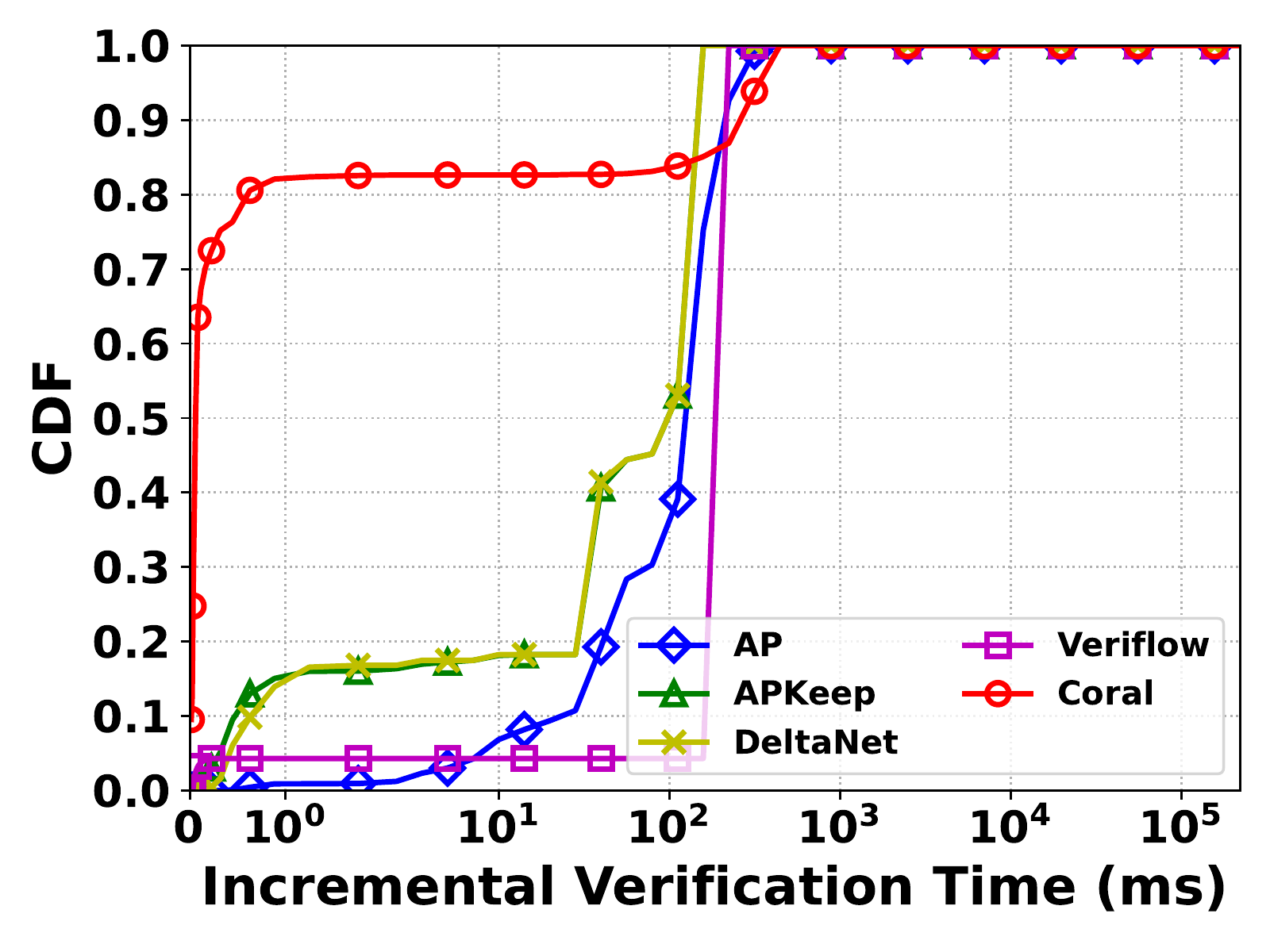}
        \caption{\label{incremental-Airtel2-1} Airtel2-1.}
    \end{subfigure}
\hfill
    \begin{subfigure}{0.24\linewidth}
\setlength{\abovecaptionskip}{0cm}
\setlength{\belowcaptionskip}{-0.cm}
        \centering\includegraphics[width=1\linewidth]{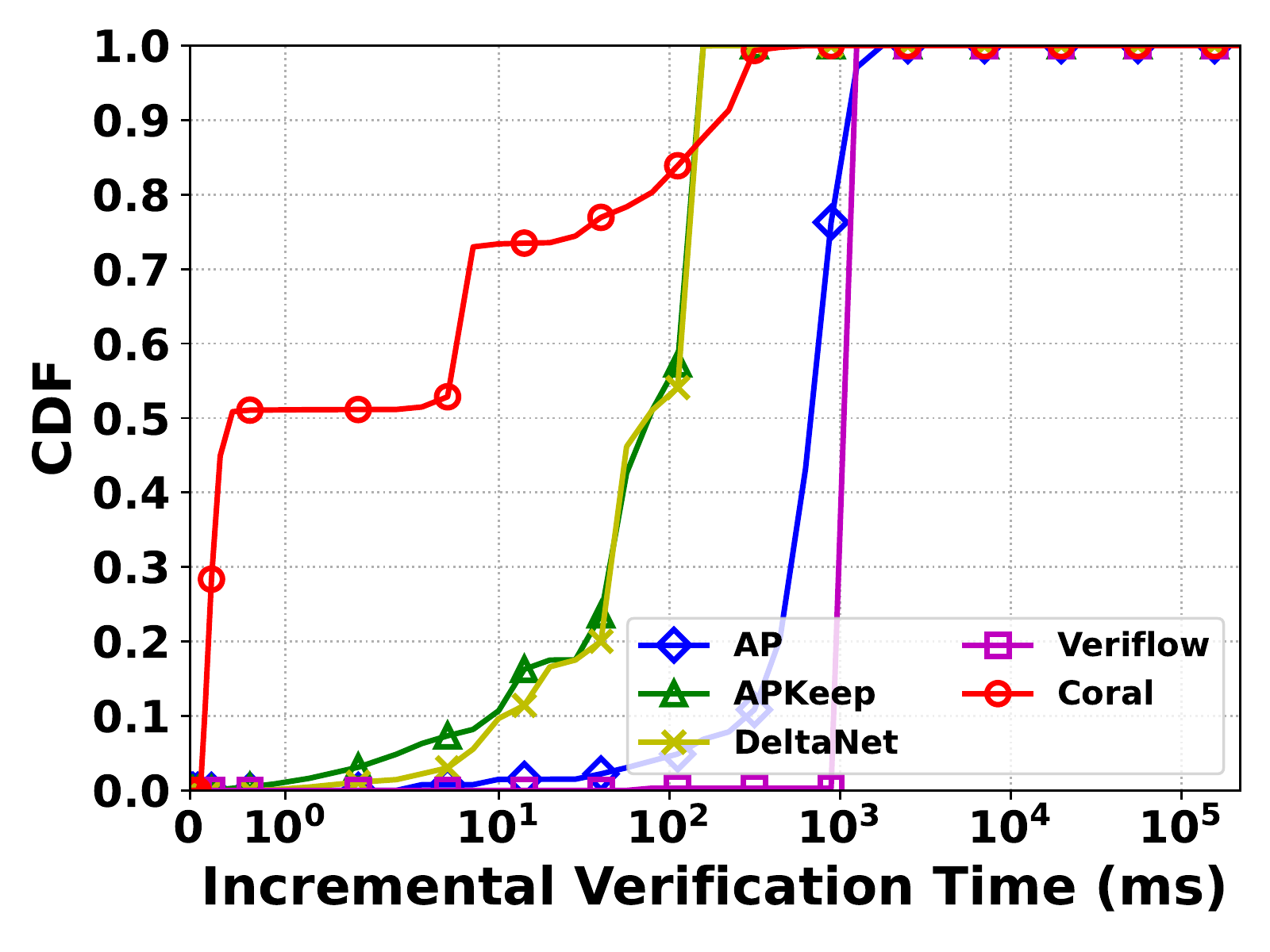}
        \caption{\label{incremental-Airtel2-2} Airtel2-2.}
    \end{subfigure}
\hfill
    \begin{subfigure}{0.24\linewidth}
\setlength{\abovecaptionskip}{0cm}
\setlength{\belowcaptionskip}{-0.cm}
        \centering\includegraphics[width=1\linewidth]{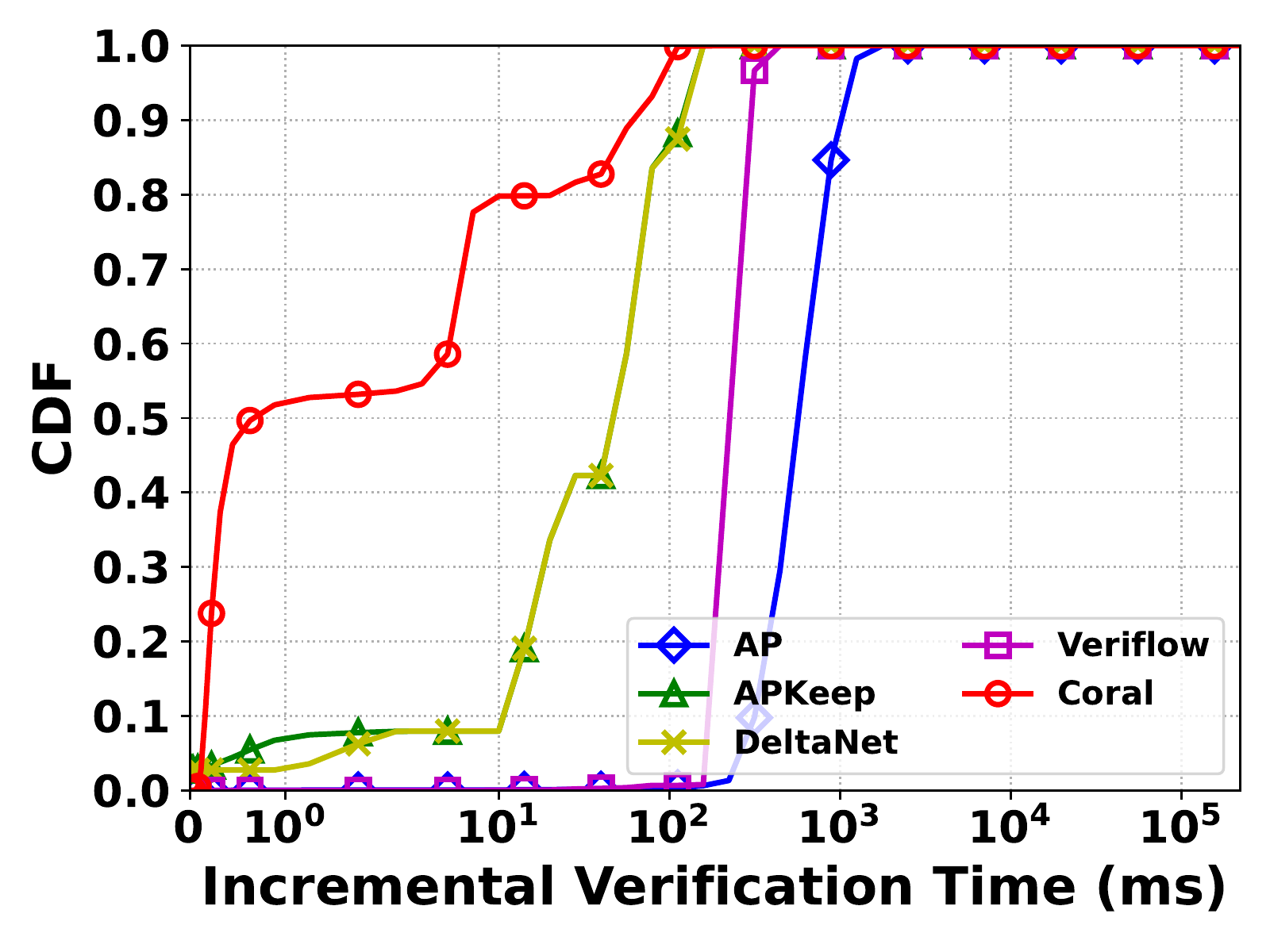}
        \caption{\label{incremental-B4-2013} B4-2013.}
    \end{subfigure}
\hfill
    \begin{subfigure}{0.24\linewidth}
\setlength{\abovecaptionskip}{0cm}
\setlength{\belowcaptionskip}{-0.cm}
        \centering\includegraphics[width=1\linewidth]{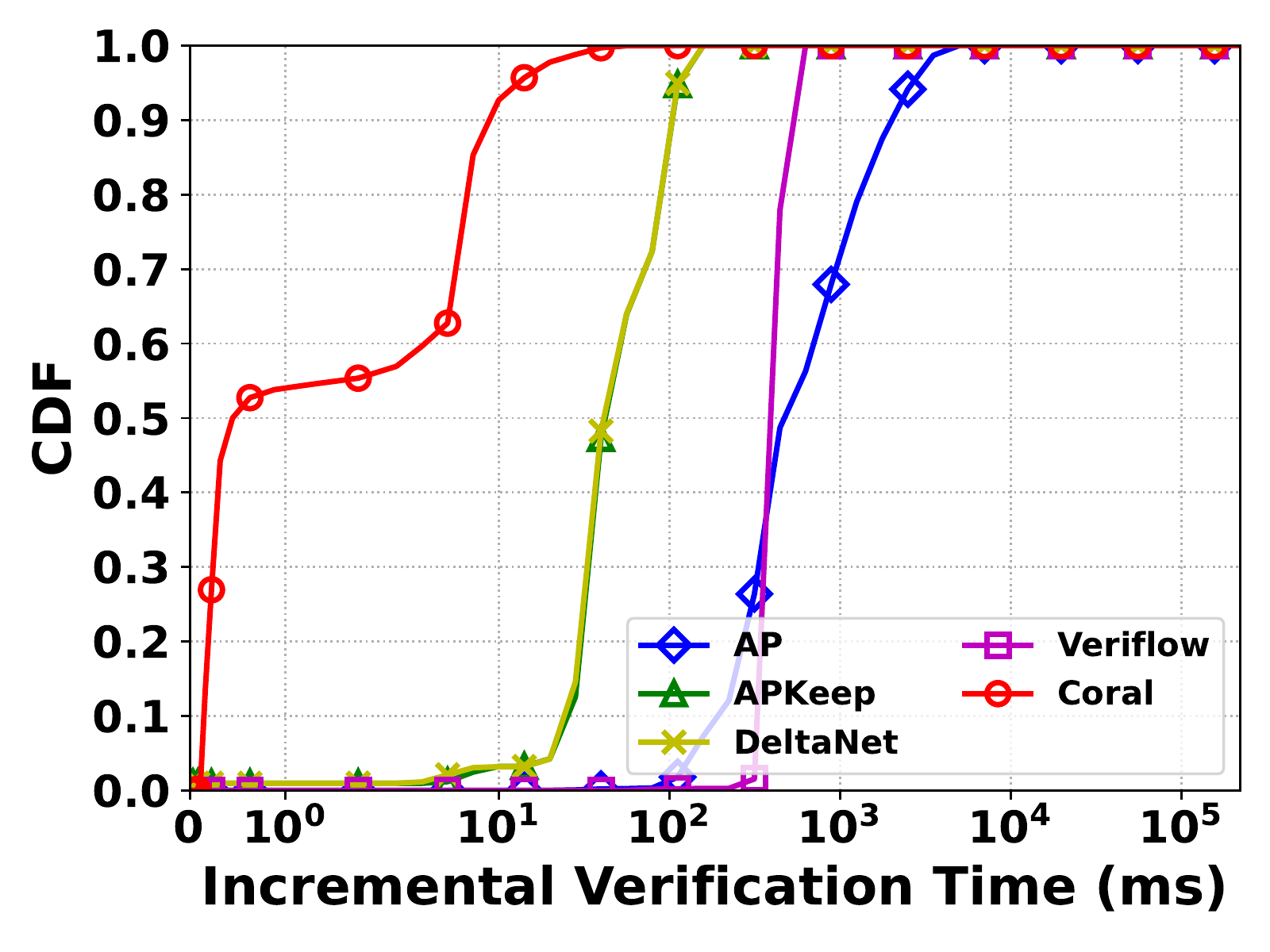}
        \caption{\label{incremental-B4-2018} B4-2018.}
    \end{subfigure}
	\caption{The CDF of incremental verification time of all datasets - part 1.}
	\label{fig:incremental-cdf-part1}
\end{figure*}

\section{CDF of Incremental Update Verification Time}\label{sec:inc-cdf}
\begin{figure*}[]
\centering
\setlength{\abovecaptionskip}{0cm}
\setlength{\belowcaptionskip}{-0.cm}
    \centering
    \begin{subfigure}{0.24\linewidth}
\setlength{\abovecaptionskip}{0cm}
\setlength{\belowcaptionskip}{-0.cm}
	    \centering\includegraphics[width=1\linewidth]{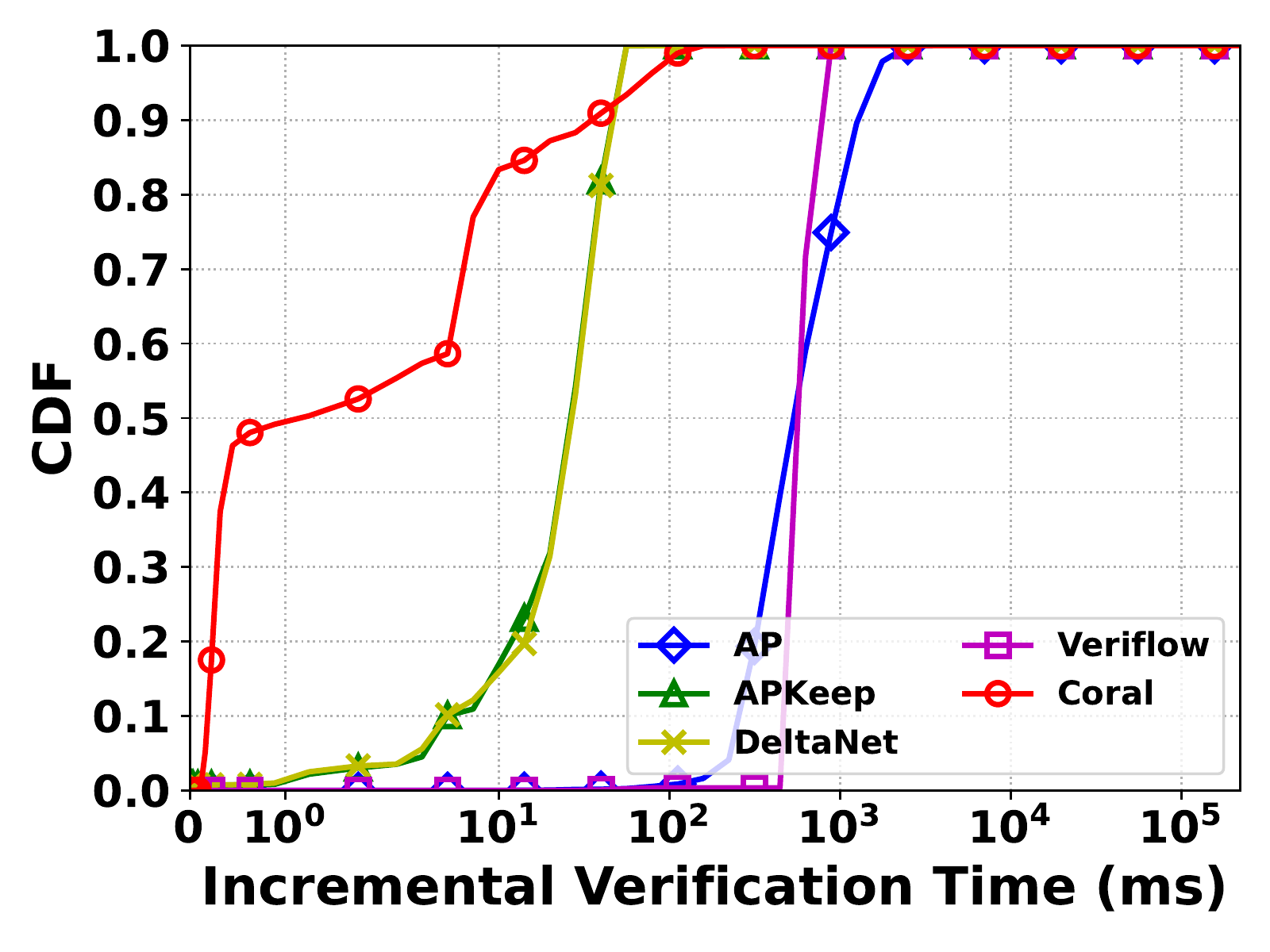}
        \caption{\label{incremental-BtNorthAmerica} BT North America.}
    \end{subfigure}
%\hfill
    \begin{subfigure}{0.24\linewidth}
\setlength{\abovecaptionskip}{0cm}
\setlength{\belowcaptionskip}{-0.cm}
	    \centering\includegraphics[width=1\linewidth]{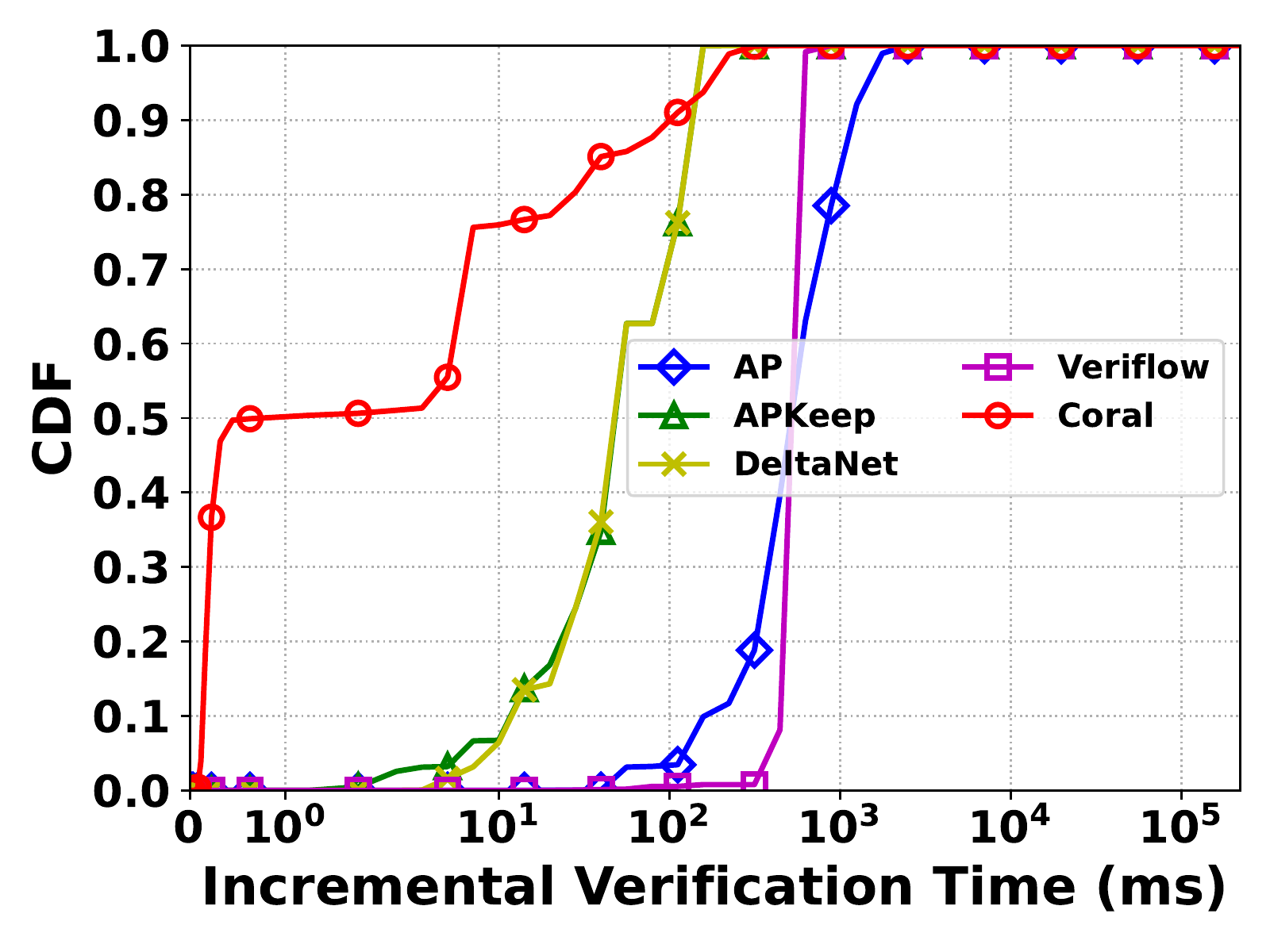}
        \caption{\label{incremental-Ntt} NTT.}
    \end{subfigure}
%\hfill
    \begin{subfigure}{0.24\linewidth}
\setlength{\abovecaptionskip}{0cm}
\setlength{\belowcaptionskip}{-0.cm}
	    \centering\includegraphics[width=1\linewidth]{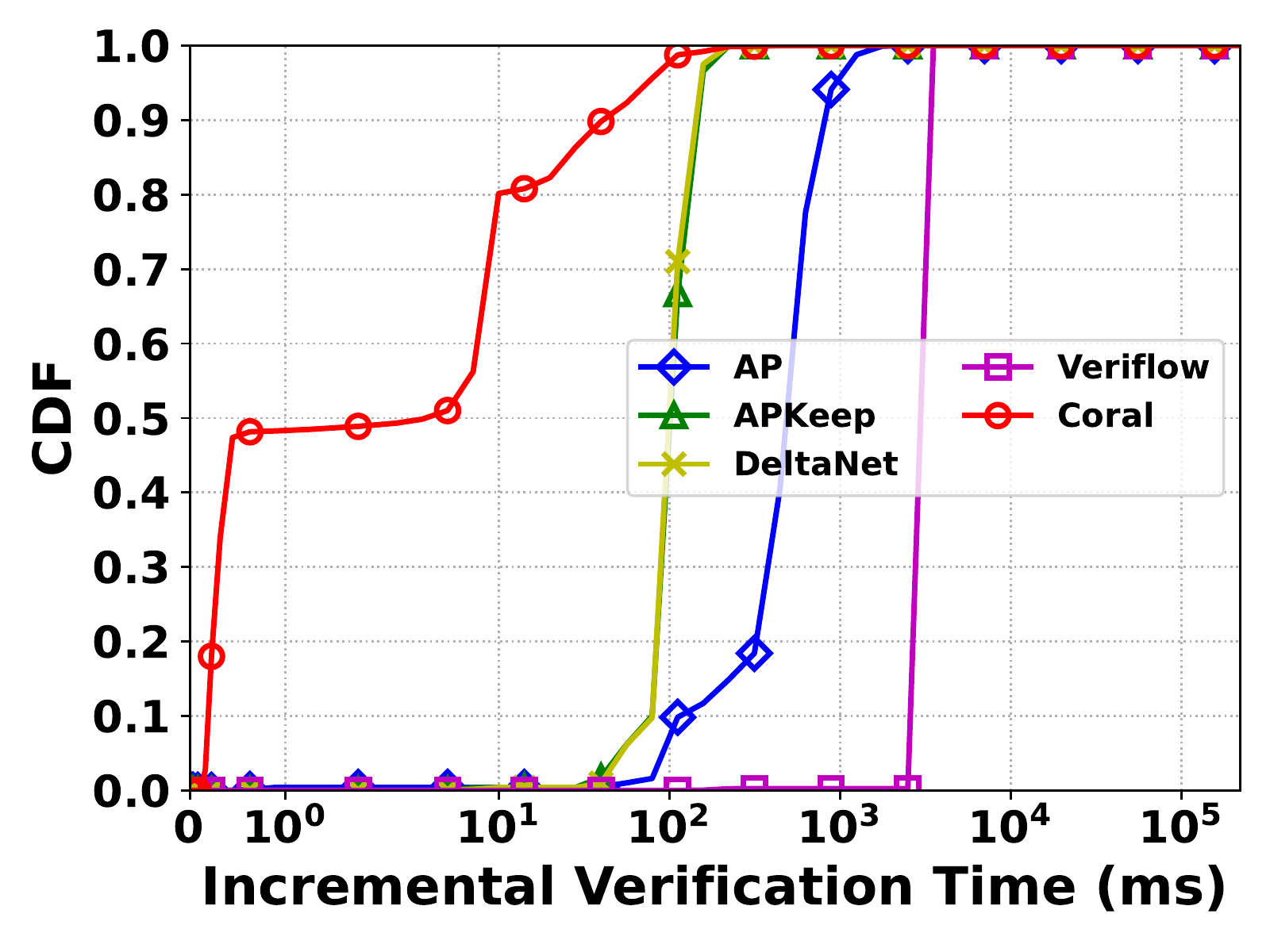}
        \caption{\label{incremental-OTEGlobe} OTEGlobe.}
    \end{subfigure}
\hfill
\\
    \begin{subfigure}{0.24\linewidth}
\setlength{\abovecaptionskip}{0cm}
\setlength{\belowcaptionskip}{-0.cm}
	    \centering\includegraphics[width=1\linewidth]{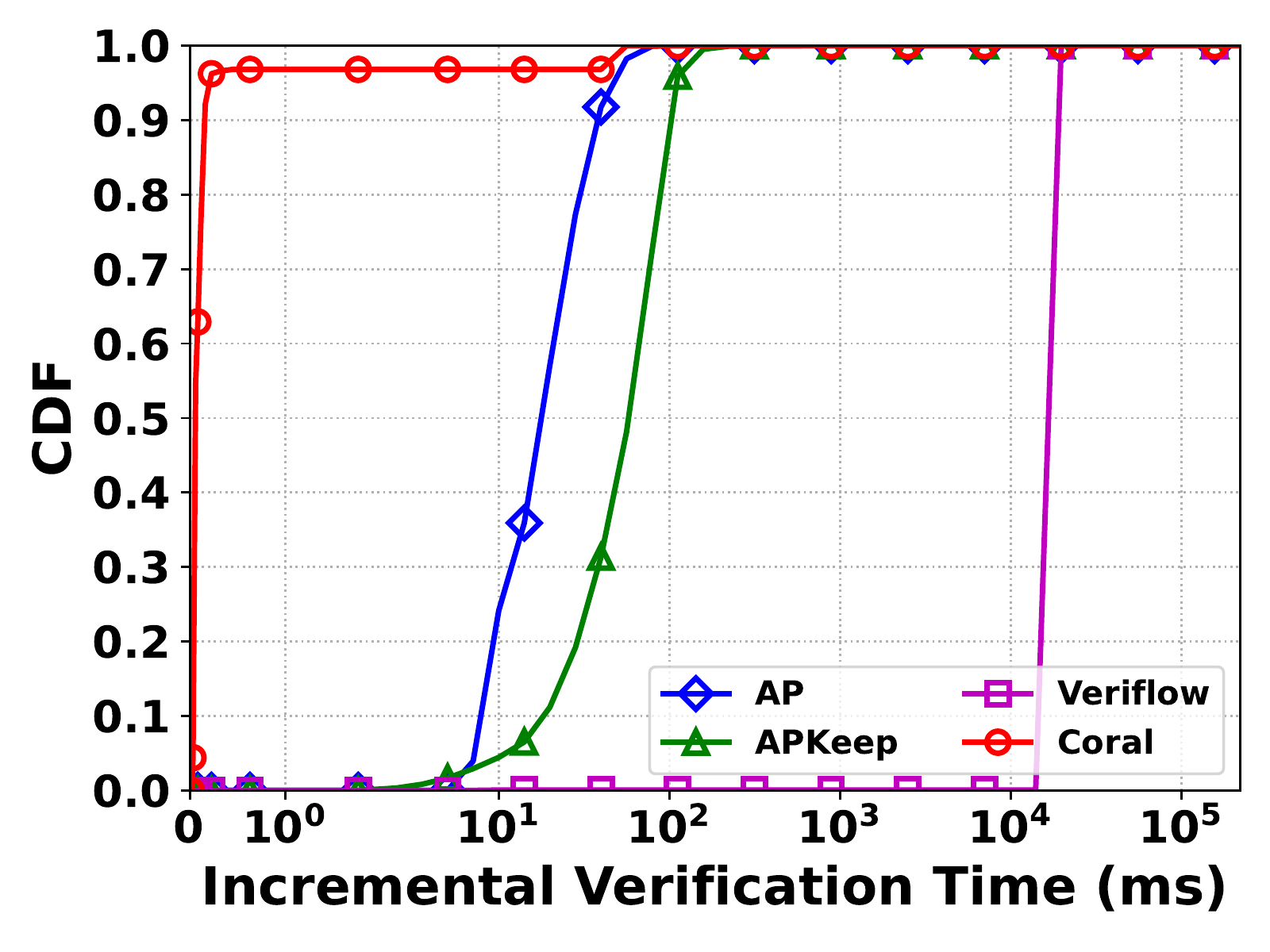}
        \caption{\label{incremental-fattree} Fattree  ($k=48$).}
    \end{subfigure}
%\hfill
    \begin{subfigure}{0.24\linewidth}
\setlength{\abovecaptionskip}{0cm}
\setlength{\belowcaptionskip}{-0.cm}
	    \centering\includegraphics[width=1\linewidth]{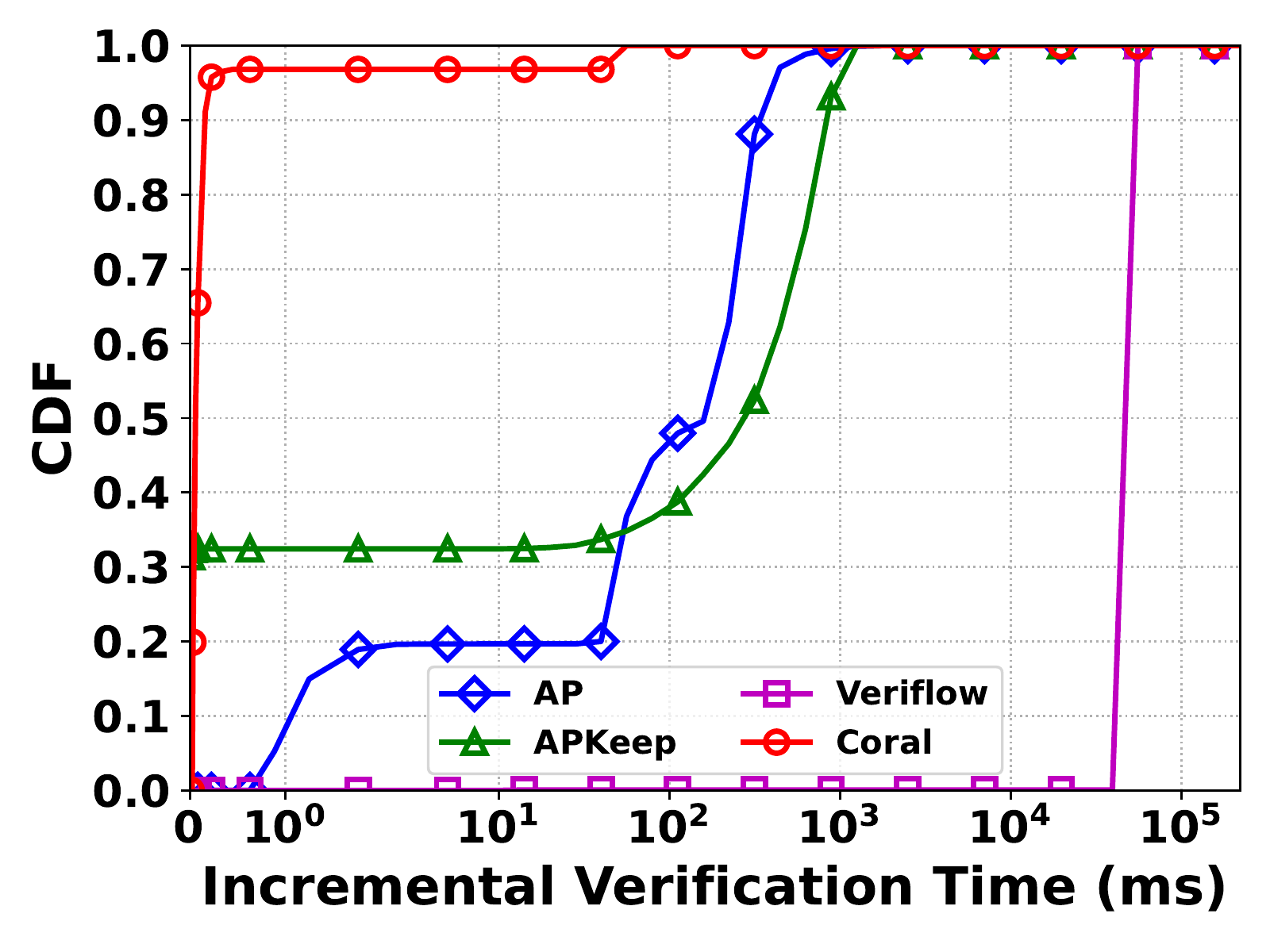}
        \caption{\label{incremental-fb} \lnet{}.}
    \end{subfigure}
	\caption{The CDF of incremental verification time of all datasets - part
	2.}
	\label{fig:incremental-cdf-part2}
\vspace{-1em}
\end{figure*}

%\qiao{TBA by Chenyang.}
As a supplementary to the statistics in Figure~\ref{fig:incremental},
Figure~\ref{fig:incremental-cdf-part1} and
Figure~\ref{fig:incremental-cdf-part2} plot the CDF of the incremental
verification time of \system{} and centralized DPV tools in comparison for each
dataset, to show that \system{} consistently outperforms state-of-the-art
centralized DPV tools for incremental update verification in all datasets.

\begin{figure*}[!h]
\centering
\setlength{\abovecaptionskip}{0cm}
\setlength{\belowcaptionskip}{-0.cm}
	\begin{subfigure}{0.32\linewidth}
\setlength{\abovecaptionskip}{0cm}
\setlength{\belowcaptionskip}{-0.cm}
	    \centering\includegraphics[width=1\linewidth]{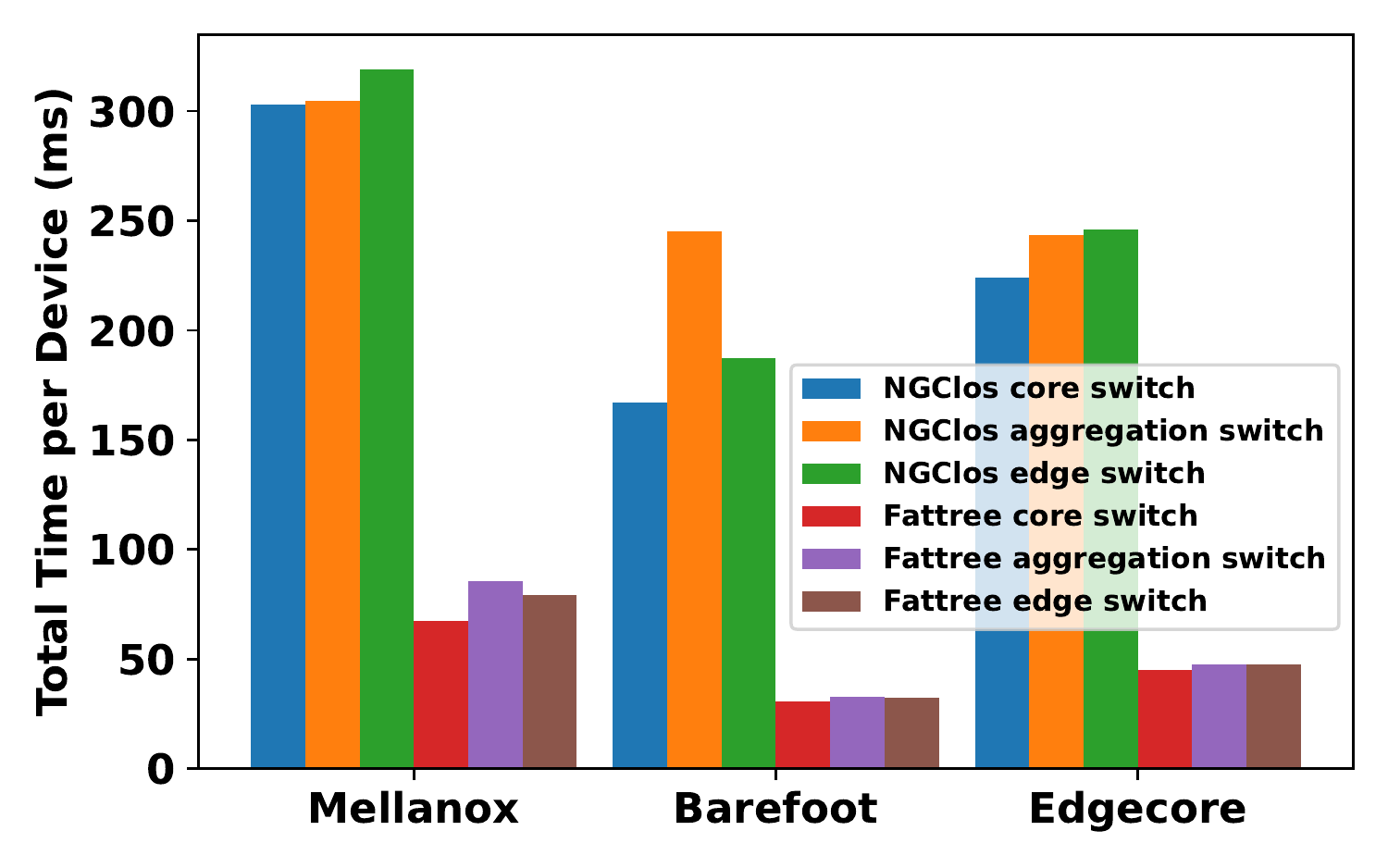}
		\caption{\label{fig:dc-time} Total time.}
	\end{subfigure}
\hfill
	\begin{subfigure}{0.32\linewidth}
\setlength{\abovecaptionskip}{0cm}
\setlength{\belowcaptionskip}{-0.cm}
	\centering\includegraphics[width=1\linewidth]{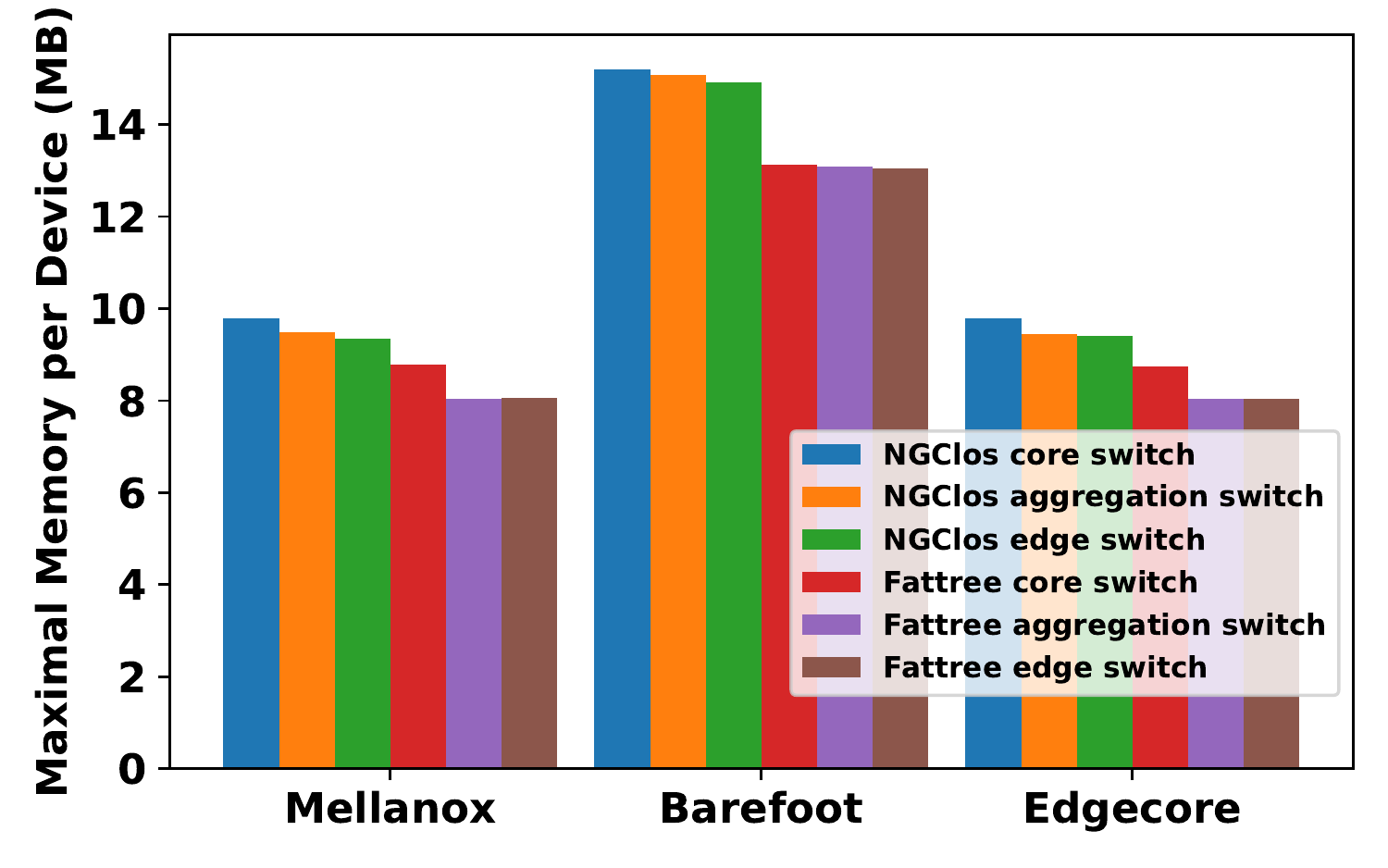}
		\caption{\label{fig:dc-memory} Maximal memory.}
	\end{subfigure}
\hfill
	\begin{subfigure}{0.32\linewidth}
\setlength{\abovecaptionskip}{0cm}
\setlength{\belowcaptionskip}{-0.cm}
	    \centering\includegraphics[width=1\linewidth]{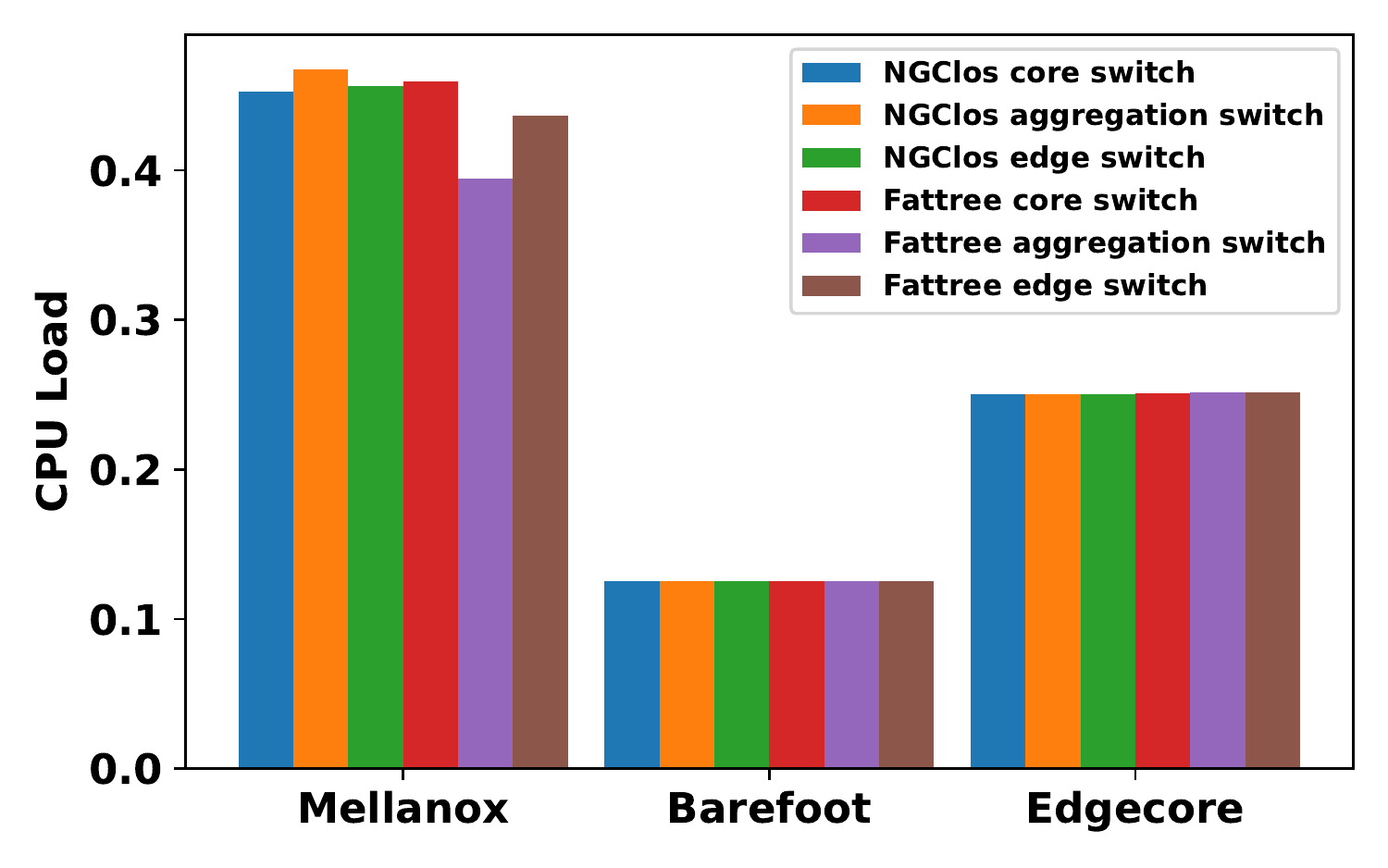}
		\caption{\label{fig:dc-load} CPU load.}
	\end{subfigure}
	\caption{Time and overhead of verifying all-shortest-path availability
	in DC networks from green start on commodity network devices.}
    \label{fig:dc-initialization}
\end{figure*}
\section{Verifying RCDC Local Contracts Using \system{}}\label{sec:rcdc}
In \S\ref{sec:dv-detail}, we have proved that the local contracts to verify
all-shortest-path availability requirement in Azure RCDC~\cite{azure} is a
special case of the counting tasks in \system{}. 
One distinction, however, is that RCDC verifies those
local contracts in centralized computation instances. In this experiment, we
study the feasibility of letting \system{} on-device planners verify these local
contracts on commodity network devices. Specifically, we pick three devices (one
edge, one aggregation and one core) in the 48-ary Fattree and the \lnet{}
datasets, respectively, and verify their local contracts on three commodity
switches. We plot the results in Figure~\ref{fig:dc-initialization}. 
Results show that all local contracts are verified on commodity switches in less
than $320ms$, with a CPU load $\leq 0.47$ and a maximal memory $\leq 15.2MB$.
The latency is consistent with the result of RCDC running in off-device
computation instances (\eg, $O(100)ms$ in Section 2.6.1 of RCDC~\cite{azure}).

\begin{figure*}
\centering
\setlength{\abovecaptionskip}{0cm}
\setlength{\belowcaptionskip}{-0.cm}
	\begin{subfigure}{0.4\linewidth}
\setlength{\abovecaptionskip}{0cm}
\setlength{\belowcaptionskip}{-0.cm}
	    \centering\includegraphics[width=1\linewidth]{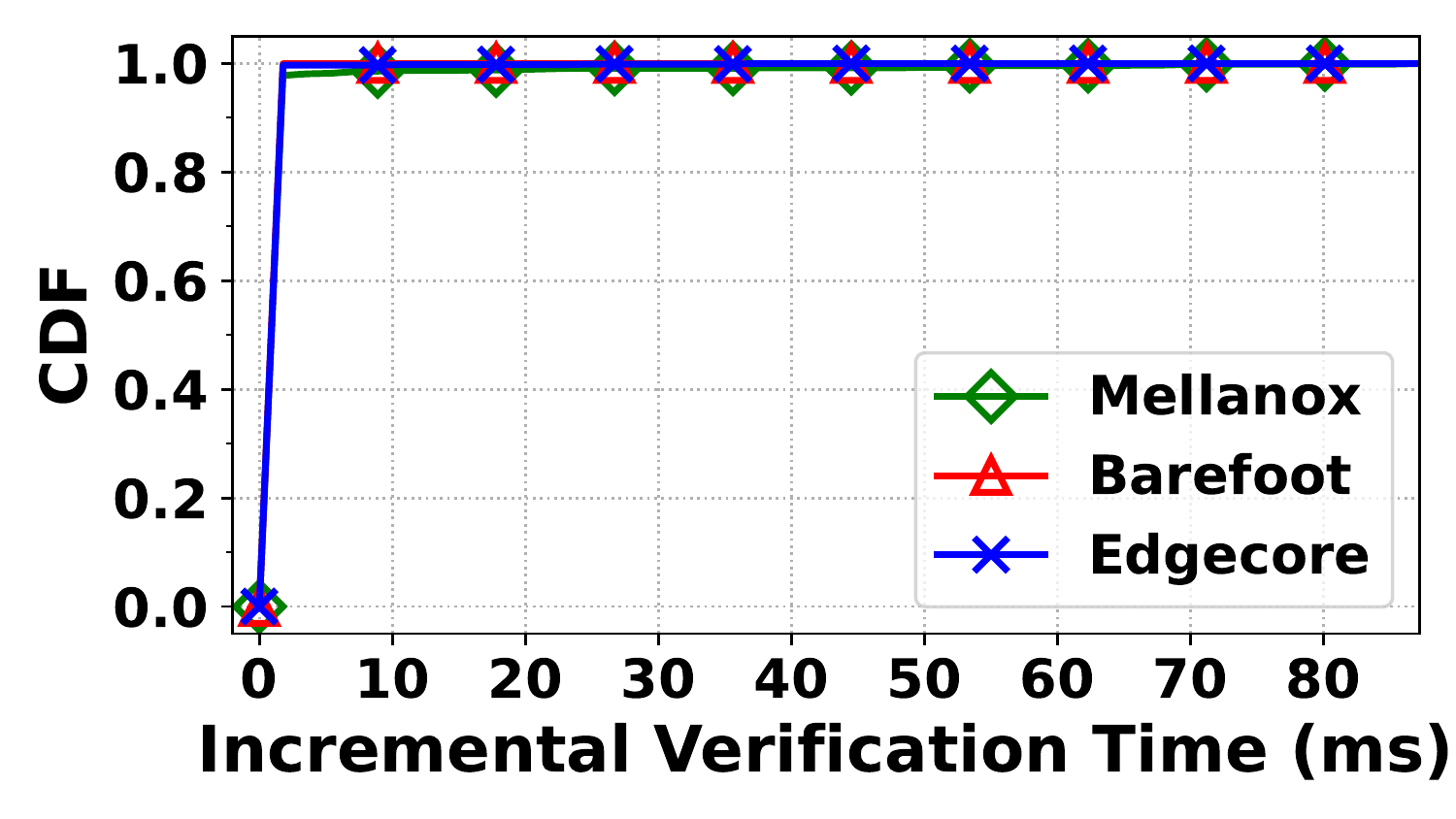}
		\caption{\label{fig:trace-total-time-ft-1000} 48-ary Fattree.}
	\end{subfigure}
%\hfill
	\begin{subfigure}{0.4\linewidth}
\setlength{\abovecaptionskip}{0cm}
\setlength{\belowcaptionskip}{-0.cm}
	    \centering\includegraphics[width=1\linewidth]{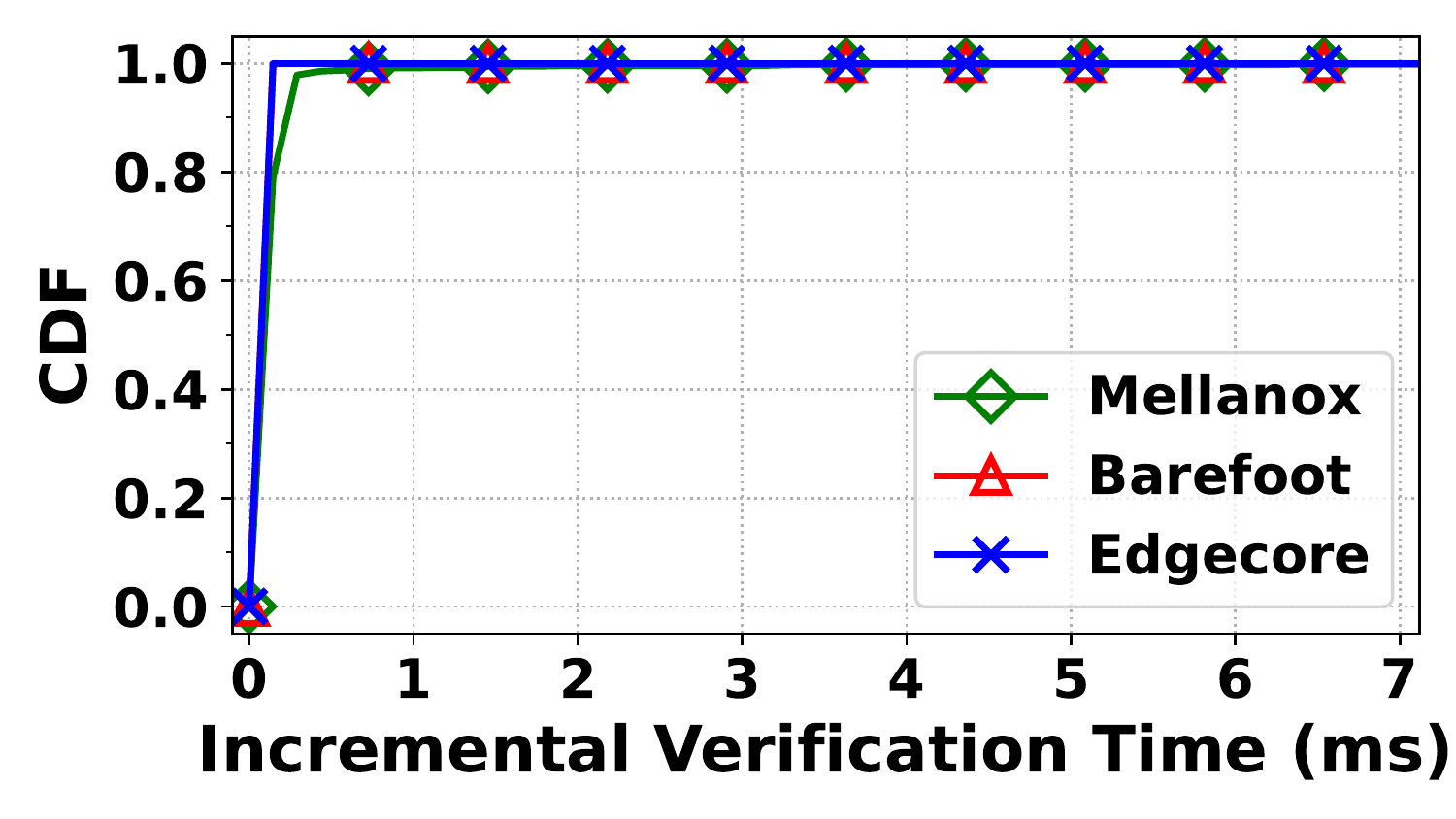}
		\caption{\label{fig:trace-message-fb-1000} \lnet{}.}
	\end{subfigure}
	\caption{Time of verifying all-shortest-path availability
	in DC networks incrementally on commodity network devices.}
    \label{fig:trace-message-1000}
\end{figure*}

We further go beyond verifying these local contracts from a green start to
verifying them incrementally. To this end, for each DC network, we randomly
generate 1,000 rule updates across the three devices, and evaluate how fast the
\system{} on-device verifiers on commodity network devices can incrementally
verify their counting tasks. Results in Figure~\ref{fig:trace-message-1000} show
that the 90\% quantile of incremental verification time on each switch model is
$0.08ms$ in 48-ary Fattree, and $0.15ms$ in \lnet{}. 

From these results, we demonstrate that \system{} can efficiently verify the
local contracts of RCDC on commodity network switches, with low overhead.

\section{Overhead of \system{} Planner}
\label{sec:dvnet-overhead}
\begin{figure*}[]
\setlength{\abovecaptionskip}{0.cm}
\setlength{\belowcaptionskip}{-0.cm}
\centering
\includegraphics[width=0.8\linewidth]{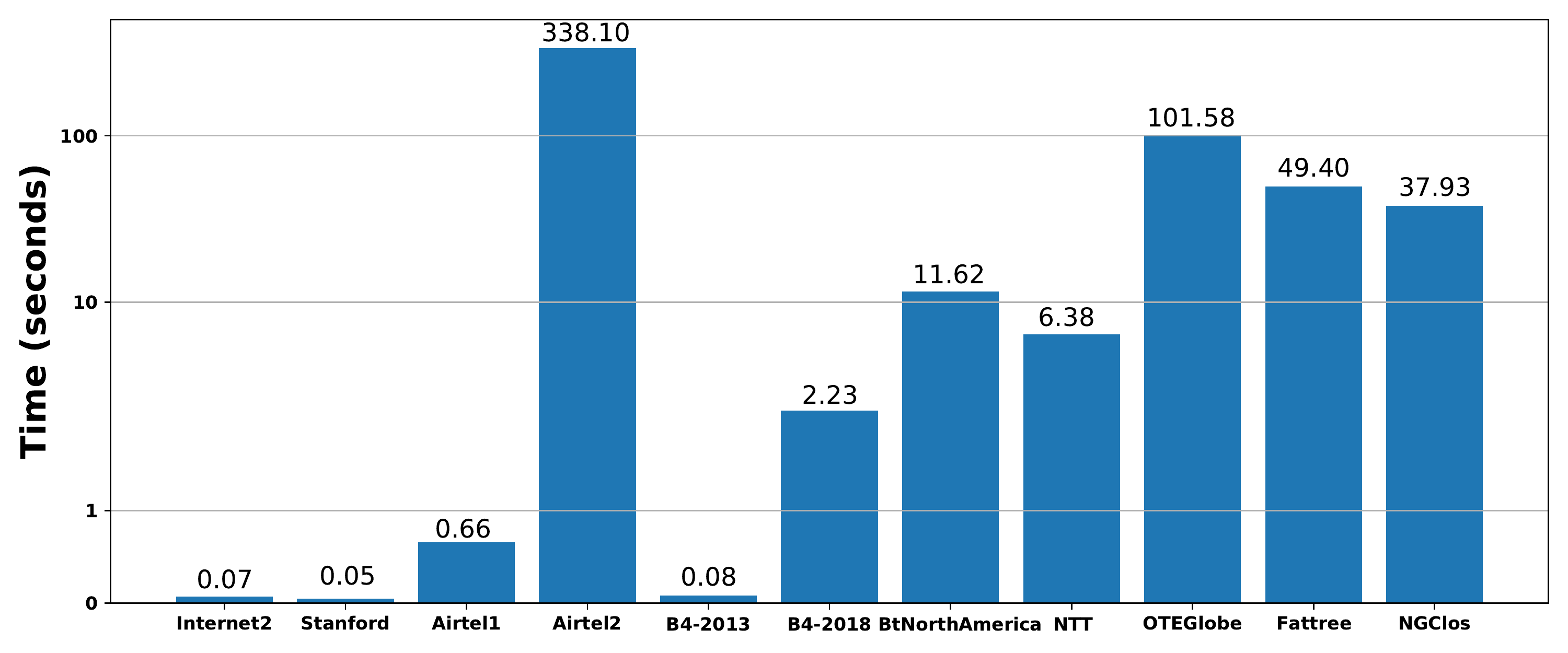}
\caption{Total time of \system{} planner in computing \dvnet{} and
	on-device tasks (seconds).}
\label{fig:dvnet-time}
%\reducespace
%\vspace{-1em}
\end{figure*}

Figure~\ref{fig:dvnet-time} shows the overhead of \system{} planner, in terms of
the total time to compute \dvnet{} and on-device tasks for each dataset
in our testbed experiments and simulations. Note that for the pair of datasets
Airtel1-1 and Airtel1-2, and the pair of Airtel2-1 and Airtel2-2, we only
generate \dvnet{} and on-device tasks once for each pair, because they each have the same
requirement and the topology.  It shows that for 9 out of 11
datasets in the evaluation, the planner finishes computing \dvnet{} and deciding
on-device tasks in less than $50s$, and that the longest time is $338.10s$. We
note that for Fattree and \lnet{}, the overhead of the planner ($49.40s$ and
$37.93s$) is even lower than some WAN networks. This is because our
implementation leverages the high symmetry of Clos-based topology to avoid
redundant computation, resulting in low overhead. Because the \system{}
planner is only needed to configure on-device verifiers before they run, we
conclude that its overhead is reasonably low and acceptable for scalable data
plane checking in various networks.

% \begin{table}[!htbp]
%     \centering
% \footnotesize
%     \begin{tabular}{|c|c|}
% 	\hline
%     \textbf{Internet2} & 0.07 \\
%     \hline
% 	\textbf{Stanford} & 0.05\\
%     \hline
%     \textbf{Airtel1} & 0.66\\
%     \hline
%     \textbf{Airtel2} & 338.10\\
%     \hline
%     \textbf{B4-2013} & 0.08\\
%     \hline
%     \textbf{B4-2018} & 2.23\\
%     \hline
%     \textbf{BT North America} & 11.62\\
%     \hline
%     \textbf{NTT} & 6.38\\
%     \hline
%     \textbf{OTEGlobe} & 101.58\\
%     \hline
%     \textbf{Fattree ($k=48$)} & 49.40\\
%     \hline
%     \textbf{LONet} & 37.93\\
%     \hline
%     \end{tabular}
%     \caption{Time (seconds)}
%     \label{tab:dvnet-time}
% \end{table}

\end{document}